\pdfoutput=1
\documentclass[10pt,aps,prd,twocolumn,amsmath,amssymb,floatfix,nofootinbib,preprintnumbers]{revtex4-1}

\usepackage{amsmath,amssymb,latexsym,times,subfigure,bm}

\usepackage[T1]{fontenc}
\usepackage{aecompl}
\usepackage{graphicx}

\newcommand{\aap}{A\&A}

\newcommand{\aj}{A.J.}
\newcommand{\mnras}{MNRAS}

\newcommand{\jcap}{JCAP}

\newcommand{\apjl}{ApJ}
\newcommand{\pasp}{PASP}
\newcommand{\pasj}{PASJ}
\newcommand{\procspie}{SPIE}

\usepackage{verbatim}
\usepackage[usenames,dvipsnames]{xcolor}
\usepackage{tabulary}
\usepackage{tabularx}
\usepackage{todonotes}
\usepackage{hyperref}

\begin{document}

\title[Redshift Uncertainty in CMB Lensing $\times$ Galaxies]{Effects of Redshift Uncertainty on Cross-Correlations of CMB Lensing and Galaxy Surveys}

\author{Ross Cawthon}
 \altaffiliation[Now at ]{Department of Physics, University of Wisconsin-Madison, 1150 University Ave, Madison, WI, 53706, USA.}
 \email{cawthon@wisc.edu}
\affiliation{%
 Department of Astronomy and Astrophysics, The University of Chicago, 5640 S. Ellis Ave, Chicago, IL, 60637, USA\\
 Kavli Institute for Cosmological Physics, The University of Chicago, Chicago, IL, 60637, USA
}%




\date{\today}

\begin{abstract}
We explore the effects of incorporating redshift uncertainty into measurements of galaxy clustering and cross-correlations of galaxy positions and cosmic microwave background (CMB) lensing maps. We use a simple Gaussian model for a redshift distribution in a redshift bin with two parameters: the mean, $z_0$, and the width, $\sigma_{\text{z}}$. We vary these parameters, as well as a galaxy bias parameter, $b_{\text{g}}$, and a matter fluctuations parameter, $\sigma_8$, for each redshift bin, as well as the parameter $\Omega_{\text{m}}$, in a Fisher analysis across 12 redshift bins from $z=0-7$. We find that incorporating redshift uncertainties degrades constraints on $\sigma_8(z)$ in the Large Synoptic Survey Telescope (LSST)/CMB-S4 era by about a factor of 10 compared to the case of perfect redshift knowledge. In our fiducial analysis of LSST/CMB-S4 including redshift uncertainties, we project constraints on $\sigma_8(z)$ for $z<3$ of less than $5 \%$. Galaxy imaging surveys are expected to have priors on redshift parameters from photometric redshift algorithms and other methods. When adding priors with the expected precision for LSST redshift algorithms, the constraints on $\sigma_8(z)$ can be improved by a factor of 2-3 compared to the case of no prior information. We also find that `self-calibrated' constraints on the redshift parameters from just the autocorrelation and cross-correlation measurements (with no prior information) are competitive with photometric redshift techniques. In the LSST/CMB-S4 era, we find uncertainty on the redshift parameters ($z_0,\sigma_{\text{z}}$) to be below 0.004(1+z) at $z<1$. For all parameters, constraints improve significantly if smaller scales can be used. We also project constraints for nearer term survey combinations, Dark Energy Survey (DES)/SPT-SZ, DES/SPT-3G, and LSST/SPT-3G, and analyze how our constraints depend on a variety of parameter and model choices.
\end{abstract}

\pacs{Valid PACS appear here}
\maketitle


\section{Introduction}
\label{sec:intro}

Large galaxy imaging surveys provide a wealth of cosmological information about the Universe. In particular, these surveys can probe the growth of structure across cosmic time. Such measurements can distinguish between different models for the mechanism causing cosmic acceleration \citep{huterer2015}. Two specific probes used by galaxy surveys to study structure growth are galaxy clustering and weak gravitational lensing. Recent and ongoing imaging surveys using these probes include the Dark Energy Survey (DES, \cite{DES}), the Kilo-Degree Survey (KIDS, \cite{kids}), the Canada-France-Hawaii Telescope Lensing Survey (CFHTLens, \cite{2012MNRAS.427..146H}) and  the Hyper-Suprime Cam survey (HSC, \cite{2012SPIE.8446E..0ZM}). The Dark Energy Survey recently produced the most comprehensive study of the growth of structure from an imaging survey \citep{keypaper} using galaxy clustering and weak lensing measurements from its first year of data (\cite{elvinpoole17}, \cite{troxel17}, \cite{prat17}). The DES Data Release 1 includes more than 300 million galaxies from the first three years of data  \citep{dr1}. In the next decade, the constraining power of imaging surveys will increase greatly when new ground-based surveys such as the Large Synoptic Survey Telescope (LSST, \cite{lsstdesc}) and space-based surveys such as Euclid \citep{euclid} and the Wide-Field Infrared Survey Telescope (WFIRST, \cite{wfirst}) begin operations. These future surveys will find significantly more galaxies and cover a much larger redshift range than current surveys. The LSST is expected to find on the order of several billion galaxies \citep{izeviclsst}.

A special case of using gravitational lensing to infer the structure of matter in the Universe is lensing of the cosmic microwave background (CMB). The CMB is made up of photons that have been free streaming since redshift $z \approx 1100$ (see e.g., \cite{2006astro.ph..4069T}). CMB lensing thus measures lensing from matter over nearly the entire lifetime of the Universe, more than 13 billion years. The first detection of CMB lensing was found by doing a cross-correlation of radio galaxies from the National Radio Astronomy Observatory Very Large Array Sky Survey and CMB data from the Wilkinson Microwave Anisotropy Probe (WMAP) \citep{2007PhRvD..76d3510S}. CMB lensing has since been detected in a number of ways including CMB-only methods and cross-correlations with several tracers of large-scale structure, including the cosmic infrared background (CIB), quasars, clusters, and galaxies detected in a number of different wavelengths (see \cite{giannantonio16} or \cite{omori18} for an extensive list). 

The cross-correlation of galaxy positions and CMB lensing is a particularly useful measurement of cosmic structure. While CMB lensing maps are impacted by matter back to $z \approx 1100$, they have the disadvantage of having no way to directly assess the redshift distribution of lenses at any particular location in the sky. All the information back to $z \approx 1100$ is stacked into one two-dimensional projection. Galaxies, having redshift measurements, provide a three-dimensional estimate of a location of matter. However, galaxy clustering alone suffers from the fact that galaxies do not directly trace the total underlying distribution of matter in the Universe, but instead are biased tracers. In galaxy clustering measurements, this galaxy bias (the relationship between the distribution of galaxies and total matter) is degenerate with the overall clumpiness of the Universe (i.e., $\sigma_8$), which provides information on competing cosmological models. The cross-correlation of galaxies and CMB lensing provides both a measurement of matter as a function of redshift and a way to break the degeneracy of galaxy bias and matter clumpiness. The cross-correlation also has the advantage of having very different systematic effects present. Galaxy surveys (of usually optical or infrared light) and CMB experiments (in the microwave band) operate in a number of different ways, making correlated systematic effects in both surveys unlikely.

These cross-correlations of galaxy clustering and CMB lensing have been measured by a number of recent experiments (see \cite{peacock18} for a recent list). In particular, \cite{giannantonio16} and \cite{omori18} cross-correlated Dark Energy Survey galaxies in tomographic redshift bins up to $z=1.2$ and $z=0.9$, respectively, with CMB lensing maps from both the South Pole Telescope (SPT) \cite{spt} and the Planck Satellite \cite{2006astro.ph..4069T}. Among current measurements, these analyses using a large optical cosmic survey out to high redshifts ($z \sim 1$) most closely mimic the type of measurements we will address in this work.  Recently a projection of the constraining power of a future measurement using LSST and the planned experiment, CMB-S4 \citep{2015APh....63...66A} was made by \cite{SS17}. However, a critical element that many of these studies do not incorporate in detail are the effects of redshift uncertainties on these measurements (though \cite{modi17} and \cite{SS17} briefly explore the issue).

While there are spectroscopic galaxy surveys (e.g., BOSS \cite{boss} and, in the future, Dark Energy Spectroscopic Instrument (DESI) \cite{desi13}), many of the best cosmological constraints (e.g., DES \cite{keypaper}) from galaxy clustering and gravitational lensing come from larger, deeper imaging surveys which suffer the downside of having only photometric redshifts from color bands. Much work goes into training these photometric redshift codes to be as accurate as possible by using spectroscopic training sets of galaxies (e.g., \cite{hoyle18des}, \cite{bonnett2016} and references therein). The method of spatially cross-correlating photometric galaxies with smaller samples of spectroscopic galaxies to infer redshift distributions (also known as `clustering redshifts') has also seen success (e.g., \cite{newman08}, \cite{cawthon17}, \cite{davis17}, \cite{gatti18} and references therein). However, even future photometric surveys like LSST expect significant uncertainty in their redshift distributions due to photometric redshift errors. Since LSST will probe higher redshifts than current surveys like the Dark Energy Survey, the issues surrounding photometric redshifts are likely to be compounded. Both the typical photometric training methods and the clustering method need spectroscopic galaxies at the same redshifts probed by the photometric survey. The photometric methods also need spectroscopic samples of galaxies with similar magnitude depth for training. Both getting the necessary number of spectroscopic measurements of galaxy redshifts and ensuring that current methods are sufficiently accurate at higher redshifts will be significant challenges.

Another interesting method to infer redshifts that has emerged is the idea of `self-calibrating' the redshift measurements from cosmological correlation functions themselves (e.g., galaxy clustering, weak lensing measurements etc.) Work by \cite{hoyle18} recently explored this idea with several types of correlation functions while holding cosmology fixed. Such methods may be needed in the future to supplement the current methods of photometric redshift calibration.

In this work, we project cosmological constraints from measurements of galaxy clustering and cross-correlations between galaxy positions and CMB lensing for current and future surveys. We use a Fisher analysis similar to that in \cite{SS17}. Unlike previous work, though, we will include redshift parameters in the Fisher analysis and highlight their impacts. Our redshift analysis will not focus on catastrophic outliers (as in, e.g., \cite{SS17}) but on the generic uncertainties of a redshift distribution for a photometric survey, represented by the mean and width of a Gaussian in each redshift bin. We show that these general (noncatastrophic) uncertainties have a significant impact on a cosmological analysis.

There are two main objectives of this work: 1. to assess how redshift uncertainties affect the expected cosmological constraints from galaxy survey and CMB lensing cross-correlations (i.e., an extension of \cite{SS17}) and 2. to assess how well the self-calibrating approach can constrain redshift distributions when cosmological parameters are allowed to vary (i.e., an extension of \cite{hoyle18}).

We focus on the cross-correlation of galaxy clustering and CMB lensing, though we note similar questions could be asked when including optical weak gravitational lensing data (i.e., cosmic shear), which can also be cross-correlated with CMB lensing (for past measurements see \cite{baxter16}, \cite{kirk16}, \cite{omori182}). As an example, \cite{schaan17} uses galaxy clustering, CMB lensing, and cosmic shear measurements together to self-calibrate the shear multiplicative bias. Since our goals are focused around the question of redshifts in galaxy surveys, we choose to focus on just galaxy clustering and CMB lensing and not focus on the interplay of shear multiplicative bias, redshifts, and other parameters (see \cite{schaan17} for a brief discussion).  For cosmological constraints, we focus on $\sigma_8$ as the main parameter that can be studied with these probes. Focusing on this parameter allows us to carefully study the impact of redshift uncertainties. Future work may incorporate a larger parameter space and set of measurements.

The setup of this paper is as follows. In Section \ref{sec:datasets}, we discuss the datasets used in this paper and their projected parameters. In Section \ref{sec:makingdndz}, we discuss how we model and parametrize redshift distributions when accounting for photometric redshift errors. In Section \ref{sec:methods}, we outline the projected power spectra measurements used in this work, and the Fisher matrix formalism we use to project constraints on cosmological and redshift parameters. In Section \ref{sec:nouncertainty}, we show Fisher constraints from an analysis without redshift uncertainties. In Section \ref{sec:fiducial}, we show our fiducial Fisher analysis incorporating redshift uncertainties. In Section \ref{sec:surveyparams}, we explore in detail how our constraints depend on various survey parameters, including priors on the redshift parameters.  In Section \ref{sec:selfcalibration}, we explore in more detail how successful our analysis is in constraining the redshift parameters. In Section \ref{sec:altmodels}, we investigate how changing the cosmological parameters we vary, including using a single $\sigma_8$ parameter across all redshifts, alters our constraints. In Section \ref{sec:conclusions}, we give our conclusions. 

\section{Datasets}
\label{sec:datasets}

\subsection{DES}

\begin{figure}
\begin{center}
\includegraphics[width=0.5 \textwidth]{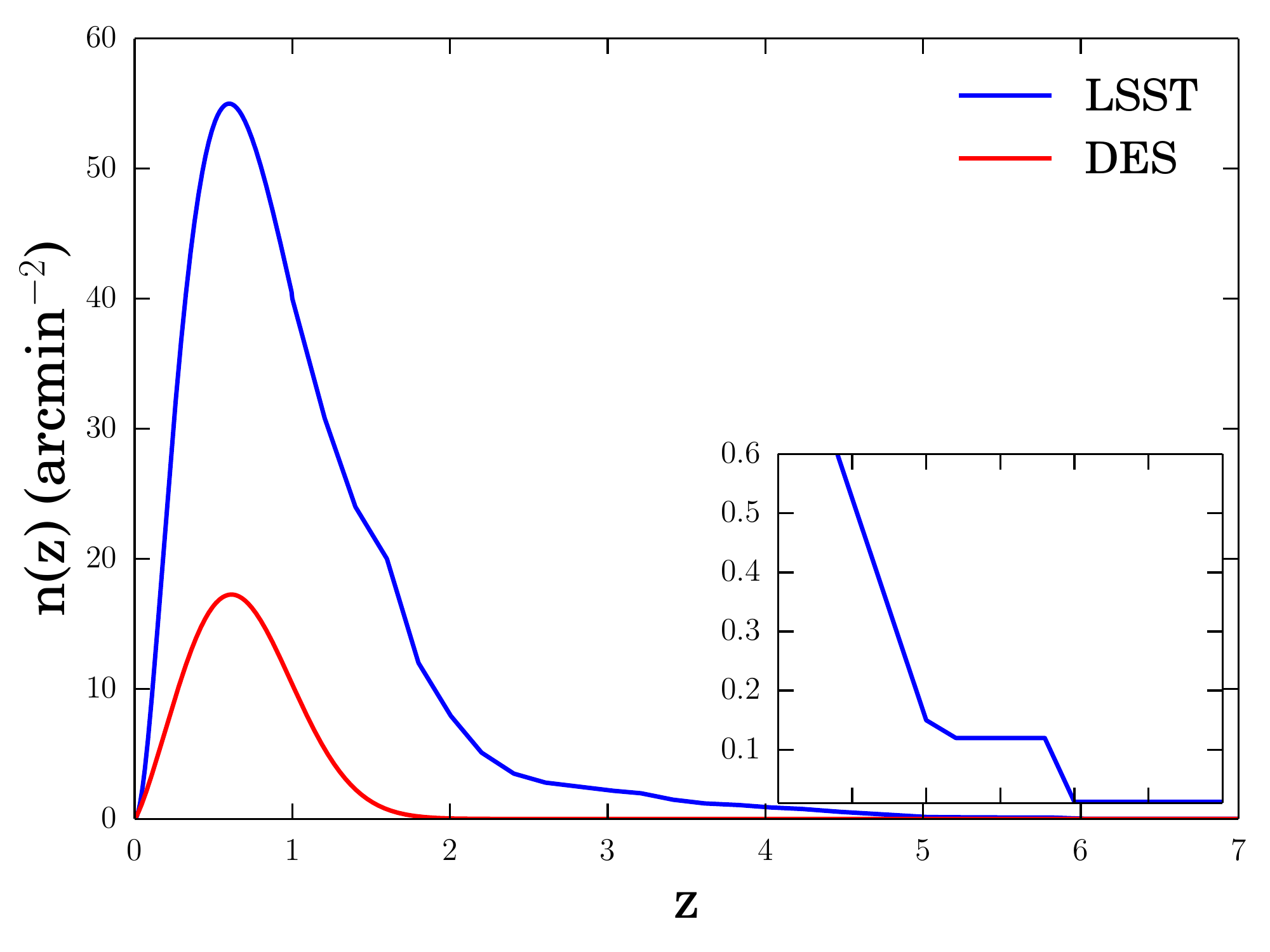}
\end{center}
\caption{The expected galaxy redshift distributions from DES (full six years of data) and LSST (first three years of data) used in this work. The redshift distributions come from \cite{nzsource} for DES and \cite{SS17} for LSST. The inset shows $4<z<7$.}
\label{fig:nzdes}
\end{figure}

The Dark Energy Survey is a six-year photometric survey covering $5000 \ \text{deg}^2$ in the $g,r,i,z, \text{and} \ y$ bands \citep{DES} which recently completed observations. The DES observed from the Blanco Telescope at the Cerro Tololo Inter-American Observatory (CTIO) in Chile. We assume a galaxy distribution for DES from \cite{nzsource} which gives

\begin{equation}
n(z) \propto (z/z_{\star})^{\alpha} \ \text{exp}[-(z/z_{\star})^{\beta}],
\label{dndzbeta}
\end{equation}

\noindent where for DES the parameters are $\alpha=1.25, \beta=2.29$, and $z_{\star}=0.88$ with the total number of galaxies having a density of $12 \ \text{arcmin}^{-2}$. This redshift distribution is shown in Figure \ref{fig:nzdes}. The full DES will cover $5000 \ \text{deg}^2$; however, the SPT, the CMB experiment that will be used for DES cross-correlations in our projections, only covers $2500 \ \text{deg}^2$, making the observed fraction of the sky $f_{\text{sky}}=0.0606$ for the power spectra in our analysis.

\subsection{LSST}
The Large Synoptic Survey Telescope is a ten-year photometric survey based at Cerro Pach\'{o}n in Chile. It is expected to start main operations in 2022. Its main Wide-Fast-Deep survey will cover $18,000 \ \text{deg}^2$ ($f_{\text{sky}}=0.45$) \citep{izeviclsst}. However, for our fiducial analysis, we will use $f_{\text{sky}}=0.5$ to more easily compare with the results in \cite{SS17}. For the galaxy distribution in LSST, we match to the prediction used in \cite{SS17} (their Figure 4) for galaxies with $i$ magnitude less than 27 after 3 years of data, shown in our Figure \ref{fig:nzdes}. When using $f_{\text{sky}}=0.5$, this prediction gives a total galaxy density of $58 \ \text{arcmin}^{-2}$.  The prediction comes from LSST simulations in \cite{gorecki14} for $0<z<4$. We note that for z<1 this $n(z)$ closely matches the LSST power law prediction from \cite{nzsource}. \cite{SS17} also adds galaxies for $4<z<7$ by extrapolating from recent results from the Subaru Hyper-Suprime Cam GOLDRUSH program \citep{goldrush} which found more than half a million candidates for $4<z<7$ galaxies based on the dropout technique \citep{steidel92}. Specifically, \cite{SS17} models the $n(z)$ from $z=4-5$ by extrapolating from the $z<4$ results of the simulations and assumes a constant number density of $0.14$ arcmin$^{2}$ from $z=5-6$ and $0.014$ arcmin$^{-2}$ for $z=6-7$. The $\ n(z)$ prediction for $z=4-7$ is noted by \cite{SS17} to perhaps be conservative, as a direct extrapolation of the limited (100 deg$^2$) GOLDRUSH results would give a factor of 2 more galaxies in this redshift range.

\subsection{SPT SZ Survey}
\begin{figure}
\begin{center}
\includegraphics[width=0.5 \textwidth]{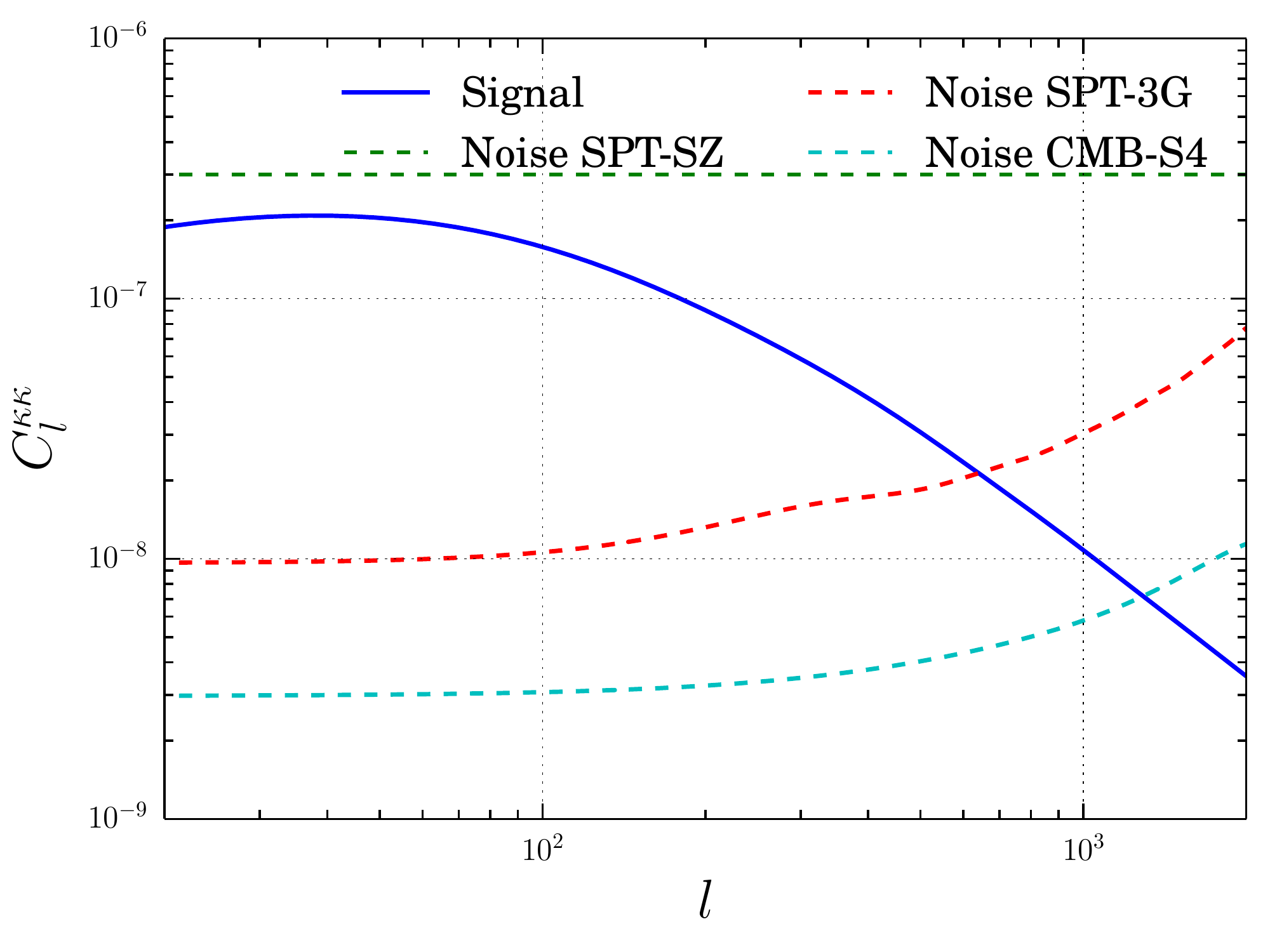}
\end{center}
\caption{The CMB lensing noise for the experiments we consider, as well as the signal of the CMB lensing autopower spectrum, $C_l^{\kappa \kappa}$. We use SPT noise estimates from \cite{giannantonio16} and the CMB-S4 estimate from \cite{SS17}.  These noise estimates enter our analysis in Equation \ref{covariance}.}
\label{fig:cmbnoise}
\end{figure}

The SPT is a 10 meter millimeter wave, wide-field telescope at the Amundsen-Scott South Pole station in Antarctica \citep{spt}. The $2500 \ \text{deg}^2$ SPT-SZ (Sunyaev-Zel'dovich) survey is described in \cite{story13}. A CMB lensing map from this survey was made in \cite{vanengelen12}. More recently, \cite{omori17} made a map covering the full survey, while also including data from the Planck Satellite \citep{planckhfi}. The lensing maps are made using the quadratic estimator technique \citep{okamotohu}. The lensing maps from SPT-SZ are made from measurements in the 150 GHz band. In this band, the temperature maps have a typical noise of $ \Delta_T=18 \mu \text{K arcmin}$.  For the expected CMB lensing noise in the autopower spectrum (i.e., $N_l^{\kappa \kappa}$), we use the noise measurement in \cite{giannantonio16}, which used a version of the maps made in \cite{vanengelen12}. The measured lensing noise of the maps in \cite{omori17} are very similar. The lensing noise for SPT-SZ as well as the projected noise for the following two experiments, SPT-3G and CMB-S4, are shown in Figure \ref{fig:cmbnoise}.

\subsection{SPT 3G Survey}
The SPT-3G survey \citep{2014SPIE.9153E..1PB} is currently in progress and is the third-generation survey on the South Pole Telescope, following the SPT-SZ survey, and the SPT-Pol survey \cite{sptpol}. We will not discuss the SPT-Pol survey due to its smaller sky coverage than SPT-SZ or SPT-3G. SPT-3G has an improved optical design allowing for more pixels in the optical plane, and uses multichroic pixels as described in \cite{2014SPIE.9153E..1PB}. These improvements should lower the temperature noise by roughly a factor of 10 compared to SPT-SZ. Like SPT-Pol, SPT-3G will also have polarization measurements. It will cover the full $2500 \ \text{deg}^2$ which was observed by SPT-SZ. For the projection of SPT-3G noise, we use an estimate by the South Pole Telescope team using a minimum-variance estimator combining T, E, and B mode measurements. This estimate is shown in \cite{giannantonio16}. We show this projected noise in Figure \ref{fig:cmbnoise}.

\subsection{CMB-S4}
The CMB-S4 experiment \citep{cmbs4} is a next-generation CMB survey expected to begin within the next decade. It is likely to have operations in both Antarctica and Chile. The sky coverage is still uncertain, though many projections have CMB-S4 covering half the sky, completely overlapping LSST. We will assume this for our fiducial analysis, giving $f_{\text{sky}}=0.5$. For the CMB lensing noise, we use the estimate in \cite{SS17} and show this in Figure \ref{fig:cmbnoise}. This estimate assumes $\Delta_{T}=1 \mu \text{K arcmin}$ noise and a minimum variance combination of multiple lensing estimators from the T, E and B mode measurements of a CMB experiment \citep{cmbs4}.

\section{Parametrizing Redshift Distributions}
\label{sec:makingdndz}

A focus of this work is to study the effects of redshift uncertainty on cosmological projections using galaxy and CMB lensing surveys. With this in mind, the observed galaxy distributions in a photometric survey like DES or LSST will never quite look like the redshift distributions mentioned in Section \ref{sec:datasets}. In a typical photometric survey, galaxies are binned by photometric redshift. High-density, faint samples of galaxies (such as the predicted distribution of galaxies with an $i$ magnitude less than 27 for LSST in Figure \ref{fig:nzdes}) typically have photometric redshift errors consistent with a Gaussian scatter. For example, LSST predicts photometric redshifts with a scatter of $\sigma_{\text{ph}}=0.05 (1+z)$ around the true redshift (\cite{lsstsciencebook}, \cite{izeviclsst}).

\begin{figure}
\begin{center}
\includegraphics[width=0.5 \textwidth]{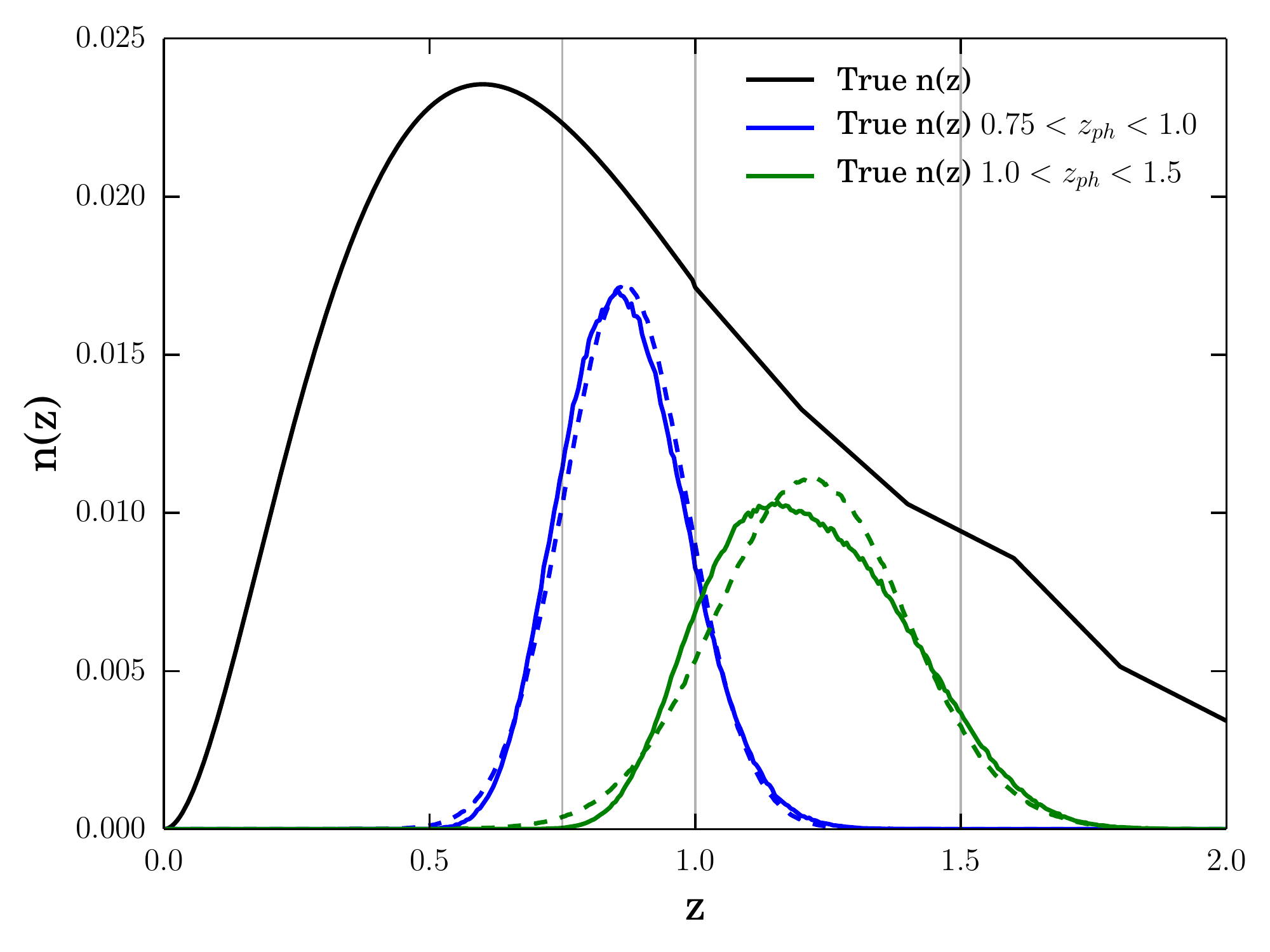}
\end{center}
\caption{The true redshift distribution and examples of photometric redshift bins we use. Shown in black is the true $n(z)$ for LSST, though it actually extends out to $z=7$. In blue and green are examples of the $n(z)$ in photometrically selected redshift bins. As seen, e.g., a bin with photometric cutoffs of $z=0.75$ and $z=1.0$ (blue) will have its true distribution extend beyond those boundaries (lined in gray). The modeled true distributions of the binned galaxies are close to being Gaussians. The dotted lines show Gaussians with the same mean and standard deviation of the true binned distributions.}
\label{fig:makingnz}
\end{figure}

To simulate what a photometrically selected and binned redshift distribution looks like, we first take the expected $n(z)$ from the references in Section \ref{sec:datasets}. We then draw galaxies from this distribution and assign them photometric redshifts, assuming the photometric redshift errors follow $\sigma_{\text{ph}}=0.05 (1+z)$, with no bias (i.e. $\bar{z}_{\text{true}}=\bar{z}_{\text{ph}})$. We then simulate what would be done for a real survey and bin the galaxies by $z_{\text{ph}}$. As can be seen in Figure \ref{fig:makingnz}, the true redshift distribution from LSST (the sum of $z_{\text{true}}$, not the sum of $z_{\text{ph}}$) after binning by photometric redshifts is nearly Gaussian in shape. To further show this, in Figure \ref{fig:makingnz} we also plot a Gaussian with the mean redshift and standard deviation of the redshifts in the binned n(z). We emphasize that Figure \ref{fig:makingnz} shows only true redshift distributions and does not mimic what a photometric redshift code would predict. This can be seen for example in the right binned sample (green) which extends beyond the photometric borders of the redshift bin, $z=1.0$ and $z=1.5$. 

In current surveys, photometric binning often produces Gaussian-like true redshift distributions in each bin (e.g., \cite{keypaper}) similar to Figure \ref{fig:makingnz}. These true redshift distributions are verified to some degree by testing photometric redshift codes on samples of galaxies with spectroscopic redshifts (e.g., \cite{hoyle18des}) or using other methods like the cross-correlations of photometric and spectroscopic galaxies to recover the redshift distribution of the photometric set (clustering redshifts, e.g., \cite{cawthon17}, \cite{davis17}). However, each of these methods has uncertainties. Exact knowledge of the redshift distribution for a photometric survey is unlikely.

Given the typical case of a Gaussian-like true redshift distribution when binning by photometric redshifts, we parametrize the redshift distributions in our main analysis (Section \ref{sec:fiducial}) with Gaussians of mean $z_0$ and width $\sigma_{\text{z}}$. This makes the redshift distribution in a bin, $i$,

\begin{equation}
n(z)_i \propto \frac{1}{\sigma_{\text{z},i}} \text{exp}[-\frac{(z-z_{0,i})^2}{2 \sigma_{\text{z},i}^2}] .
\label{gaussiandndz}
\end{equation}

\begin{figure}
\begin{center}
\includegraphics[width=0.5 \textwidth]{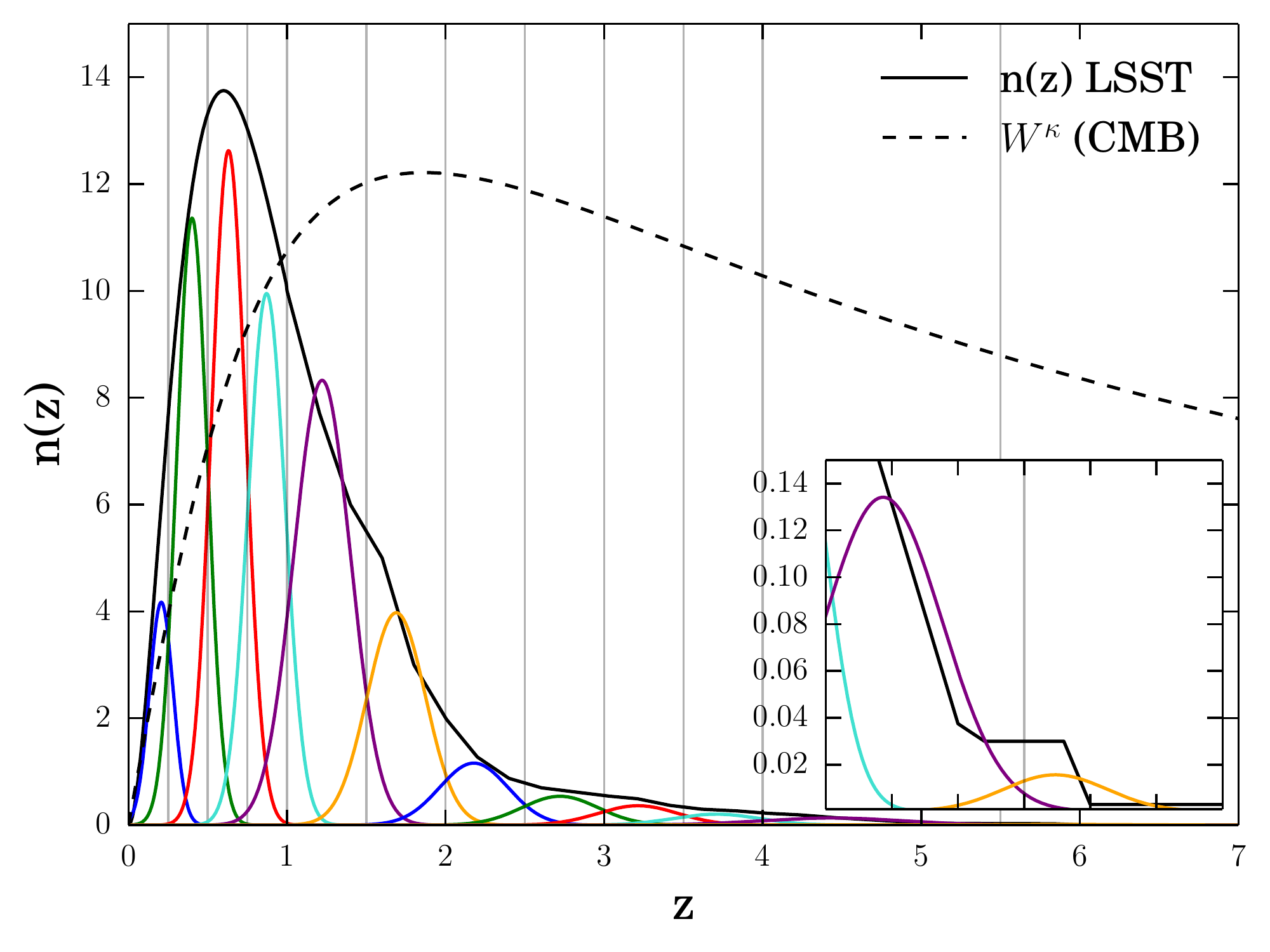}
\end{center}
\caption{The 12 Gaussian redshift distributions for our tomographic redshift bins which will be used in our analysis starting in Section \ref{sec:fiducial}. We list all of the means, $z_{0,i}$, and widths, $\sigma_{\text{z},{i}}$, of these bins in Appendix \ref{sec:appendix0}. Also shown is the full $n(z)$ prediction for LSST from which the Gaussian distributions are estimated in the manner described in Section \ref{sec:makingdndz} and Figure \ref{fig:makingnz}.  The inset shows $4<z<7$. We also show $W^{\kappa}$ from Equation \ref{wkappa}, which is the CMB lensing kernel. This parameter weights the redshifts that most efficiently lens the CMB. The curve for $W^{\kappa}$ is normalized to the full $n(z)$ curve.}
\label{fig:nz12}
\end{figure}

For our fiducial analysis beginning in Section \ref{sec:fiducial}, we use 12 tomographic redshift bins with a Gaussian redshift distribution in each bin. These redshift distributions are shown in Figure \ref{fig:nz12}, along with the full $n(z)$ prediction for LSST from \cite{SS17}. The parameters, $z_{0,i}$ and $\sigma_{\text{z},{i}}$ are shown in Appendix \ref{sec:appendix0}, Table \ref{table1} for both LSST and DES. Figure \ref{fig:nz12} also shows the CMB lensing kernel (described in Equation \ref{wkappa}) which shows what redshifts most efficiently lens the CMB. The lensing kernel peaks at about $z \approx 2$. In Section \ref{sec:fiducial} and later, we allow the parameters $z_{0,i}$ and $\sigma_{\text{z},i}$ of each of the Gaussians in Figure \ref{fig:nz12} to vary in our Fisher analysis (Section \ref{sec:fisher}). This gives a simple framework for accounting for redshift uncertainties in the Fisher analysis and should be accurate in the limit that the binned redshift distributions are Gaussian.

\section{Methods}
\label{sec:methods}
\subsection{Power Spectra}
The CMB lensing convergence, $\kappa$, in a given line of sight, $\hat{n}$, is the integral over all the matter fluctuations that will cause gravitational lensing,

\begin{equation}
\kappa (\hat{n}) =\int dz W^{\kappa} (z) \delta (\chi(z) \hat{n}, z),
\label{kappa}
\end{equation}

\noindent where $\delta (\chi(z) \hat{n}, z)$ is the overdensity of matter at comoving distance, $\chi$, and redshift, $z$. The distance kernel, $W^{\kappa}$, is given by

\begin{equation}
W^{\kappa}(z)=\frac{3}{2} \Omega_{\text{m}} H_0^{2} \frac{(1+z)}{H(z)} \frac{\chi(z)}{c} [\frac{\chi(z_\text{cmb}) - \chi(z)}{\chi(z_{\text{cmb}})}],
\label{wkappa}
\end{equation}

\noindent where $\Omega_{\text{m}}$ is the fraction of the matter density today compared to the present critical density of the Universe, $H_0$ is the Hubble parameter today, $H(z)$ is the Hubble parameter as a function of redshift, $c$ is the speed of light, and $\chi_{\text{cmb}}$ is the comoving distance to the surface of last scattering where the CMB was emitted \citep{2012ApJ...753L...9B}.

As galaxies are expected to be biased tracers of matter fluctuations, the galaxy overdensity in a given line of sight is

\begin{equation}
g (\hat{n}) =\int dz W^{g} (z)  \delta (\chi(z) \hat{n}, z).
\label{gals}
\end{equation}

\noindent The kernel, $W^{g}$, is given by

\begin{equation}
W^g(z)=b_{\text{g}}(z) \frac{1}{n_{\text{tot}}} \frac{dn(z)}{dz},
\label{wg}
\end{equation}

\noindent where $b_{\text{g}}(z)$ is the galaxy bias, the ratio of the overdensity of galaxies to the overdensity of matter, assumed here to be independent of scale; $n_{\text{tot}}$ is the total number of galaxies in the sample; and $\frac{dn(z)}{dz}$ is the redshift distribution of those galaxies.

At small angular scales, we can use the Limber approximation (\cite{1953ApJ...117..134L}, \cite{1992ApJ...388..272K}, see Appendix \ref{sec:appendix2}) to write the cross-power spectrum of two of our fields, $i$ and $j$, where $i,j \in {\kappa_{\text{cmb}}, g_{z=0-0.25},g_{z=0.25-0.5},...}$  at multipole $l$ as: 

\begin{equation}
C_l^{i j}=\int \frac{dz}{c} \frac{H(z)}{\chi(z)^2} W^{i}(z) W^{j}(z) P(k=\frac{l}{\chi(z)}, z)
\label{clequation}
\end{equation}

\noindent where $P(k=\frac{l}{\chi(z)}, z)$ is the matter power spectrum at wave number $k$ for a given redshift $z$. We calculate all of the power spectra using the Planck 2015 flat-$\Lambda \text{CDM}$ cosmological parameters including external data \citep{planck15}. These parameters are $h=0.6774$, $\Omega_{\text{m}}=0.3089$, $\Omega_{\text{b}}=0.04860$, $\tau=0.066$, $n_{\text{s}}=0.9667$, and $A_{\text{s}}=2.1413 \times 10^9$ at a pivot scale of $k=0.05 \ \text{Mpc}^{-1}$, corresponding to $\sigma_8(z=0)=0.8159$. The matter power spectrum, $P(k, z)$, is calculated using the Boltzmann code in the CAMB program (\cite{2012JCAP...04..027H}, \cite{2000ApJ...538..473L}) with the program Halofit (\cite{2003MNRAS.341.1311S}) to calculate the nonlinear regime of clustering.

The Gaussian covariances for the power spectra, $C_l$, are

\begin{equation}
\text{cov}(C_l^{i j}, C_l^{i' j'}) = \frac{\delta_{l l'}}{f_{\text{sky}}(2l+1)} (\hat{C}_l^{i i'} \hat{C}_l^{j j'} + \hat{C}_l^{i j'} \hat{C}_l^{j i'}),
\label{covariance}
\end{equation}

\noindent where the upper indices $i$ and $j$ again refer to the different fields. The power spectra denoted by $\hat{C}$ include noise,

\begin{equation}
\hat{C}_l = C_l (\text{theory}) + N_l,
\label{chat}
\end{equation}

\noindent where for galaxy autocorrelations, the shot noise term is $N_l=1/\rho$, where $\rho$ is the galaxy density per steradian, and for the CMB lensing autocorrelation, the predicted $N_l$ for different CMB experiments are shown in Figure \ref{fig:cmbnoise}. For cross-correlations, $N_l=0$. 

We note that Equation \ref{covariance} ignores the non-Gaussian corrections for galaxy clustering and CMB lensing covariance (see, e.g., \cite{krauseeifler} and \cite{motlochhulevy} for calculations of these terms). In \cite{giannantonio16}, the amplitude of non-Gaussian corrections to the covariance is estimated by comparing measurements on mock catalogs from an N-body simulation and mock catalogs from a Gaussian random realization of galaxy and CMB lensing fields. The different covariance estimates from these two tests had negligible impact on their amplitude parameter (similar to $\sigma_8$) constraints. This suggests that the non-Gaussian contributions of the covariance are minor for these probes at the scales they used, which were $l=30-2000$, nearly identical to scales we use.

\begin{figure}
\begin{center}
\includegraphics[width=0.5 \textwidth]{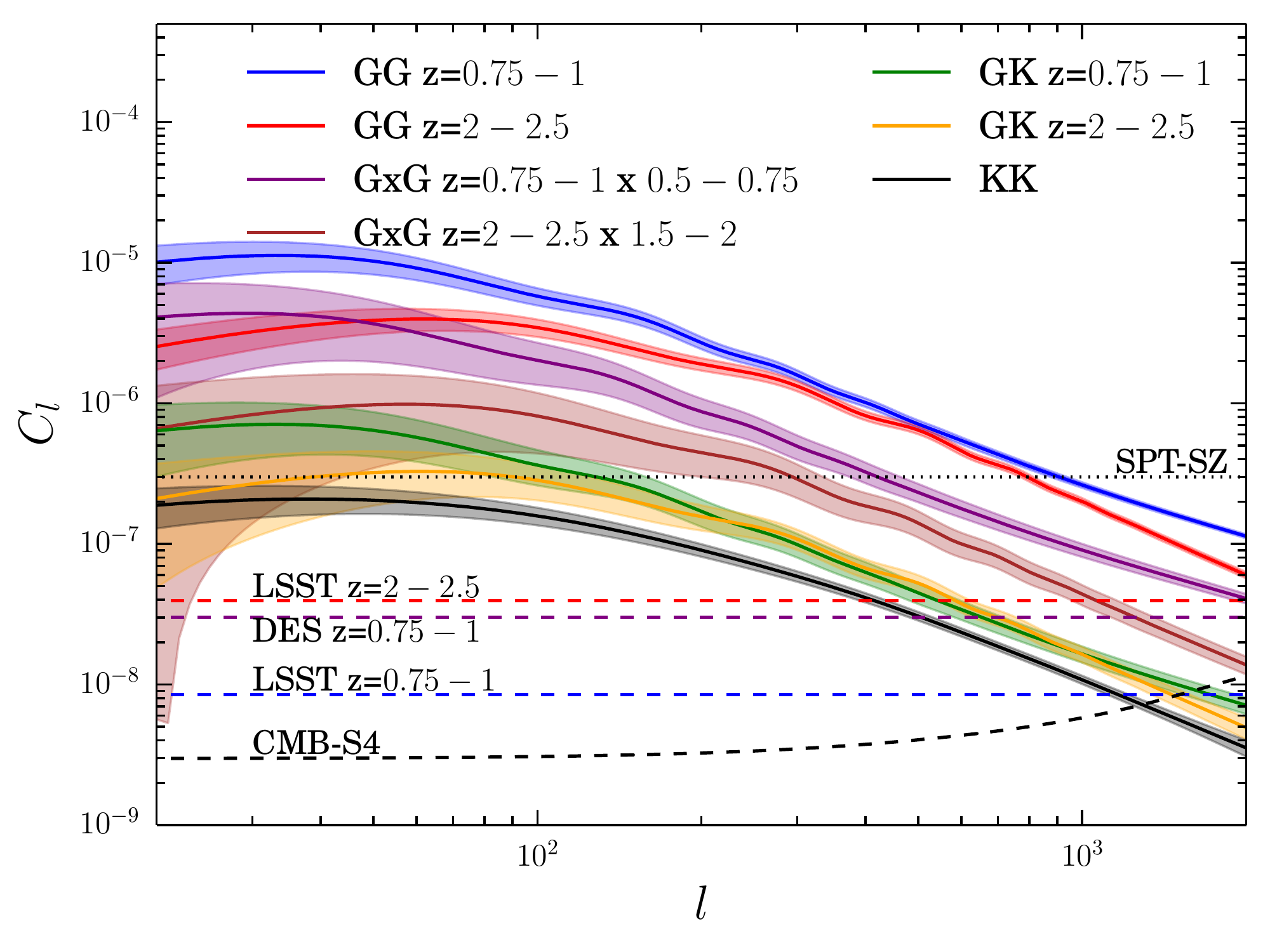}
\end{center}
\caption{Sample theoretical angular power spectra ($C_l$) for different redshift bins (Figure \ref{fig:nz12}) in our analysis. Shown are measurements for the bins with photometric redshifts between $z=0.75-1$ and $z=2-2.5$. These measurements include galaxy autocorrelations (GG), cross-correlations between adjacent galaxy redshift bins (GxG), and cross-correlations of galaxies and CMB lensing (GK). The amplitudes of the $C_l$ curves do not include the noise terms (Equation \ref{clequation}). However, we use noise levels for the LSST/CMB-S4 era in the error bands given by the covariance, $\text{cov}(C_l^{i j},C_l^{i j})$, in Equation \ref{covariance}. Also shown is the CMB lensing autocorrelation (KK); the noise levels ($N_l$) for the CMB experiments, SPT-SZ and CMB-S4; and the shot noise for the two galaxy bins in LSST and the $z=0.75-1$ bin in the DES. The noise levels are represented by dotted or dashed lines.}
\label{fig:allcl}
\end{figure}

We show some sample power spectra in Figure \ref{fig:allcl} for two of the 12 redshift bins used in the fiducial analysis (Figure \ref{fig:nz12}). Shown are galaxy autocorrelations, cross-correlations between galaxy bins, cross-correlations between galaxies and CMB lensing, and the CMB lensing autocorrelation. The error bands represent the covariance (Equation \ref{covariance}) estimates of the LSST/CMB-S4 era. Also shown are some of the relevant noise levels, $N_l$, for the different experiments. We can see that many more of the multipoles of the cross-correlation between galaxies and CMB lensing are signal dominated in the LSST/CMB-S4 era compared to the DES/SPT-SZ era.

\subsection{Fisher Matrix}
\label{sec:fisher}

We use a Fisher matrix formalism similar to \cite{SS17} (their Section VI) to derive constraints on parameters. The Fisher formalism assumes all the cosmological information is contained in the power spectra, which is true in the limit that the fields are Gaussian. In our fiducial analysis (Section \ref{sec:fiducial}), we have 12 tomographic redshift bins of galaxies (Figure \ref{fig:nz12}), and the CMB lensing field, $\kappa$. This gives us $N=13$ fields, which means there are 13 autospectra and $N(N-1)/2=78$ cross-spectra, for a total of 91 spectra. However, we assume the cross-spectra of non-neighboring redshift bins are zero. \footnote{We note that tests suggest our methodology would notably gain precision by using cross-correlations between non-neighboring redshift bins. However, we believe our methodology likely overestimates the information from such correlations. These correlations are completely sourced by the tails of the redshift distributions in Figure \ref{fig:nz12}. In our strict two-parameter Gaussian model for the redshift distribution in each bin, the tails correlate with $\sigma_z$ and are informative. In a real dataset, though, the Gaussian approximation will not be so accurate that information in the tails could tell you  much about the whole distribution (i.e., $\sigma_z$). Therefore, we believe assuming zero information from non-neighboring bin correlations is a more realistic model. We leave research on a more flexible method for utilizing the tails to future work.} This reduces the total number of nonzero spectra to $3(N-1)=36$.

Following \cite{SS17}, we define a large one-dimensional vector containing all the spectra:

\begin{equation}
\bm{d}=(\bm{d}_{l_{\text{min}}}, \bm{d}_{l_{\text{min+1}}},...,\bm{d}_{l_{\text{max}}}) .
\label{d}
\end{equation}

\noindent For each $l$,

\begin{equation}
\bm{d}_l=(C_l^{1 1}, C_l^{1 2}, ..., C_l^{N N} )
\label{dl}
\end{equation}

\noindent with N being the number of fields. Since $C_l^{i j} = C_l^{j i}$, $\bm{d}_l$ has $N(N+1)/2$ spectra, 91 spectra when $N=13$ fields, with only 36 of these being nonzero as mentioned previously. The Fisher matrix is then

\begin{equation}
F_{a b}=\sum\limits_{l=l_{\text{min}}}^{l_{\text{max}}} \frac{\partial \bm{d}_l}{\partial \theta_a}[\text{cov}(\bm{d}_l,\bm{d}_l)]^{-1} \frac{\partial \bm{d}_l}{\partial \theta_b}
\label{fisher}
\end{equation}

\noindent where $\theta$ is a parameter that depends on the measurements, $\bm{d}_l$, and $a,b$ index the parameters. For our fiducial analysis, we use $l_{\text{min}}=20$ and $l_{\text{max}}=1000$ (similarly as in \cite{SS17}), though we test other $l_{\text{max}}$ values. In all cases, we do not bin in $l$ in this work. This Fisher setup assumes that the fields overlap (i.e., the CMB and galaxy experiments overlap completely on the sky), which is the case in the projected experiments of Section \ref{sec:datasets}. The projected error on a parameter, $\theta$, is then

\begin{equation}
\sigma(\theta_a) \geq \sqrt[]{(F^{-1})_{a a}} .
\label{uncequation}
\end{equation}

In Section \ref{sec:priors}, we analyze the effects of adding priors to our analysis. We add priors by substituting

\begin{equation}
F_{a a} \rightarrow F_{a a} + \frac{1}{p(\theta_a)}, 
\label{priorequation}
\end{equation}

\noindent where $p(\theta_a)$ is the prior on the parameter. When applying priors, Equation \ref{priorequation} is applied before the Fisher matrix is inverted in Equation \ref{uncequation}.

In our fiducial analysis, there are five types of parameters varied. These include the redshift parameters, $z_{0,i}$ and $\sigma_{\text{z},i}$, defined in Equation \ref{gaussiandndz} for each of the redshift bins indexed by $i$. We also vary for each redshift bin, $b_{\text{g},i}$, the amplitude of the galaxy bias and $\sigma_{8,i}$, which measures the amplitude of the matter power spectrum on scales of $8 \ h^{-1}$ Mpc, where $h=H_0/(100 \ \text{km/sec/Mpc})$. We use parametrizations similar to those in \cite{SS17} for these latter two parameters. In Equation \ref{gals}, we model the galaxy bias, $b_{\text{g}}(z)$, as

\begin{equation}
b_{\text{g}}(z)=b_{\text{g},i} (1+z) .
\label{biasg}
\end{equation}

\noindent This formalism matches the general linear behavior of galaxy bias with $(1+z)$ often seen in flux-limited photometric galaxy samples (e.g., \cite{2016MNRAS.455.4301C}) while including uncertainty in a single amplitude parameter in each bin. We implement $\sigma_{8,i}$ into the power spectra (Equation \ref{clequation}) by substituting

\begin{equation}
P(k,z) \rightarrow (1+s_i)^2 P(k,z) 
\label{sigma8}
\end{equation}

\noindent where $s_i \equiv (\sigma_{8,i}/\sigma_{8,\text{fid}} -1)$ is the fractional difference of $\sigma_{8}$ in that bin compared to the fiducial cosmology. The Fisher analysis is centered on $b_{\text{g},i}=1$ and $\sigma_{8,i}=\sigma_{8,\text{fid}}$ for all redshift bins, $i$.  We note that when we calculate the CMB lensing autocorrelation, $C_l^{\kappa_{\text{cmb}} \kappa_{\text{cmb}}}$, we apply $s_i$ from the minimum and maximum photometric redshift boundaries for each bin $i$, though this does not map perfectly to the redshifts of the galaxies in bin $i$. \footnote{Since, unlike in \cite{SS17}, our redshift bins overlap, we must make a choice whether to tie the definition $\sigma_{8,i}$ to a specific redshift range or to a specific redshift binned sample. We choose the latter as that is how many photometric redshift binned samples are analyzed (e.g., \cite{giannantonio16}). However, this does make how to specifically calculate $\partial C_l^{\kappa_{\text{cmb}} \kappa_{\text{cmb}}}/ \partial \sigma_{8,i}$ ill defined since we are not defining $\sigma_{8,i}$ over a precise redshift range. In any case, the contributions of $C_l^{\kappa_{\text{cmb}} \kappa_{\text{cmb}}}$ are very minor in the analysis, so we do not think this impacts the results significantly.} 

The fifth type of parameter we allow to vary is the matter density of the Universe, $\Omega_{\text{m}}$. This parameter enters into $P(k,z)$, the calculation for $H(z)$, as well as in the CMB lensing kernel, $W^\kappa(z)$ (Equation \ref{wkappa}). When we vary $\Omega_{\text{m}}$, we also vary $\Omega_{\Lambda}$, the cosmological constant energy density in $\Lambda \text{CDM}$, to keep the Universe flat.

The parameters $z_{0,i},\sigma_{\text{z},i},b_{\text{g},i}$, and $\sigma_{8,i}$ are measured in each redshift bin. Along with $\Omega_{\text{m}}$, this gives a total of $n=4(N-1)+1$ parameters, which is 49 in the case of $N=13$ fields. The Fisher matrix will be $n \ \text{x} \ n$ in size.

\section{Results with No Redshift Uncertainty}
\label{sec:nouncertainty}
\begin{figure}
\begin{center}
\includegraphics[width=0.5 \textwidth]{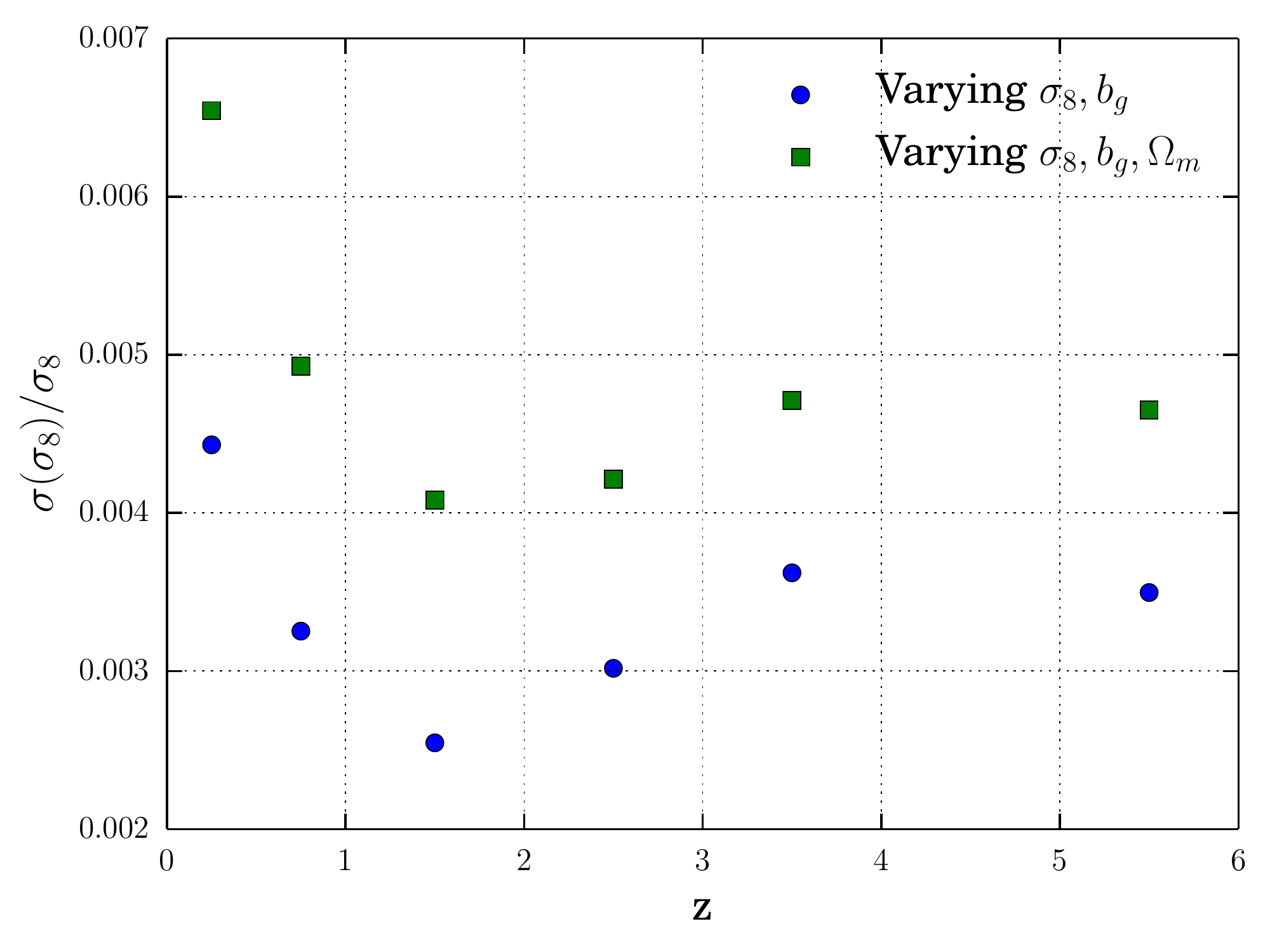}
\end{center}
\caption{The constraints on $\sigma_8$ in the scenario with no redshift uncertainty for the LSST/CMB-S4 era. We plot the case with only $\sigma_8$ and $b_{\text{g}}$  allowed to vary in each bin (to be able to compare with the analysis in \cite{SS17}) as well as the case with also $\Omega_{\text{m}}$ being allowed to vary. For this analysis, we set $l_{\text{max}}$=1000. }
\label{fig:SS_s8_6bin}
\end{figure}

\begin{figure}
\begin{center}
\includegraphics[width=0.5 \textwidth]{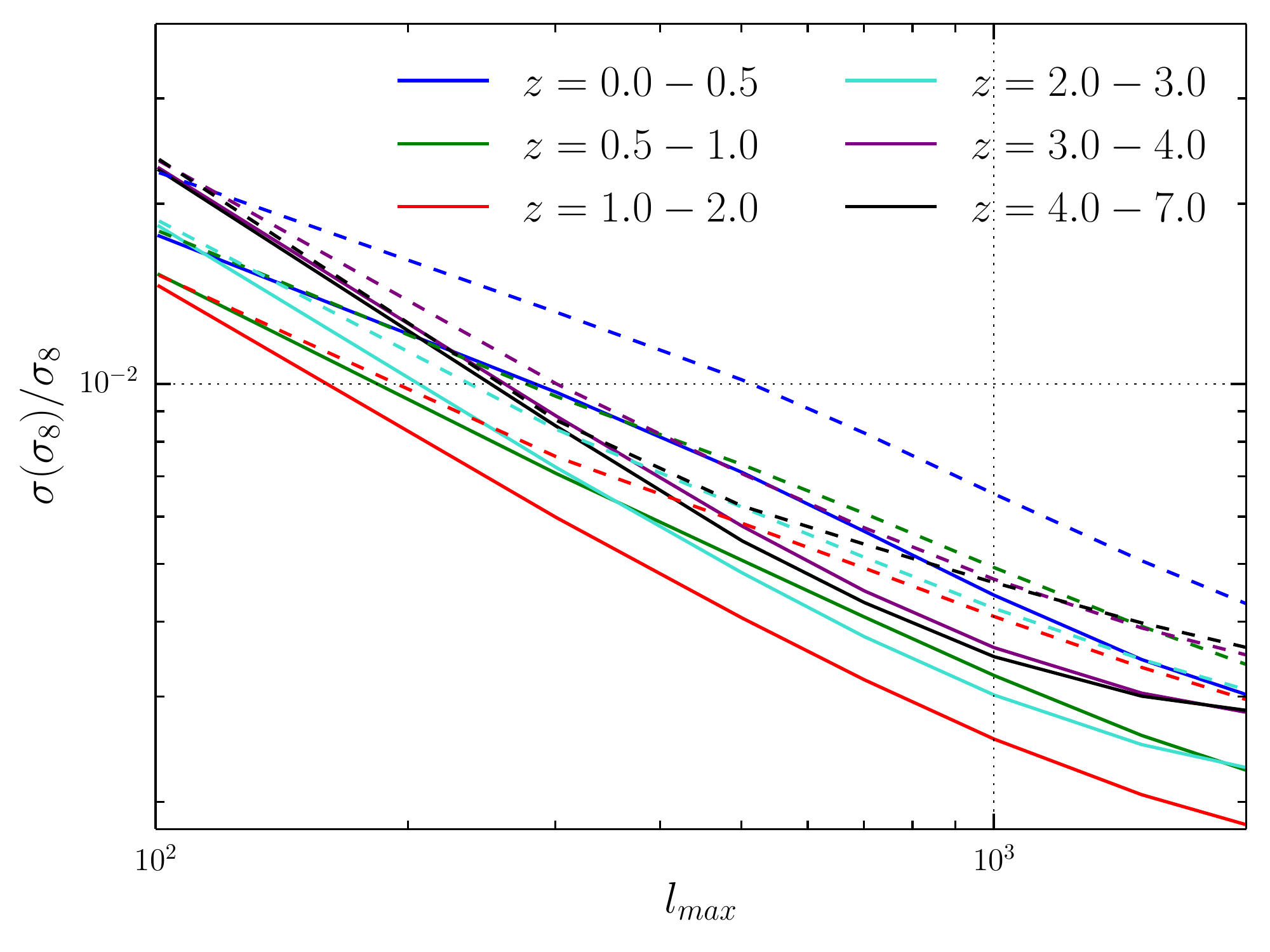}
\end{center}
\caption{Same analysis of $\sigma_8$ constraints while having no redshift uncertainty as in Figure \ref{fig:SS_s8_6bin}, but with varying $l_{\text{max}}$ in the LSST/CMB-S4 era. This allows a direct comparison with the analysis of \cite{SS17} (its Figure 9). Our results are very similar. The dotted lines are for the case in which we allow $\Omega_{\text{m}}$ to vary as well, while the solid lines are with only varying $\sigma_8$ and $b_{\text{g}}$.}
\label{fig:SS_s8_6bin_lmax}
\end{figure}

We first analyze the results of a Fisher matrix analysis when there is no redshift uncertainty. We briefly do an analysis that allows us to compare most directly to the results in \cite{SS17}. We use the full $n(z)$ of LSST (black line in Figure \ref{fig:nz12}) and not the Gaussian redshift distributions as will be used in Section \ref{sec:fiducial}. We divide this $n(z)$ into the six tomographic bins used in \cite{SS17} with boundaries at $z=[0,0.5,1,2,3,4,7]$. Since there is no redshift uncertainty, here our Fisher setup has 6 values for $\sigma_{8,i}$ and $b_{\text{g},i}$ and $\Omega_{\text{m}}$ for 13 parameters. \cite{SS17} does not vary $\Omega_{\text{m}}$, so we also show results without this parameter.  We show the constraints on $\sigma_8$ for this setup in Figure \ref{fig:SS_s8_6bin} when we set $l_{\text{max}}$=1000. We show how the results change as a function of $l_{\text{max}}$ in Figure \ref{fig:SS_s8_6bin_lmax}. These constraints are nearly identical to those in \cite{SS17} (its Figure 9) when not including $\Omega_{\text{m}}$ and about $30-60 \%$ larger when varying $\Omega_{\text{m}}$. The largest difference in our analysis here compared to \cite{SS17} is that we do not include any Sloan Digital Sky Survey or DESI galaxies at low redshifts as its authors do. This makes their constraints in the two lowest redshift bins better. 

\begin{figure*}
\begin{center}
\includegraphics[width=1.0 \textwidth]{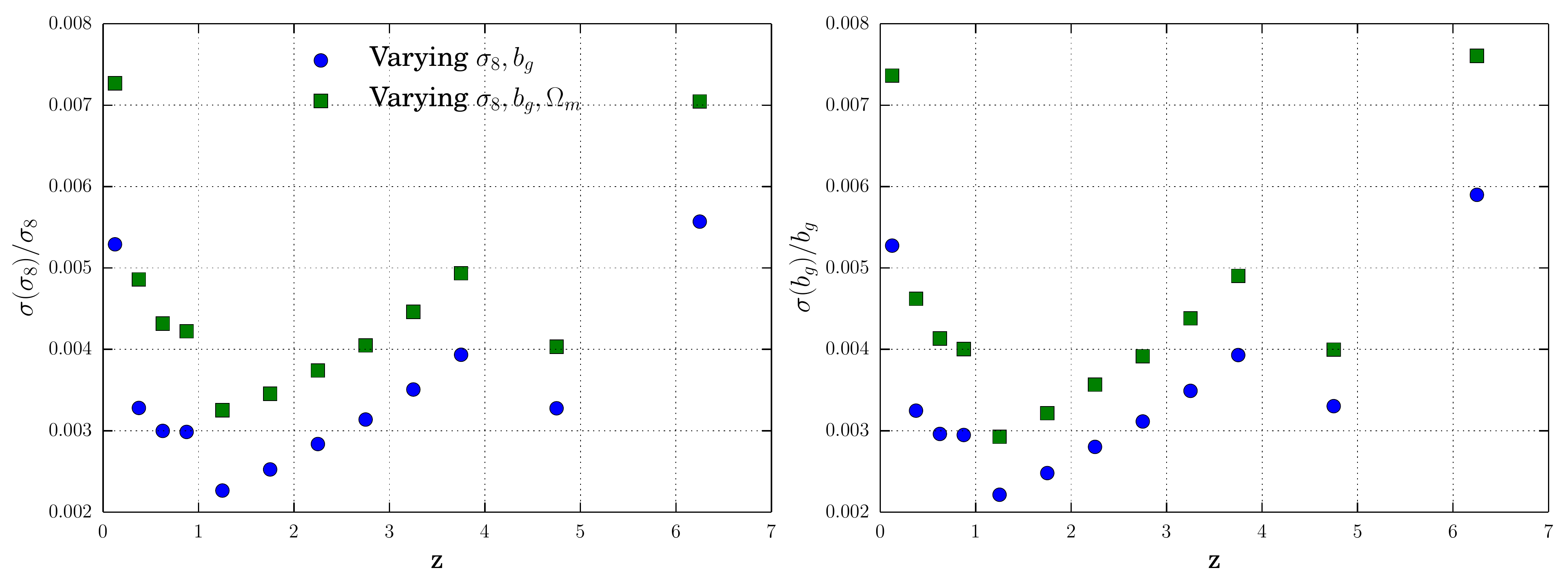}
\end{center}
\caption{Fractional constraints on $\sigma_8$ and $b_{\text{g}}$ as a function of redshift for the case of 12 tomographic redshift bins and no redshift uncertainty. Plotted are the cases in which $\Omega_{\text{m}}$ is fixed or allowed to vary. The constraints largely weaken with higher redshift as the number density drops, however at $z=1.25$ and $z=4.75$, the redshift width  of the bin increases, leading to larger numbers of galaxies in the bin and smaller constraints.}
\label{fig:SS_s8_12bin}
\end{figure*}

For our fiducial analysis in Section \ref{sec:fiducial}, we will use smaller redshift bins, splitting each of the bins used in \cite{SS17} in half, giving us the 12 redshift bins shown in Figure \ref{fig:nz12}. These smaller redshift bins are more similar to current analyses on data, such as from the Dark Energy Survey (e.g., \cite{giannantonio16}, \cite{keypaper}). The approximation of a Gaussian redshift distribution as a result of photometric redshift binning (Section \ref{sec:makingdndz}) is also more accurate for smaller redshift bins. We first test the effect of the smaller bins while still having no redshift uncertainty. We divide the LSST $n(z)$ distribution directly into 12 tomographic bins with boundaries at $z=[0,0.25,0.5,0.75,1,1.5,2,2.5,3,3.5,4,5.5,7]$. Again, we assume all redshifts can be known directly from the black line in Figure \ref{fig:nz12}, and do not use the Gaussian distributions of that figure yet. In this setup, our Fisher analysis has 12 values for $\sigma_{8,i}$ and $b_{\text{g},i}$ and $\Omega_{\text{m}}$ for 25 parameters. The constraints on $\sigma_8$ and $b_\text{g}$ in this setup are shown in Figure \ref{fig:SS_s8_12bin}. We again show the case with and without $\Omega_{\text{m}}$ in the figures as well. Compared to Figure \ref{fig:SS_s8_6bin}, the constraints on $\sigma_8$ from doubling the number of redshift bins when not using $\Omega_{\text{m}}$ are a little worse, as expected from shrinking the number of galaxies (and thus increasing the shot noise) in each bin. The constraints for the average of two smaller bins are about $25 \%-50 \%$ worse than the larger bin of the same redshift range (e.g., comparing the average constraint between of $z=0-0.25$ and $z=0.25-0.5$ to the constraint on $z=0-0.5$). Of course, the benefit of more bins is gaining more precise information of the full $\sigma_8(z)$. Interestingly, when also varying $\Omega_{\text{m}}$, the constraints on $\sigma_8$ have very little degradation when switching from 6 bins to 12 bins. With $\Omega_{\text{m}}$ varying, the constraints on the smaller bins approximately equal the constraints on the larger bins (Figures \ref{fig:SS_s8_6bin} and \ref{fig:SS_s8_12bin}). The typical degradation of constraints when dividing into smaller bins is offset in this case by having a better constraint on $\Omega_{\text{m}}$ due to more measurements (12 instead of 6) constraining $\Omega_{\text{m}}$. The constraints go from $\sigma(\Omega_{\text{m}})=0.0007$ with 6 bins to $\sigma(\Omega_{\text{m}})=0.0003$ with 12 bins. 

\section{Results with Redshift Uncertainty}
\label{sec:fiducial}

In this section, we show the fiducial analysis of varying five parameters in the Fisher analysis of Section \ref{sec:fisher}: $\sigma_{8,i}$,$ \ b_{\text{g},i}$,$\ z_{0,i}$,$\ \sigma_{\text{z},i}$ in each redshift bin, and $\Omega_{\text{m}}$. With 12 redshift bins for our main analysis of an LSST-like sample, we have 49 parameters. The redshift distributions with central values of $z_{0,i}$ and Gaussian width $\sigma_{\text{z},i}$ are shown in Figure \ref{fig:nz12}, and listed in Appendix \ref{sec:appendix0}.

\begin{figure*}
\begin{center}
\includegraphics[width=1.0 \textwidth]{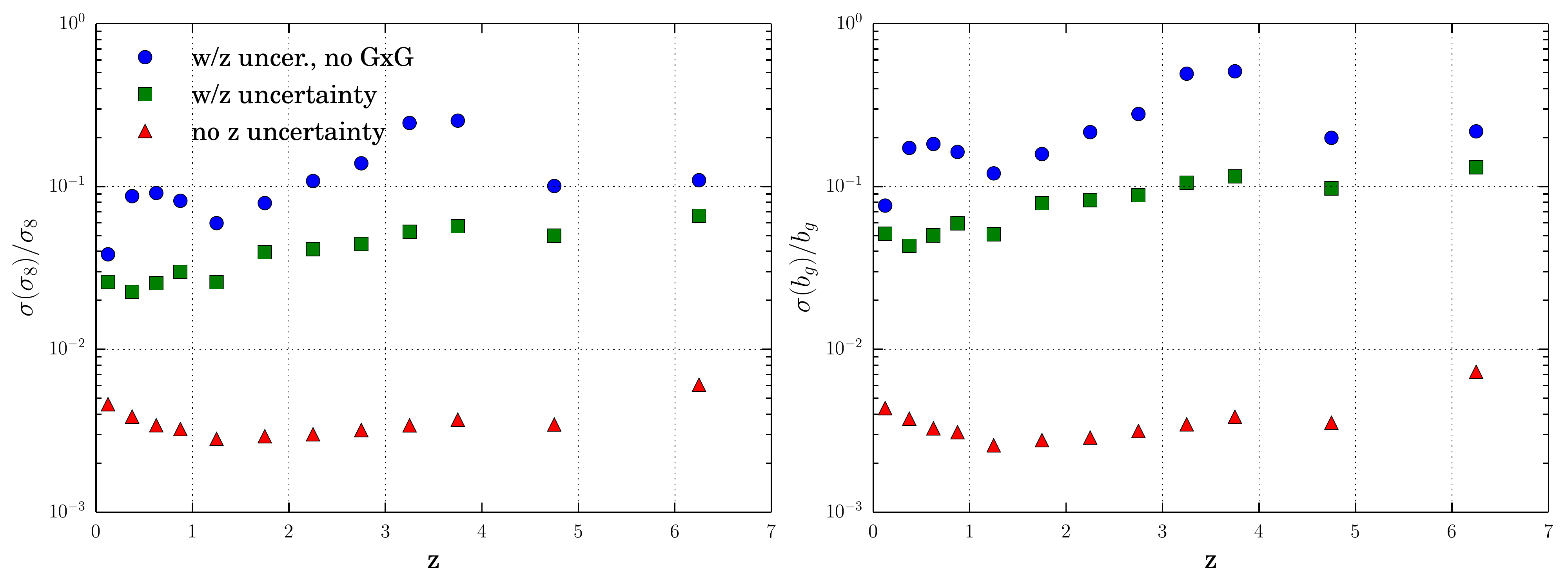}
\end{center}
\caption{Fractional constraints on $\sigma_8$ and $b_{\text{g}}$ for the fiducial case of LSST+CMB-S4, with $l_{\text{max}}=1000$, and $f_{\text{sky}}=0.5$. Shown are the results with no redshift uncertainty, varying $b_{\text{g},i}$ and $\sigma_{8,i}$ for each redshift bin, as well as $\Omega_{\text{m}}$. Also shown is the fiducial analysis in which we include redshift uncertainty by allowing the parameters $z_{0,i}$ and $\sigma_{\text{z},i}$ to vary in each bin. We also show the case in whcih we have redshift uncertainties, but do not use the cross-correlations of two adjacent galaxy bins in redshift space (no GxG). }
\label{fig:s4_s8}
\end{figure*}

\begin{figure*}
\begin{center}
\includegraphics[width=1.0 \textwidth]{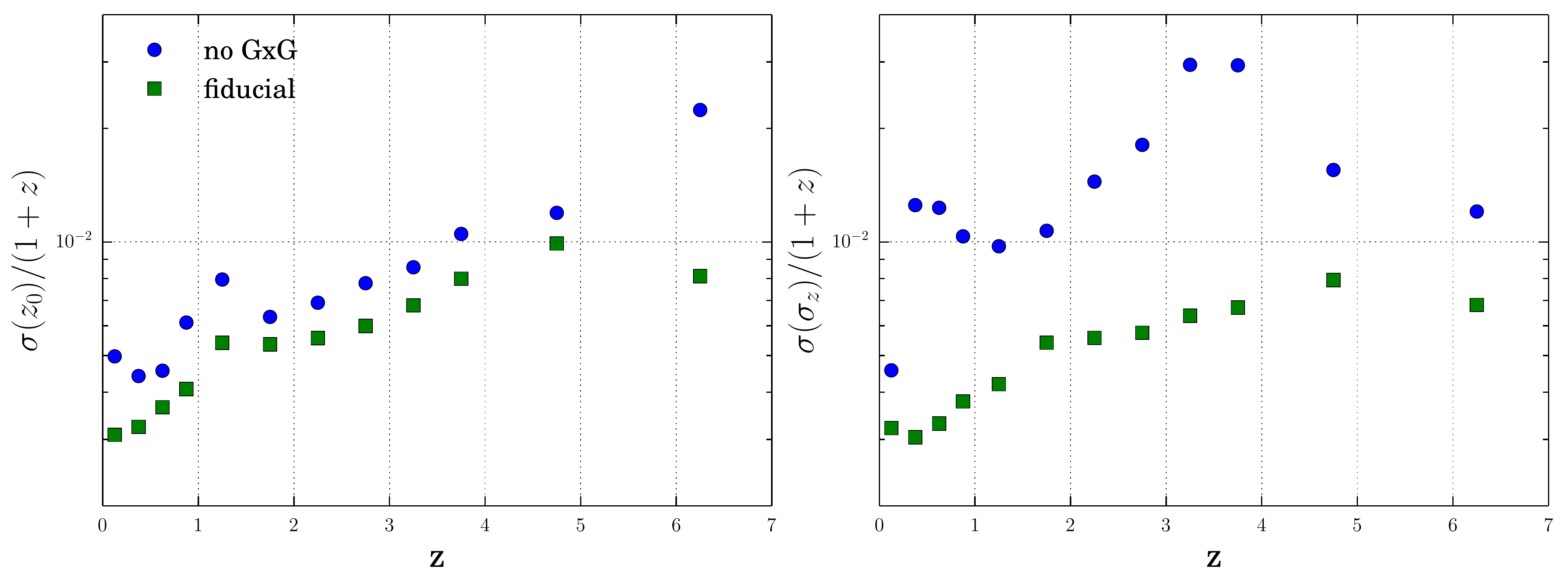}
\end{center}
\caption{The constraints on the mean redshift, $z_{0,i}$, and the width of the redshift distribution, $\sigma_{\text{z},i}$, in each redshift bin for our fiducial analysis of Figure \ref{fig:s4_s8}. We also show the constraints when the cross-correlations of two adjacent galaxy bins in redshift space are not used (no GxG).}
\label{fig:s4_z0}
\end{figure*}

For our fiducial analysis of the LSST/CMB-S4 era including redshift uncertainties, we show our constraints on the various parameters in Figures \ref{fig:s4_s8} and \ref{fig:s4_z0}. Figure \ref{fig:s4_s8} shows the constraints on $\sigma_{8,i}$ and $b_{\text{g},i}(z)$, in the cases with and without redshift uncertainty (i.e., with $z_{0,i}$ and $\sigma_{\text{z},i}$ fixed.) \footnote{We note that the results for no redshift uncertainty in Figure \ref{fig:s4_s8} differ slightly from those in Figure \ref{fig:SS_s8_12bin}. This is due to the fact that the underlying galaxy distributions are slightly different in these two cases. In Figure \ref{fig:SS_s8_12bin}, the underlying galaxy distribution is the true distribution binned by redshift (i.e., the black line in Figure \ref{fig:nz12} separated by the gray lines) similar to that in \cite{SS17}, while in Figure \ref{fig:s4_s8}, the galaxy distribution in each bin is a Gaussian (colored lines in Figure \ref{fig:nz12}) with parameters known exactly in the no redshift uncertainty case.} We can see that the addition of redshift uncertainty in these parameters increases errors on the other parameters by roughly a factor of 10. We also show the results for the parameters when cross-correlations of adjacent galaxy bins are not used (labeled as `no GxG'). In this case, errors on parameters tend to increase by another factor of 2 or more. This highlights the importance of the cross-correlations between galaxy bins, a measurement that in principle is not necessary when galaxy redshifts are known perfectly, and galaxy bins do not overlap in redshift space. 

Figure \ref{fig:s4_z0} shows the constraints on the redshift parameters $z_{0,i}$ and $\sigma_{\text{z},i}$ in each of the 12 photometric bins. We again also plot the results when not using the galaxy-galaxy cross-correlations of adjacent redshift bins. As seen in the figure, the galaxy-galaxy cross-correlations are of particular importance for $\sigma_{\text{z}}$. The cross-correlations break degeneracies between $\sigma_8$, $b_{\text{g}}$, and $\sigma_{\text{z}}$ that remain when only having galaxy autocorrelations and galaxy-CMB lensing cross-correlations for each bin (see Appendix \ref{sec:appendix} for more discussion). 

We also note that the constraints on $\Omega_{\text{m}}$ in the scenarios of no galaxy-galaxy cross-correlations, the fiducial analysis, and the no redshift uncertainty case are $\sigma(\Omega_{\text{m}})=[0.00075, 0.00060, 0.00025]$, respectively. The improvement on $\Omega_{\text{m}}$ with more redshift information is more mild than on $\sigma_8$ due to $\Omega_{\text{m}}$ not being part of the degeneracy of $\sigma_8$, $b_{\text{g}}$, and $\sigma_{\text{z}}$ (Appendix \ref{sec:appendix}).

\section{Dependence on Survey Parameters}
\label{sec:surveyparams}

\subsection{Example: DES-SPT}
In this section, we vary different survey parameters that affect the precision of the constraints on the five types of parameters. We first look at a specific example of varying the survey parameters, using the expected galaxy density and redshift distributions from the full Dark Energy Survey and CMB lensing noise from SPT-SZ and the future SPT-3G. This represents a nearer term projection for parameters using our methodology compared to the fiducial analysis of LSST/CMB-S4.

\begin{figure*}
\begin{center}
\includegraphics[width=1.0 \textwidth]{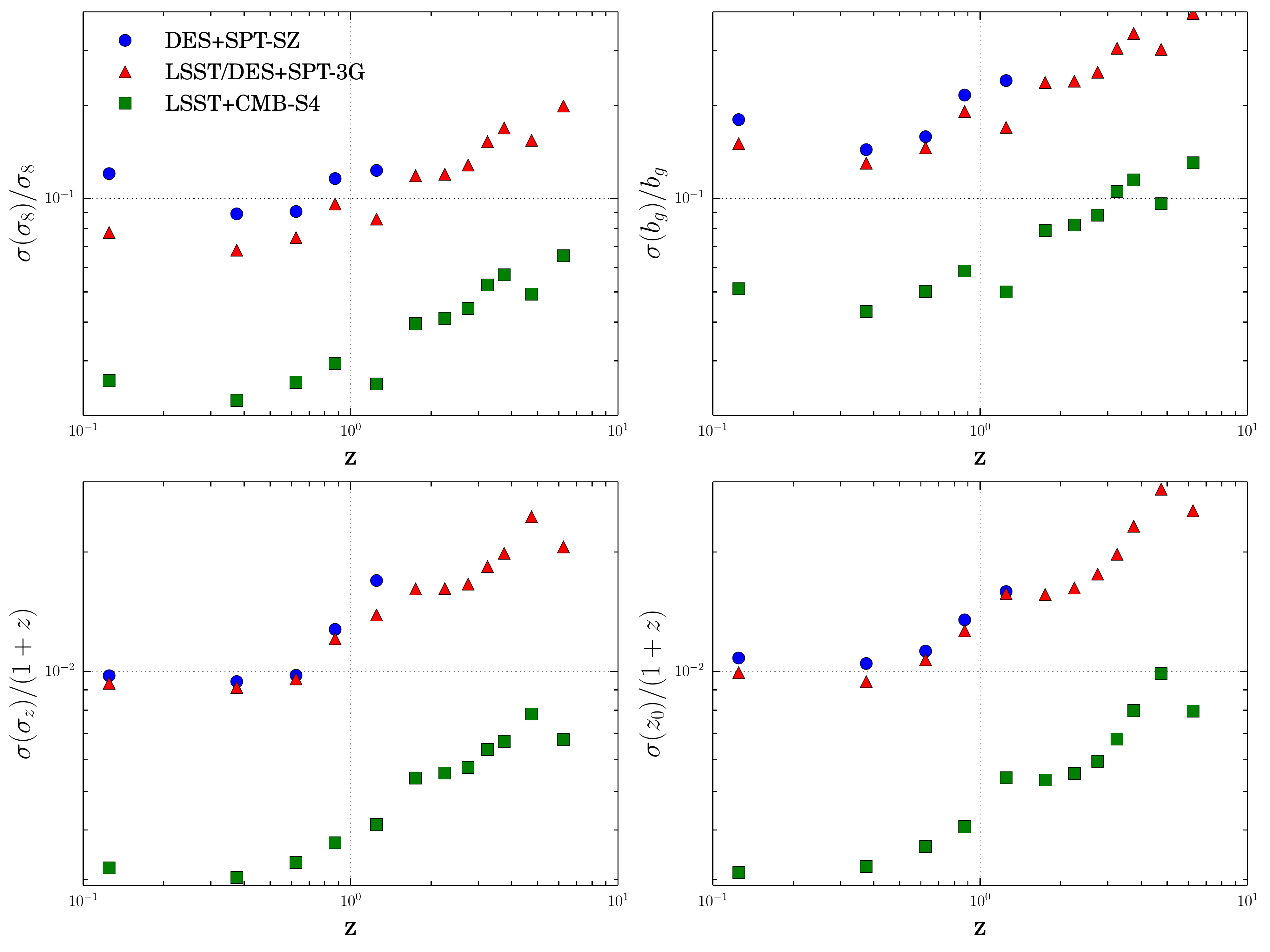}
\end{center}
\caption{The constraints on all four of our redshift bin parameters, $\sigma_{8,i}$, $b_{\text{g},i}$, $z_{0,i}$, and $\sigma_{\text{z},i}$ for each of the survey combinations for our fiducial analysis including redshift uncertainties. Each of the surveys use $l_{\text{max}}=1000$. The correlations with SPT have $f_{\text{sky}}=0.061$ ($2500 \ \text{deg}^2$) and the correlation of LSST and CMB-S4 has $f_{\text{sky}}=0.5$ ($20,000 \ \text{deg}^2$). The results for DES+SPT-3G and LSST+SPT-3G are within $5 \%$ of each other for the redshift bins DES goes up to, so only LSST+SPT-3G is plotted. In general, constraints are better with higher densities and thus decrease with redshift, though there are exceptions in which the bin size is increased in redshift width from the previous bin (e.g., $z=1.25$). We plot here logarithmically on the x axis to give more space in showing the DES constraints while still showing the full redshift range of LSST. }
\label{fig:s8_allsurveys}
\end{figure*}

\begin{figure}
\begin{center}
\includegraphics[width=0.5 \textwidth]{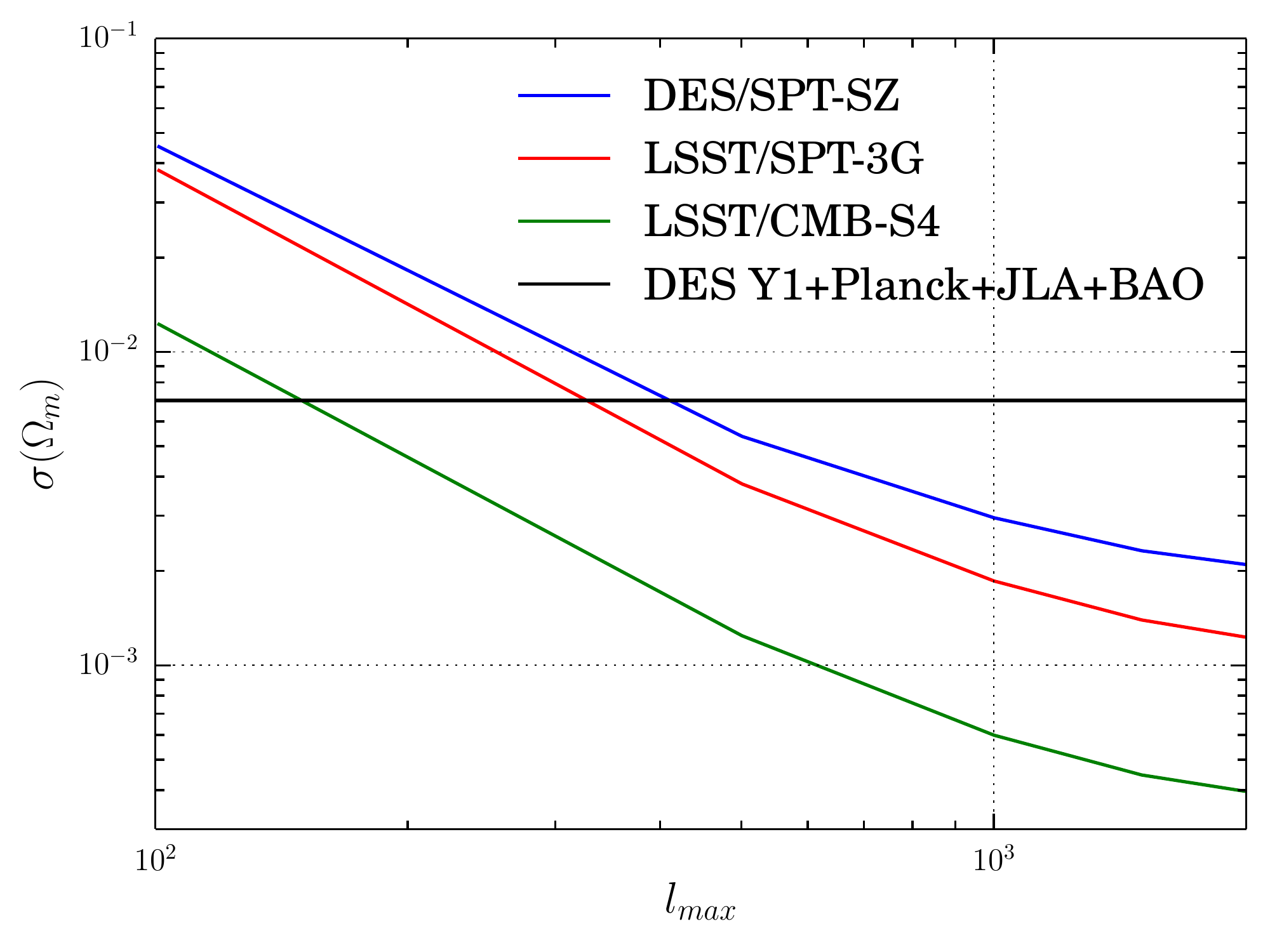}
\end{center}
\caption{The constraints on $\Omega_{\text{m}}$ for each of the survey combinations for our fiducial analysis including redshift uncertainties (same analysis as in Figure \ref{fig:s8_allsurveys}). For an example of current constraints on $\Omega_{\text{m}}$ from photometric surveys, we show the constraint from the recent DES year 1 analysis \citep{keypaper} when combined with CMB data from Planck, Type Ia supernovae data from the Joint Light-Curve Analysis (JLA), and baryon acoustic oscillation (BAO) data from multiple surveys (see references in \cite{keypaper}).}
\label{fig:om_allsurveys}
\end{figure}

Figure \ref{fig:s8_allsurveys} shows the constraints for the four parameters that exist in each redshift bin for DES+SPT-SZ, LSST+SPT-3G, and our fiducial analysis on LSST+CMB-S4. Not shown are the constraints for the combination of DES+SPT-3G. These constraints are within $5 \%$ of the constraints for LSST+SPT-3G, in the bins where DES has data (the first five data points, up to $z<1.5$), so we do not show them. While the DES/SPT-SZ constraints are approximately factors of 2-3 weaker than LSST/CMB-S4, an approximately $10 \%$ constraint on $\sigma_8$ is still possible in all of our bins and should be achievable with these surveys in the next few years. We show the constraints on $\Omega_{\text{m}}$ for the different era analyses in Figure \ref{fig:om_allsurveys}. We see that the constraints on $\Omega_{\text{m}}$ improve by a factor of about 3-5 from the DES/SPT-SZ era to the LSST/CMB-S4 era depending on the $l_{\text{max}}$ used. We also see that all eras of measuring the power spectra used in this work should improve upon the constraints from the recent DES year 1 analysis of galaxy clustering and weak lensing plus other datasets in \cite{keypaper}.

\subsection{Dependence on $l_{\text{max}}$}
\label{sec:lmax}
The largest multipole, $l_{\text{max}}$ (smallest scale), to which these measurements can be used and modeled is a parameter with still a fair bit of uncertainty. In \cite{giannantonio16}, $l_{\text{max}}=2000$ was used for correlations of DES science verification data and SPT-SZ. However, in \cite{baxter18}, the authors realize that a newer version (and perhaps older versions) of the SPT lensing map are significantly impacted by thermal Sunyaev-Zel'dovich bias. This leads them to only use real space angular separations of $15$ arcminutes or greater, roughly equivalent to using an $l_{\text{max}}=720$. In \cite{SS17}, the authors use $l_{\text{max}}=1000$ for their fiducial projections but also vary $l_{\text{max}}$ out to 2000. They cite the issues of modeling nonlinear galaxy bias at small scales (large $l$) as a concern. \cite{modi17} also looks extensively at the effects of modeling small-scale nonlinear bias on galaxy-CMB lensing cross-correlations. On the other hand, \cite{giannantonio16} and \cite{2016MNRAS.455.4301C} find for DES science verification galaxies that linear galaxy bias is a good approximation in most cases down to $l_{\text{max}}=2000$, even though this can be a factor of 4 smaller than where the matter power spectrum becomes nonlinear.

\begin{figure}
\begin{center}
\includegraphics[width=0.5 \textwidth]{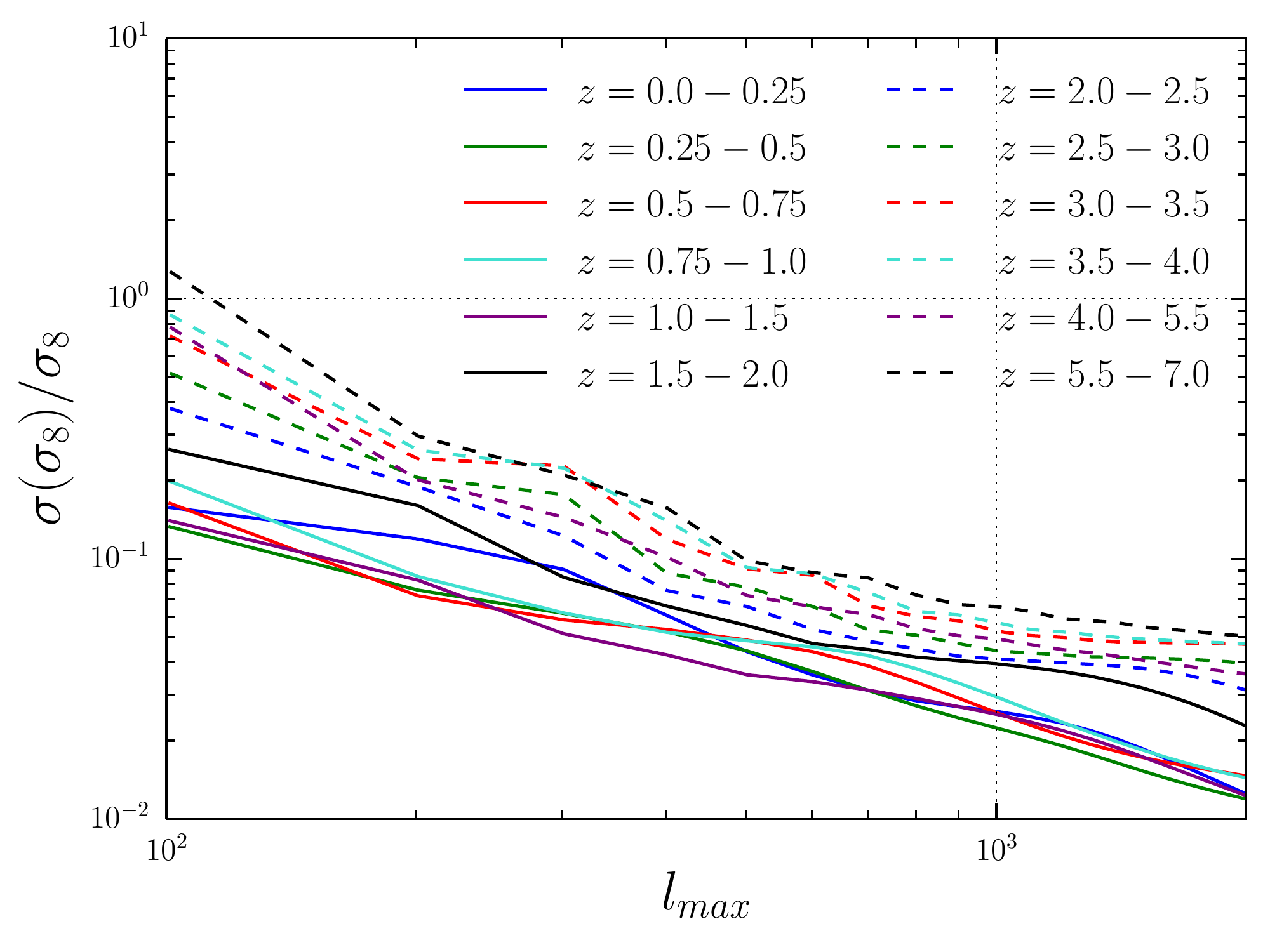}
\end{center}
\caption{Dependence on $l_{\text{max}}$ for the $\sigma_8$ constraints. We use the fiducial parameters of LSST and CMB-S4, including $f_{\text{sky}}=0.5$.}
\label{fig:lmaxplot}
\end{figure}

We chose $l_{\text{max}}=1000$ for our fiducial analysis but vary it in this section, much like the treatment in \cite{SS17}. Figure \ref{fig:lmaxplot} shows the $\sigma_8$ constraints for varying $l_{\text{max}}$ values for the LSST/CMB-S4 measurement. We can see that $l_{\text{max}}$ can significantly impact the constraints. Increasing $l_{\text{max}}$ from 1000 to 2000 approximately doubles the constraining power for the $z<1.5$ bins, though this makes less of a difference in the higher redshift bins.

\subsection{Dependence on $f_{\text{sky}}$}

\begin{figure}
\begin{center}
\includegraphics[width=0.5 \textwidth]{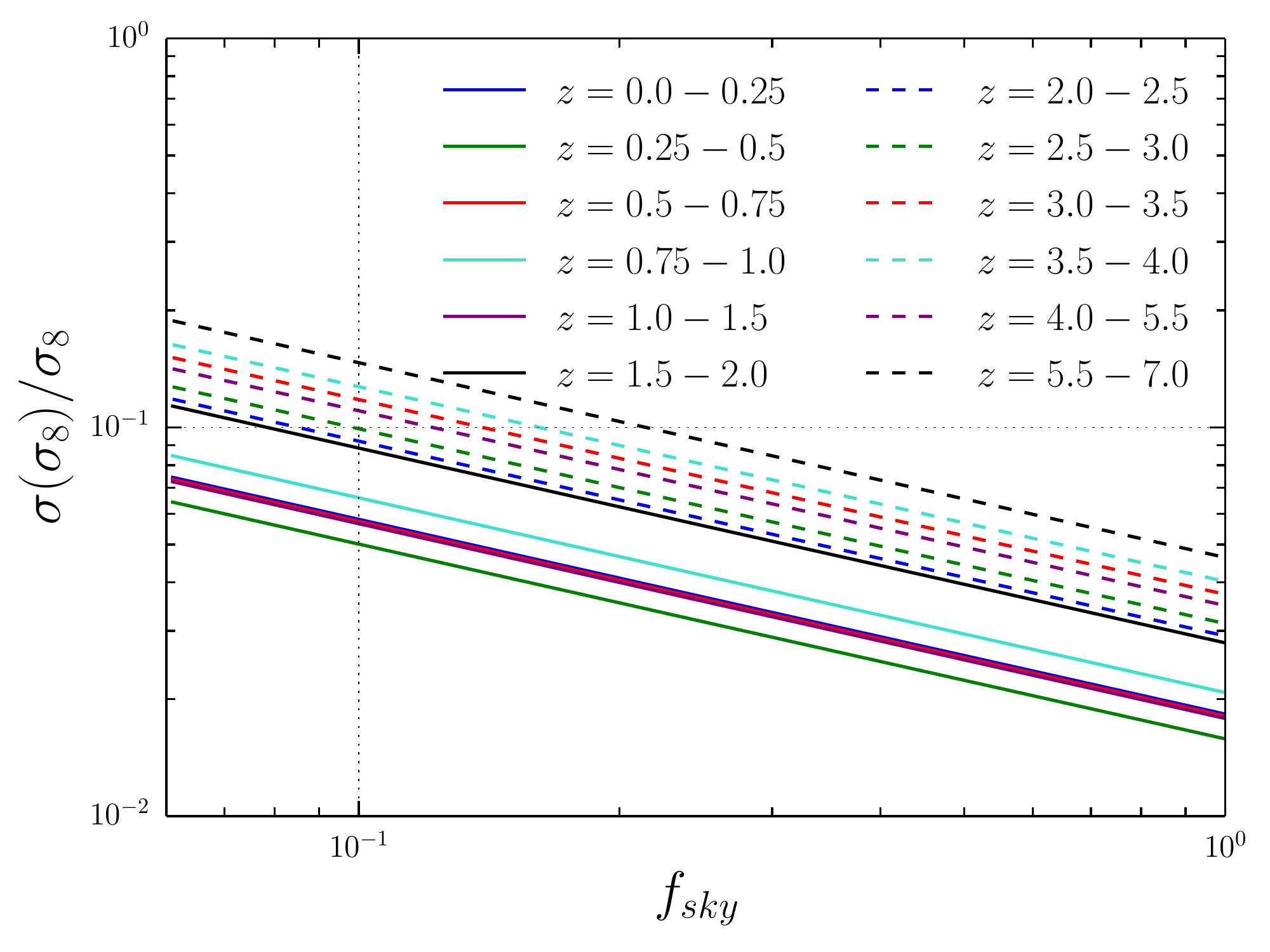}
\end{center}
\caption{Dependence on $f_{\text{sky}}$ for the $\sigma_8$ constraints. We use the fiducial parameters of LSST and CMB-S4, including $l_{\text{max}}=1000$. The dependence scales as approximately $1/\sqrt{f_{\text{sky}}}$. The current observations using DES and SPT-SZ cover 2500 $\text{deg}^2$ which is $f_{\text{sky}}=0.06$, the x limit of the plot. The combination of LSST and CMB-S4 is expected to approach 5000 $\text{deg}^2$, which is $f_{\text{sky}}=0.5$.}
\label{fig:fskyplot}
\end{figure}

Another important parameter to study is the overlapping sky fraction of the surveys, $f_{\text{sky}}$. We show our fiducial analysis of LSST/CMB-S4 for a range of $f_{\text{sky}}$ values in Figure \ref{fig:fskyplot}. The constraints on parameters scale as approximately $1/\sqrt{f_{\text{sky}}}$ due to the factor of $f_{\text{sky}}$ in Equation \ref{covariance}. On the far left of the plot is the value $f_{\text{sky}}=0.061$, which is the overlap of the DES and SPT. Keeping all other parameters the same, the increase from this overlap, to our fiducial value of $f_{\text{sky}}=0.5$ with LSST and CMB-S4, improves constraints on $\sigma_8$ by almost a factor of 3. This highlights the importance of having maximal overlap between CMB-S4, which is still in the planning phases, and LSST. We also note that based on this scaling, a possibly more realistic value of $f_{\text{sky}}=0.45 \ (18,000 \ \text{deg}^2)$ for LSST will degrade constraints by approximately $5 \%$ compared to the results in our fiducial analyses using $f_{\text{sky}}=0.5$.

\subsection{Dependence on Measurement Noise}

In this section, we study how the $\sigma_8$ constraints change when varying the measurement noise, $N_l$, in Equation \ref{chat} for both CMB lensing and galaxy clustering. The galaxy clustering noise is determined by the galaxy density of the sample: $N_l^{g g}=1/\rho$, where $\rho$ has units of galaxies per steradian. The CMB lensing noise expectations for the three CMB experiments are shown in Figure \ref{fig:cmbnoise}. 

\begin{figure*}
\begin{center}
\includegraphics[width=1.0 \textwidth]{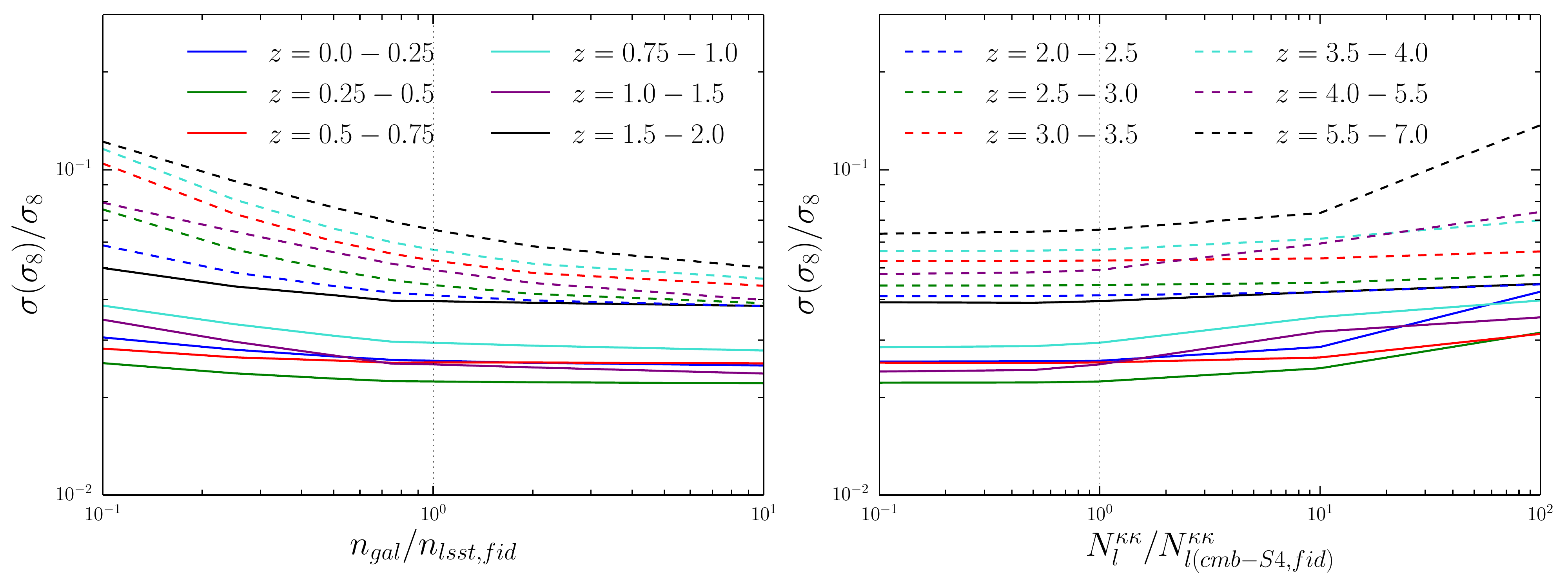}
\end{center}
\caption{(Left) Dependence of the $\sigma_8$ constraints on galaxy density in units of the fiducial LSST prediction. The LSST has about three to five as many galaxies as DES in the bins that they overlap in. (Right) Dependence of the $\sigma_8$ constraints on CMB lensing noise in units of the fiducial prediction for CMB-S4 lensing noise. SPT-SZ has about a factor of 50-100 times the lensing noise as CMB-S4. SPT-3G has about 3-8 times the noise of CMB-S4. All other parameters match the fiducial LSST+CMB-S4 analysis, including $l_{\text{max}}=1000$.}
\label{fig:noiseplot}
\end{figure*}

In Figure \ref{fig:noiseplot}, we show the constraints on $\sigma_8$ for LSST/CMB-S4 when varying the galaxy density (left) and CMB lensing noise (right). We vary the galaxy density at all redshifts by multiplying LSST $n(z)$ from Figure \ref{fig:nz12} by a constant factor. For reference, at the redshifts where the DES and LSST overlap ($z~<1$), LSST has greater density by about a factor of 3-5. We vary the lensing noise by multiplying the fiducial CMB-S4 noise curve (Figure \ref{fig:cmbnoise}) by a constant factor. SPT-SZ has approximately 50-100 times more noise than CMB-S4, and SPT-3G has about 3-8 times more noise than CMB-S4, with the factor changing with $l$.

\begin{figure*}
\begin{center}
\includegraphics[width=1.0 \textwidth]{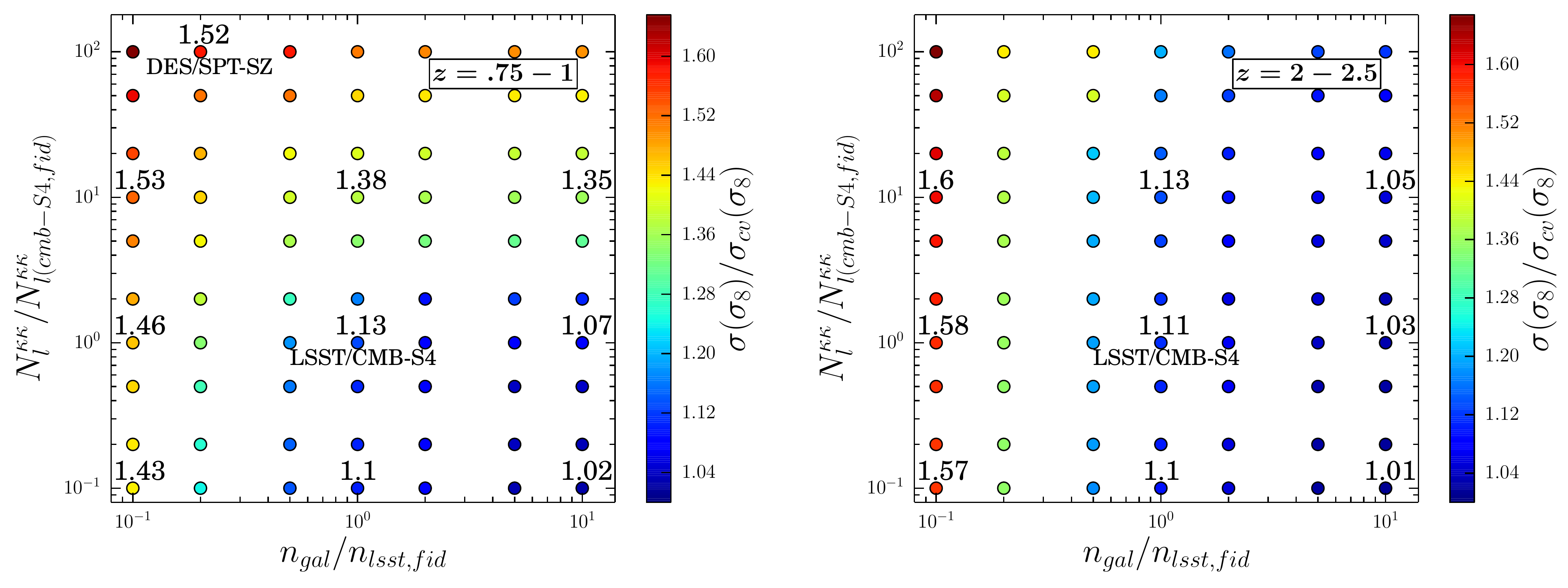}
\end{center}
\caption{Uncertainty on $\sigma_8$ as a function of galaxy density and CMB lensing noise in units of the fiducial values for LSST and CMB-S4 respectively for two redshift bins, $z=0.75-1$ (left) and $z=2-2.5$ (right). The color values are plotted with respect to the cosmic variance limit, $\sigma_{\text{cv}} (\sigma_8)$ which is the constraint found when setting galaxy density to infinity and CMB noise to 0. Increasing the density or decreasing the lensing noise beyond the fiducial values makes the constraints approach the cosmic variance limit (bottom right of each plot). The fiducial values for density and noise for LSST/CMB-S4 in each redshift bin, and roughly the noise level for DES/SPT-SZ in the lower redshift bin are marked with the names of the surveys. We note that the marked point for DES/SPT-SZ is only noting a change in the noise levels of those surveys. The actual constraints from those surveys (e.g., Figure \ref{fig:s8_allsurveys}) also include the difference in $f_\text{sky}$, which changes results considerably. This plot uses $f_\text{sky}=0.5$ throughout. The higher redshift bin shows greater dependence on the galaxy density. This is due to the lower galaxy density in that bin, leaving more room for improvement in lowering the galaxy shot noise ($1/\rho$.) We use $l_{\text{max}}=1000$ for this comparison.}
\label{fig:noisecolorbar}
\end{figure*}

We can see that the constraints in Figure \ref{fig:noiseplot} only modestly depend on the measurement noise, particularly in going to lower CMB noise or higher galaxy density than the fiducial LSST/CMB-S4 prediction. The fiducial LSST and CMB-S4 noise levels are low enough that the measurements approach the cosmic variance limit, where $N_l \to 0$. Lowering the noise level further cannot gain much more information on these measurements, particularly at low redshift. We show this explicitly in Figure \ref{fig:noisecolorbar}, in which we show the fractional difference of the $\sigma_8$ constraints with a variety of noise estimates to $\sigma_{\text{cv}}(\sigma_8)$, the cosmic variance limit of uncertainty on $\sigma_8$ when the measurement noise $N_l=0$, for two different redshift bins. LSST/CMB-S4 approaches this limit in most of the redshift bins [i.e., $\sigma_{\text{fid}} (\sigma_8)/\sigma_{\text{cv}} (\sigma_8) <1.2$ for $z<3$]. At higher redshifts, the lower density of galaxies is a more significant limitation.

In summary, once we are in the LSST/CMB-S4 era, improvements on measurement noise will yield very modest gains compared to improvements on $l_{\text{max}}$ and $f_{\text{sky}}$.

\subsection{Dependence on Redshift Priors}
\label{sec:priors}

So far, our analysis has assumed no prior information on any of the cosmological or redshift parameters we vary. In this section, we see how our results change when adding priors on the redshift parameters. As mentioned, photometric surveys like DES and LSST put considerable effort into calibrating photometric redshift methods, so any real analysis will have some level of prior on quantities like $z_0$ and $\sigma_{\text{z}}$. We apply a range of plausible priors for LSST redshifts to our analysis. The most recent LSST Dark Energy Science Collaboration (DESC) Science Requirements Document \citep{newlsst} provides some targets for redshift priors on galaxy samples. In it, the precision on the mean redshift of photometric bins to be used in large-scale structure measurements (in the full ten-year analyses), $\sigma(z_0)$, is required to be $0.003(1+z)$ in order to not significantly degrade cosmological measurements. Similarly, the precision on the width of the redshift distribution, $\sigma(\sigma_{\text{z}})$, is required to be $0.03(1+z)$ for the same samples of galaxies. The precision for samples of galaxies to be used as weak lensing sources are tighter, $0.001(1+z)$ and $0.003(1+z)$, for the mean and width of the redshift distributions respectively. For some redshift ranges, the priors on redshifts may be significantly better than these numbers for LSST. In \cite{newman15}, it is shown that the spatial cross-correlation of photometric and spectroscopic galaxies (clustering redshifts) could yield constraints on both the mean and width of photometric redshift bins of approximately $0.0004(1+z)$ for $z=0.5-1.5$. The exact priors available in the LSST era will depend on a number of factors, including the number, redshift range, and magnitude depth of spectroscopic samples, the number density of the photometric samples, the types of galaxies in the photometric samples, and the width of the photometrically selected bins ($\sigma_{\text{z}}$). Each of these factors can make constraints significantly weaker at higher redshifts. 

\begin{figure*}
\begin{center}
\includegraphics[width=1.0 \textwidth]{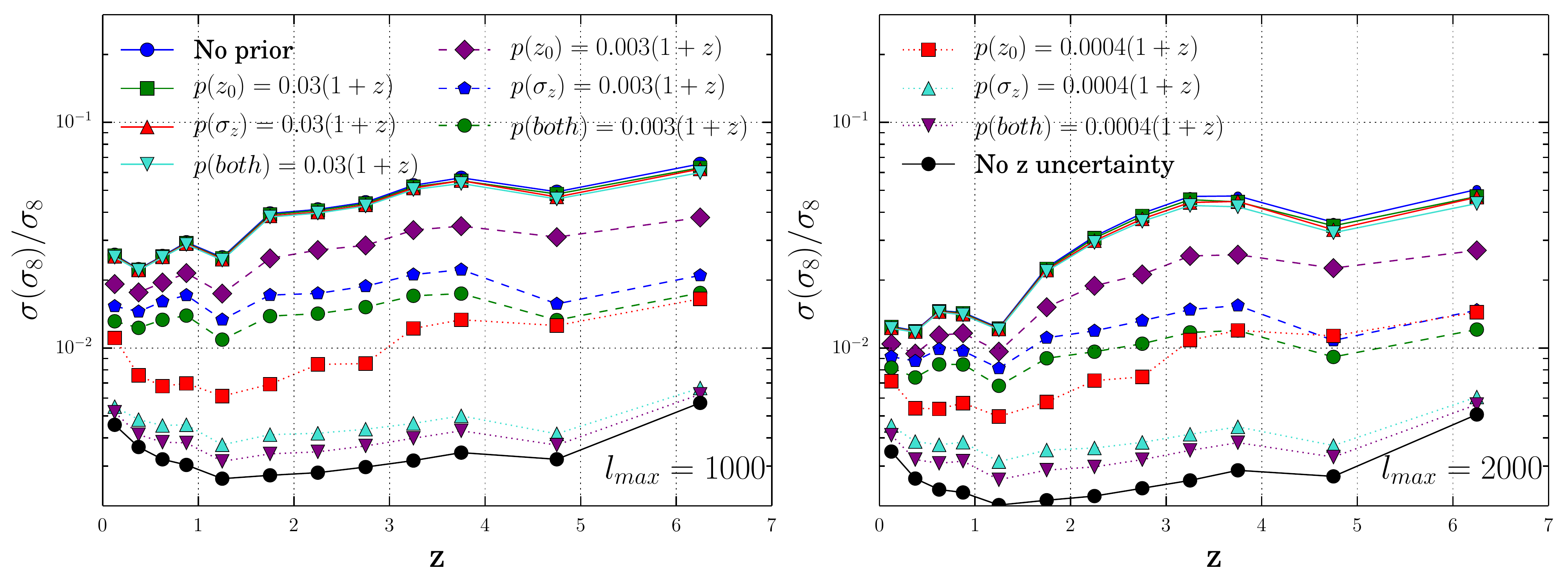}
\end{center}
\caption{Constraints on $\sigma_8$ for our fiducial LSST/CMB-S4 analysis when adding priors on redshift parameters. Left: constraints when having $l_{\text{max}}=1000$. Right: constraints when having $l_{\text{max}}=2000$. Each curve adds either a prior on $z_0$, on $\sigma_{\text{z}}$ or an equal prior on each. We compare the curves with priors to the fiducial case of no prior information, and the opposite extreme of no redshift uncertainty with $z_0$ and $\sigma_{\text{z}}$ fixed in the Fisher analysis. The priors of $0.003(1+z)$ and $0.03(1+z)$ come from the LSST DESC Science Requirements Document (SRD) requirements \cite{newlsst} for $z_0$ and $\sigma_{\text{z}}$, respectively. The prior of $0.0004(1+z)$ is a plausible future achievement by clustering redshifts at low $z$ found in \cite{newman15}.}
\label{fig:lsstpriors1}
\end{figure*}

We use the numbers mentioned in the previous paragraph as a broad range of possible priors available in the LSST era. In Figure \ref{fig:lsstpriors1}, we plot how the constraints on $\sigma_{8,i}$ change for a range of prior assumptions on $z_{0,i}$ and $\sigma_{\text{z},i}$. We plot the different scenarios for both $l_{\text{max}}=1000$ and $2000$. We use the simple model of having just $z_0$ priors, just $\sigma_{\text{z}}$ priors, or priors on each of the same magnitude. We make the broad assumption of having the priors scale as $(1+z)$. We can see in Figure \ref{fig:lsstpriors1} that the priors on $\sigma_{\text{z}}$ are more important than the priors on $z_0$ for constraining $\sigma_8$. This makes sense, as $\sigma_{\text{z}}$ and $\sigma_8$ both provide an overall scaling to the galaxy autocorrelations, which have the highest signal to noise ratio (S/N) of any of the power spectra. Meanwhile, the dependence on $z_0$ is less degenerate with $\sigma_8$ (see Appendix \ref{sec:appendix}).

Figure \ref{fig:lsstpriors1} shows that redshift priors can improve the constraints on $\sigma_8$ considerably. For the case of priors of $0.003(1+z)$ on both $z_0$ and $\sigma_{\text{z}}$, the constraints on $\sigma_8$ improve by about a factor of 2-3 from the no prior information case. When adding priors of $0.0004(1+z)$ predicted from clustering redshifts in \cite{newman15}, the constraining power is within $50 \%$ of the no redshift uncertainty scenario ($z_0$ and $\sigma_{\text{z}}$ fixed). We thus see that redshift priors from techniques like clustering redshifts are very beneficial. This model of priors, however, is almost certainly too optimistic for $z>1.5$, where there will be fewer spectroscopic galaxies available for the clustering redshift method. We also see that a prior of $0.03(1+z)$ (the current LSST DESC requirement for $\sigma_{\text{z}}$) adds nearly zero constraining power for our fiducial analysis of LSST/CMB-S4.

\begin{figure*}
\begin{center}
\includegraphics[width=1.0 \textwidth]{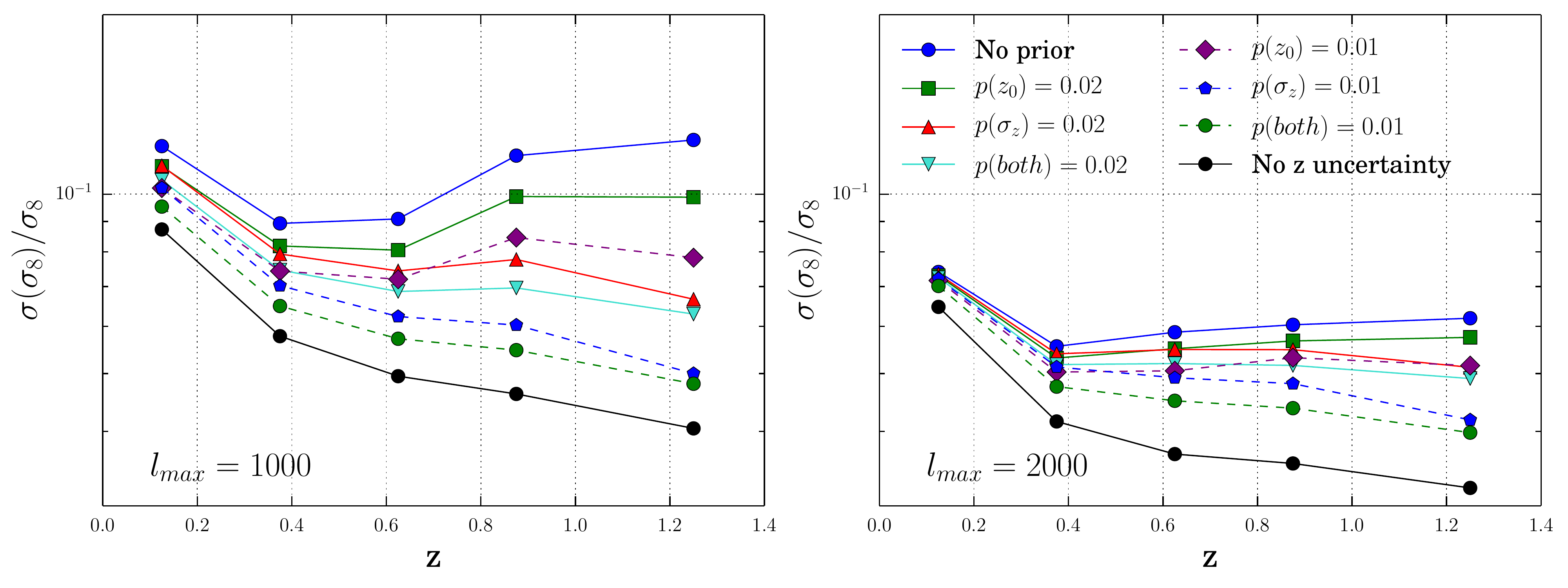}
\end{center}
\caption{Constraints on $\sigma_8$ for the surveys DES/SPT-SZ with priors on the redshift parameters. Left: constraints with $l_{\text{max}}=1000$. Right: constraints with $l_{\text{max}}=2000$. We compare priors on $z_0$, $\sigma_{\text{z}}$, and both parameters with the case of no prior information and the case of no redshift uncertainty. We base the priors on recent DES results and do not vary them with $(1+z)$, unlike Figure \ref{fig:lsstpriors1}.}
\label{fig:despriors1}
\end{figure*}

In Figure \ref{fig:despriors1}, we show a similar analysis for the DES+SPT-SZ era for $l_{\text{max}}=1000$ and 2000. We project DES redshift parameter priors on the order of 0.01-0.02 based on recent calibrations of redshift bins in DES year 1 cosmological analyses. The weak lensing source galaxies used in \cite{keypaper} and \cite{troxel17} are separated into photometrically selected bins. The mean redshift of these bins is constrained to about an accuracy of 0.02 both in tests of photometric redshift methods on samples of spectroscopically measured galaxies \cite{hoyle18des} and in using spatial cross-correlations with spectroscopic galaxies (clustering redshifts,  \cite{davis17}). These results were fairly constant across redshift, so we do not vary our priors with the factor $(1+z)$ here. The brighter redMaGiC galaxies used in DES year 1 results (\cite{rozo16}, \cite{elvinpoole17}) had tighter constraints on their mean redshifts from clustering redshift measurements in \cite{cawthon17}. However, the modeled galaxy densities in our work are much higher than this brighter sample, making the weak lensing source sample a more appropriate sample to use for plausible redshift priors.

We see a similar dependence overall on redshift priors for the DES/SPT-SZ era as in the future LSST/CMB-S4 era. Tightening the redshift priors brings results closer to the case of no redshift uncertainty. We again see that $\sigma_{\text{z}}$ is more important than $z_0$ for constraining $\sigma_8$. In the DES year 1 analysis (\cite{keypaper} and the others mentioned above), only $z_0$ was constrained. Figure \ref{fig:despriors1} (left) shows that adding a $0.02$ prior on $\sigma_{\text{z}}$ to the already achieved $0.02$ prior on $z_0$ would improve constraints on $\sigma_8$ for the highest two redshift bins by about $30 \%$. If $l_{\text{max}}$ can be extended to 2000 (right side of Figure \ref{fig:despriors1}), the gains of a $0.02$ prior on $\sigma_{\text{z}}$ only go up to $15 \%$.

\section{Constraints on Redshift Parameters}
\label{sec:selfcalibration}

We move from our discussion in Section \ref{sec:priors} on the effect of redshift information back now to constraints on redshift parameters themselves. We explore the ability of galaxy clustering and galaxy-CMB lensing correlation measurements to self-calibrate redshifts (without prior redshift information) and compare those constraints to photometric redshift techniques. The idea of calibrating redshifts strictly from correlation functions was studied in more detail recently in \cite{hoyle18}. A significant difference in this work, though, is not fixing the cosmology while solving for redshift parameters. 

\begin{figure*}
\begin{center}
\includegraphics[width=1.0 \textwidth]{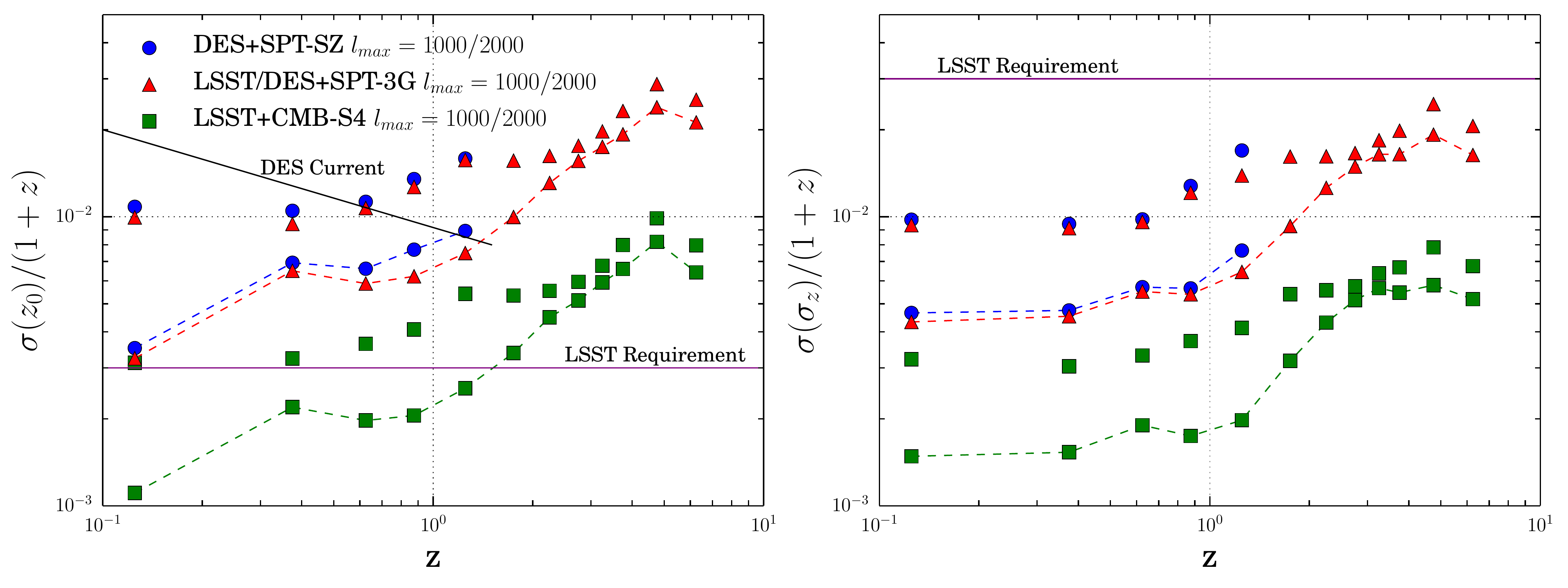}
\end{center}
\caption{The constraints on the mean of a binned redshift distribution (left) and the width of the distribution (right) for our fiducial analyses of different surveys with no prior information on any parameters. The unconnected points are with $l_{\text{max}}=1000$ and the dotted lines are with $l_{\text{max}}=2000$. We compare these constraints to the current approximate DES constraints for mean redshifts in a bin from photometric redshift methods \cite{davis17} and target constraints for both redshift parameters in LSST. The LSST requirements are from the LSST DESC Science Requirements Document \cite{newlsst}. We note that DES has not tried to constrain $\sigma_{\text{z}}$ directly in data, so we only show a constraint on $z_0$. These constraints from DES are roughly constant with redshift, so we see a decrease with redshift when plotting $\sigma(z_0)/(1+z)$.}
\label{fig:comparez0}
\end{figure*}

As mentioned in Section \ref{sec:priors}, the Dark Energy Survey is already calibrating the mean redshift of bins to an uncertainty of about 0.02. The Large Synoptic Survey Telescope broadly has a requirement of constraining the mean of redshift bins to a precision of $0.003 (1+z)$, though likely that number can be improved upon at low redshifts as mentioned in Section \ref{sec:priors}. In Figure \ref{fig:comparez0}, we compare the LSST DESC SRD \cite{newlsst} required redshift constraints and the current DES redshift constraints to our Fisher analysis of $\sigma_{\text{z}}$ and $z_0$ with no prior information applied. We show results for both $l_{\text{max}}=1000 \ \text{and} \ 2000$ in Figure \ref{fig:comparez0}. The projections on DES from correlations with SPT beat the current threshold of 0.02 constraints on the redshift parameters in the first three redshift bins, even if only $l_{\text{max}}=1000$ can be used. As mentioned previously, currently DES has only constrained the mean redshift of bins, $z_0$ and not the width, $\sigma_{\text{z}}$. Work in, e.g., \cite{newman08} suggests constraints on each parameter should be comparable, though, from clustering redshift measurements with spectroscopic galaxies. For LSST, the constraints for $l_{\text{max}}=2000$ at low redshifts ($z<1.5$) are stronger than the goal 0.003(1+z) uncertainty on $z_0$. For $l_{\text{max}}=1000$, the constraints are weaker than this goal, though within a factor of 2 for $z<3$. All of the constraints for both $l_{\text{max}}$ values are better than the LSST requirement on $\sigma_{\text{z}}$ of $0.03(1+z)$ for large-scale structure analyses.  

This result of getting competitive redshift constraints from only the self-calibration of power spectra measurements is significant. The results in Figure \ref{fig:comparez0} show that most of the current LSST DESC SRD requirements can be beaten with this method, particularly if small scales out to $l_{\text{max}}=2000$ can be used. Even if the constraints of self-calibrating redshifts from power spectra measurements end up merely comparable to traditional methods of photometric redshift estimation, though, this could add significant information to cosmic surveys. A discrepancy could point to systematics in either the photometric redshift or power spectra measurements. We note that our methodology is not strictly independent of a photometric redshift code, as it does implicitly assume the use of a photo-z (or some other) method to bin the galaxies in the first place, in particular creating bins with a $\sigma_z$ smaller than the bin size. (See \cite{tanoglidis} for a study on the effects of various bin widths and $\sigma_z$ values for a galaxy clustering analysis.)

\section{Constraints with Alternative Models}
\label{sec:altmodels}

In this section, we look at how our results vary with a couple simple changes to our fiducial model of keeping all cosmological parameters fixed, except for $\sigma_{8,i}$ in 12 redshift bins, and $\Omega_{\text{m}}$. We consider two alternative models. The first uses a single $\sigma_8$ parameter instead of a $\sigma_{8,i}$ in each of the 12 redshift bins. Specifically, this generalizes Equation \ref{sigma8} to have one value $s$ rather than 12 $s_i$ values. This model thus allows only a constant scaling of $\sigma_8$ with respect to the $\Lambda$CDM prediction across all redshifts, rather than the more flexible 12 $\sigma_{8,i}$ values.

The second modification we explore is allowing more cosmological parameters to vary. We include five extra parameters: the dark energy equation of state parameters, $w_0$ and $w_{\text{a}}$; the Hubble constant, $H_0$; the density parameter for baryons, $\Omega_\text{b}$; and the primordial spectral index for curvature perturbations at wave number k=0.05 Mpc$^{-1}$, $n_{\text{s}}$ (see, e.g., \cite{planck15} for more details on parameters). We again vary parameters from their fiducial values set to the Planck 2015 flat-$\Lambda \text{CDM}$ cosmological parameters including external data \citep{planck15} as shown in Section \ref{sec:methods}. The dark energy parameters are varied from their fiducial $\Lambda \text{CDM}$ values of $w_0=-1$ and $w_{\text{a}}=0$. This set of parameters is similar to those used for exploring weak lensing surveys in \cite{schaan17}. As with other parameters, we include no prior information, and just allow the data (galaxy and CMB lensing correlations) to constrain all parameters simultaneously. For both of these analyses, we continue to vary parameters in each of the 12 redshift bins for $b_{\text{g}}, \sigma_{\text{z}}$ and $z_0$, or just $b_{\text{g}}$ in results labeled `no z uncertainty,' as well as $\Omega_{\text{m}}$.

\begin{figure}
\begin{center}
\includegraphics[width=0.5 \textwidth]{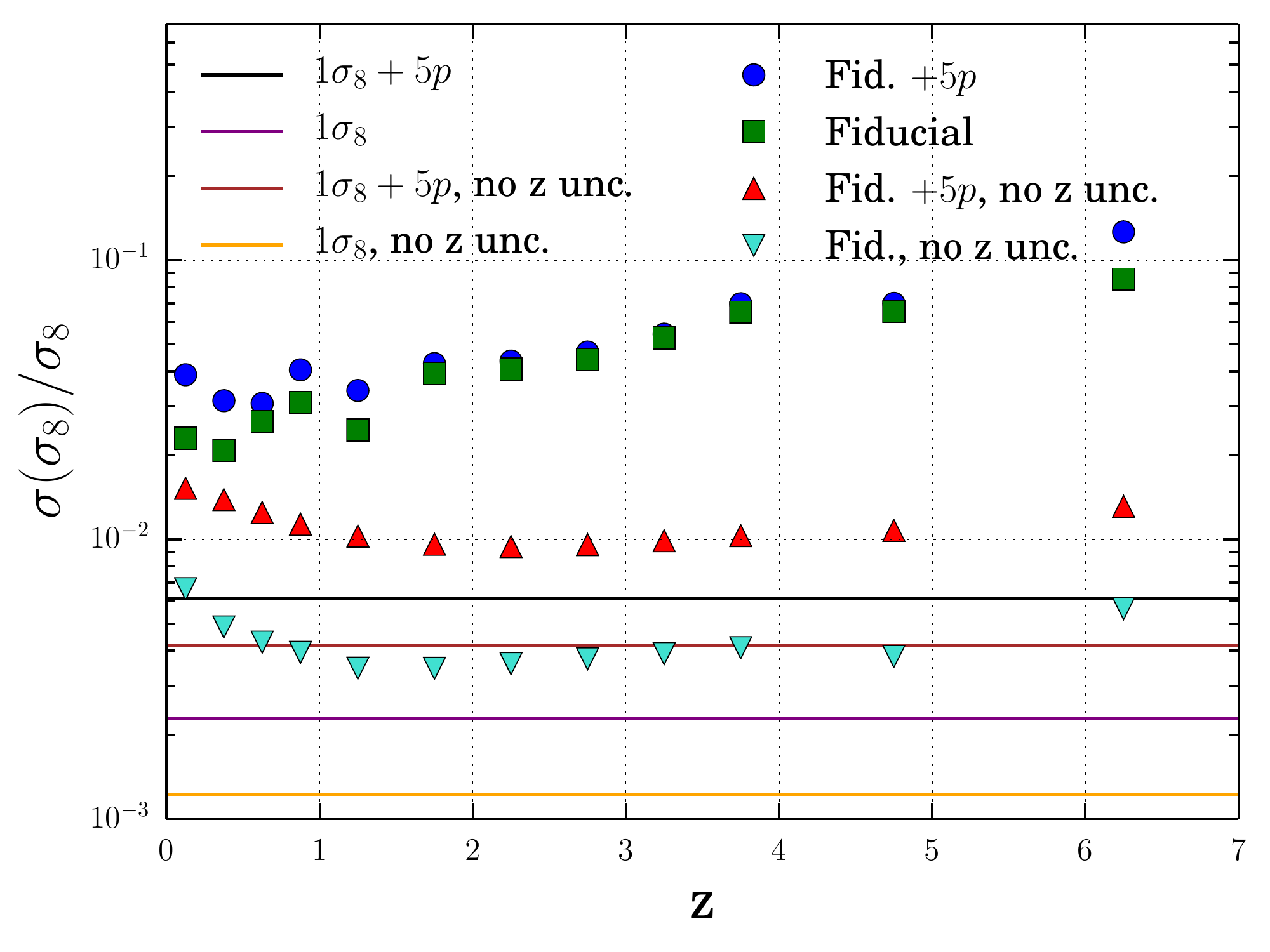}
\end{center}
\caption{Constraints on $\sigma_8$ for a range of alternative models. `Fiducial' signifies our main analysis with 12 redshift bins and a $\sigma_{8,i}$ parameter in each of them. `$5 p$' signifies a model in which we add five cosmological parameters which can vary in our Fisher analysis: $w_0, w_{\text{a}}, H_0, \Omega_{\text{b}}$ and $n_{\text{s}}$. The horizontal lines are constraints in models in which we have just a single $\sigma_8$ amplitude that affects all redshifts ($1 \sigma_8$). For each model, we also show the case in which redshifts are perfectly known (no z uncertainty) meaning $\sigma_{\text{z}}$ and $z_0$ are not allowed to vary in the Fisher analysis. Compared to the fiducial analysis, adding cosmological parameters impacts the results far less than changing the amount of redshift information or to a redshift-independent $\sigma_8$ model does. We use $l_{\text{max}}=1000$ for this part of the analysis.}
\label{fig:altmodels}
\end{figure}

Our results for $\sigma_8$ constraints with these two types of modifications are shown in Figure \ref{fig:altmodels}. Of note, the constraints from the fiducial analysis of $\sigma_8$ across the 12 redshift bins change by only an average of 24 $\%$ when adding the five extra parameters (comparing the blue and green data points). Similar to $\Omega_{\text{m}}$, since these extra parameters do not have the same degenerate scaling of $\sigma_8, b_{\text{g}}$, and $\sigma_{\text{z}}$, their inclusion has a relatively minor effect (see Figure \ref{fig:dclextra} in Appendix \ref{sec:appendixextra}). In comparison, removing the part of the degeneracy either by fixing $\sigma_{\text{z}}$ and $z_0$ (the no z uncertainty labeled points) or eliminating the $\sigma_{8,i}$ in favor of a single $\sigma_8$ scaling across all redshift bins (the four solid lines in Figure \ref{fig:altmodels}) makes a much larger difference in constraining power of $\sigma_8$.

As expected, when we reduce the parameter space to a single $\sigma_8$ value ($1 \sigma_8$ in Figure \ref{fig:altmodels}) rather than 12 $\sigma_{8,i}$ parameters (fiducial), we see significantly increased precision. The constraint for the single $\sigma_8$ parameter is about a factor of 20 smaller than the constraints on the low-z redshift bins in the fiducial analysis.

On the other hand, these results also highlight that measurements of galaxies and CMB lensing are particularly sensitive to measuring $\sigma_8$ as a function of $z$. As mentioned, when adding five additional cosmological parameters to the fiducial analysis, the $\sigma_{8,i}$ are only marginally affected, by about 30$ \%$. For the single-value $\sigma_8$ model, though, adding five parameters degrades constraints by almost a factor of 3. Thus, in the single-value model, priors on a number of different cosmological parameters are impactful to using these galaxy and CMB measurements. In the $\sigma_{8,i}$ model, though, these measurements give comparable constraints with or without prior information on these five extra parameters. Instead, the key factor to improved $\sigma_{8,i}$ constraints is the accuracy of the redshifts, particularly $\sigma_{\text{z}}$. We note that, even for a single-value $\sigma_8$ model, a combination of a lensing observable (CMB or otherwise) is still needed with a galaxy clustering sample to break the $\sigma_8 -b_g$ (galaxy bias) degeneracy.

\section{Conclusions}
\label{sec:conclusions}

In this work, we sought to answer two questions: 1. how are analyses of galaxy clustering and CMB lensing affected by uncertainties in redshift parameters, and 2. can redshift parameters be self-calibrated by galaxy and CMB lensing correlations? We found in Section \ref{sec:fiducial} that the presence of redshift uncertainties can increase errors on, e.g., $\sigma_8(z)$ by an order of magnitude. We showed the importance of using the cross-correlations of different galaxy bins ($C_l^{g_i g_j}$), which in the assumption of perfect redshift knowledge is not a necessary measurement.

Though the redshift uncertainties degrade the analysis, the projected cosmological constraints are still fairly impressive. Our fiducial analysis (Figure \ref{fig:s8_allsurveys}) constrains $\sigma_8$ in each redshift bin in the DES/SPT-SZ era to about $10 \%$. For LSST/CMB-S4, the constraints get down to $2 \%-3 \%$ at low redshifts ($z<1.5$) and are still below $7 \%$ higher at higher redshifts. Constraints of this level should help in distinguishing between e.g., $\Lambda \text{CDM}$ and models of modified general relativity as the cause of cosmic acceleration. As a comparison, \cite{lsstsciencebook} predicts $2 \%$ measurements on $\sigma_8(z)$ from $z=0.5-3$ from LSST weak lensing and baryon acoustic oscillation data plus Planck CMB results and finds that these constraints could decisively rule out, e.g., a Dvali-Gabadadze-Porrati modified general relativity model \cite{dgp}.

In Section \ref{sec:surveyparams}, we explored what survey parameters most affect these measurements of cosmological and redshift parameters. Among different survey parameters explored individually, we found the largest dependences on $f_{\text{sky}}$, $l_{\text{max}}$, and priors on the redshift parameters. The constraining power can be doubled or better by increasing $l_{\text{max}}$ from 1000 to 2000 (Figure \ref{fig:lmaxplot}) or with good priors on the redshift parameters from other data sources (e.g., Figure \ref{fig:lsstpriors1}). The analysis of $f_{\text{sky}}$ (Figure \ref{fig:fskyplot}) shows that the significant increase in overlap of surveys in the future (LSST/CMB-S4 will have eight times as much overlapping area as DES/SPT) accounts for much of the increased precision on $\sigma_8$. In contrast, we found that increasing the galaxy density or reducing the CMB lensing noise (Figure \ref{fig:noiseplot}) beyond expectations for LSST/CMB-S4 only yields marginal improvement since the constraints approach the cosmic variance limits.

We also showed in Section \ref{sec:selfcalibration} the constraints on redshift parameters from the Fisher analysis and compared them to current and expected constraints on redshift parameters from photometric redshift techniques (Figure \ref{fig:comparez0}). The constraints projected in this work are comparable to the photometric techniques. This suggests that self-calibration of redshift parameters from cosmological measurements themselves can be competitive with other techniques. That such constraints can be achieved simultaneously with cosmological constraints (i.e., $\sigma_8$) is an important finding for the feasibility of this method as a redshift probe. While we did also show in Section \ref{sec:surveyparams} that priors expected from clustering redshifts may be comparable in performance to perfect redshift knowledge, such priors will likely only be achieved for low redshifts (e.g. \cite{newman15} explores $z<1.5$). Thus, the self-calibration redshifts found in our study may be vital to higher redshift analyses, and at least a check for low redshifts.

Finally, in Section \ref{sec:altmodels}, we explored some extended models using different sets of parameters compared to our fiducial analysis. We found that adding multiple cosmological parameters only marginally impacts the results, since most of the parameters are not degenerate with $\sigma_{8,i}$ and the redshift width, $\sigma_{\text{z},i}$, in the different redshift bins. We also found that simplifying to a single-$\sigma_8$ parameter across redshifts unsurprisingly leads to smaller constraints, but such a model is also impacted more by adding extra parameters into the analysis (Figure \ref{fig:altmodels}).

A number of assumptions that may need more study in the future were made in this work. The largest element that was a focus of this work was the redshift distribution modeling. A two-parameter Gaussian model may not be sufficient for accurately incorporating redshift distributions and their uncertainties into analyses on data. More work on the resilience of this model and extensions to make the model more flexible should be done. An advantage of the simple model we use is the strong dependence of the power spectra on the redshift parameters. This allows for self-calibration of redshift parameters from just the power spectra measurements. A risk in having too many redshift parameters is creating degeneracies in which multiple redshift parameters may impact the power spectra in similar ways. Another effect we do not address is that of `catastrophic redshift errors' (see, e.g., \cite{hearin2010}), in which galaxies are placed into a photometric bin significantly offset from their true redshift. This is unlike our model of an unbiased Gaussian noise added to the redshift estimates in Section \ref{sec:makingdndz}. Adding such errors to our analysis would also significantly add to the modeling parameter space. We leave such investigations for future work, though we note that \cite{SS17} finds that these types of errors can be constrained from the galaxy clustering and galaxy-CMB lensing correlations, so their impact may be qualitatively different from the redshift-related degeneracies studied here. We also note that there may be inaccuracies in our analysis due to using the Limber approximation (Equation \ref{clequation}) at low $l$. This is discussed in Appendix \ref{sec:appendix2}.

There are several other possibly impactful parameters not addressed in this work which are mentioned in \cite{SS17}, whose analysis we broadly followed in order to isolate the effects of adding redshift uncertainty. These factors include nonlinear galaxy bias, non-Gaussian terms in the covariance, redshift space distortions, biases in the CMB lensing map, and differences between a Monte Carlo analysis and a Fisher analysis. \cite{SS17} also notes that bispectra could add useful information to an analysis like this.

This work should highlight the importance of incorporating redshift uncertainty and modeling into cosmological analyses using galaxies and CMB lensing, as well as inspire more work on self-calibrating redshifts with these and other measurements. While we did not use weak gravitational lensing of galaxies (cosmic shear), similar concerns about redshift uncertainties and modeling should be studied for using that probe, and many of the techniques in this work could be applied. The issue of how to address redshift uncertainty has never been more important than the upcoming era of LSST, in which we will be probing redshift regimes currently still sparse in available spectroscopic measurements for calibrating photometric redshift techniques.

\section*{Acknowledgments}

R.C. thanks Josh Frieman, Scott Dodelson, Sam Passaglia, Chihway Chang, Eric Baxter, Ami Choi, and Ben Hoyle for useful conversations related to this work. R.C. is supported by the Kavli Institute for Cosmological Physics at the University of Chicago through Grant No. NSF PHY-1125897 and an endowment from the Kavli Foundation and its founder Fred Kavli.

\addcontentsline{toc}{chapter}{Bibliography}
\bibliographystyle{apsrev4-1}
\bibliography{references_cmb_pz4.bib}

\begin{thebibliography}{68}%
\makeatletter
\providecommand \@ifxundefined [1]{%
 \@ifx{#1\undefined}
}%
\providecommand \@ifnum [1]{%
 \ifnum #1\expandafter \@firstoftwo
 \else \expandafter \@secondoftwo
 \fi
}%
\providecommand \@ifx [1]{%
 \ifx #1\expandafter \@firstoftwo
 \else \expandafter \@secondoftwo
 \fi
}%
\providecommand \natexlab [1]{#1}%
\providecommand \enquote  [1]{``#1''}%
\providecommand \bibnamefont  [1]{#1}%
\providecommand \bibfnamefont [1]{#1}%
\providecommand \citenamefont [1]{#1}%
\providecommand \href@noop [0]{\@secondoftwo}%
\providecommand \href [0]{\begingroup \@sanitize@url \@href}%
\providecommand \@href[1]{\@@startlink{#1}\@@href}%
\providecommand \@@href[1]{\endgroup#1\@@endlink}%
\providecommand \@sanitize@url [0]{\catcode `\\12\catcode `\$12\catcode
  `\&12\catcode `\#12\catcode `\^12\catcode `\_12\catcode `\%12\relax}%
\providecommand \@@startlink[1]{}%
\providecommand \@@endlink[0]{}%
\providecommand \url  [0]{\begingroup\@sanitize@url \@url }%
\providecommand \@url [1]{\endgroup\@href {#1}{\urlprefix }}%
\providecommand \urlprefix  [0]{URL }%
\providecommand \Eprint [0]{\href }%
\providecommand \doibase [0]{http://dx.doi.org/}%
\providecommand \selectlanguage [0]{\@gobble}%
\providecommand \bibinfo  [0]{\@secondoftwo}%
\providecommand \bibfield  [0]{\@secondoftwo}%
\providecommand \translation [1]{[#1]}%
\providecommand \BibitemOpen [0]{}%
\providecommand \bibitemStop [0]{}%
\providecommand \bibitemNoStop [0]{.\EOS\space}%
\providecommand \EOS [0]{\spacefactor3000\relax}%
\providecommand \BibitemShut  [1]{\csname bibitem#1\endcsname}%
\let\auto@bib@innerbib\@empty
\bibitem [{\citenamefont {{Huterer}}\ \emph {et~al.}(2015)\citenamefont
  {{Huterer}} \emph {et~al.}}]{huterer2015}%
  \BibitemOpen
  \bibfield  {author} {\bibinfo {author} {\bibfnamefont {D.}~\bibnamefont
  {{Huterer}}} \emph {et~al.},\ }\href {\doibase
  10.1016/j.astropartphys.2014.07.004} {\bibfield  {journal} {\bibinfo
  {journal} {Astroparticle Physics}\ }\textbf {\bibinfo {volume} {63}},\
  \bibinfo {pages} {23} (\bibinfo {year} {2015})},\ \Eprint
  {http://arxiv.org/abs/1309.5385} {arXiv:1309.5385} \BibitemShut {NoStop}%
\bibitem [{\citenamefont {{Flaugher}}(2005)}]{DES}%
  \BibitemOpen
  \bibfield  {author} {\bibinfo {author} {\bibfnamefont {B.}~\bibnamefont
  {{Flaugher}}},\ }\href {\doibase 10.1142/S0217751X05025917} {\bibfield
  {journal} {\bibinfo  {journal} {International Journal of Modern Physics A}\
  }\textbf {\bibinfo {volume} {20}},\ \bibinfo {pages} {3121} (\bibinfo {year}
  {2005})}\BibitemShut {NoStop}%
\bibitem [{\citenamefont {{de Jong}}\ \emph {et~al.}(2013)\citenamefont {{de
  Jong}} \emph {et~al.}}]{kids}%
  \BibitemOpen
  \bibfield  {author} {\bibinfo {author} {\bibfnamefont {J.~T.~A.}\
  \bibnamefont {{de Jong}}} \emph {et~al.},\ }\href {\doibase
  2013Msngr.154...44D} {\bibfield  {journal} {\bibinfo  {journal} {The
  Messenger}\ }\textbf {\bibinfo {volume} {154}},\ \bibinfo {pages} {44}
  (\bibinfo {year} {2013})}\BibitemShut {NoStop}%
\bibitem [{\citenamefont {{Heymans}}\ \emph {et~al.}(2012)\citenamefont
  {{Heymans}} \emph {et~al.}}]{2012MNRAS.427..146H}%
  \BibitemOpen
  \bibfield  {author} {\bibinfo {author} {\bibfnamefont {C.}~\bibnamefont
  {{Heymans}}} \emph {et~al.},\ }\href {\doibase
  10.1111/j.1365-2966.2012.21952.x} {\bibfield  {journal} {\bibinfo  {journal}
  {\mnras}\ }\textbf {\bibinfo {volume} {427}},\ \bibinfo {pages} {146}
  (\bibinfo {year} {2012})},\ \Eprint {http://arxiv.org/abs/1210.0032}
  {arXiv:1210.0032} \BibitemShut {NoStop}%
\bibitem [{\citenamefont {{Miyazaki}}\ \emph {et~al.}(2012)\citenamefont
  {{Miyazaki}} \emph {et~al.}}]{2012SPIE.8446E..0ZM}%
  \BibitemOpen
  \bibfield  {author} {\bibinfo {author} {\bibfnamefont {S.}~\bibnamefont
  {{Miyazaki}}} \emph {et~al.},\ }in\ \href {\doibase 10.1117/12.926844} {\emph
  {\bibinfo {booktitle} {Ground-based and Airborne Instrumentation for
  Astronomy IV}}},\ \bibinfo {series} {\procspie}, Vol.\ \bibinfo {volume}
  {8446}\ (\bibinfo {year} {2012})\ p.\ \bibinfo {pages} {84460Z}\BibitemShut
  {NoStop}%
\bibitem [{\citenamefont {{Abbott}}\ \emph
  {et~al.}(2018{\natexlab{a}})\citenamefont {{Abbott}} \emph
  {et~al.}}]{keypaper}%
  \BibitemOpen
  \bibfield  {author} {\bibinfo {author} {\bibfnamefont {T.~M.~C.}\
  \bibnamefont {{Abbott}}} \emph {et~al.} (\bibinfo {collaboration} {Dark
  Energy Survey Collaboration}),\ }\href {\doibase 10.1103/PhysRevD.98.043526}
  {\bibfield  {journal} {\bibinfo  {journal} {\prd}\ }\textbf {\bibinfo
  {volume} {98}},\ \bibinfo {pages} {043526} (\bibinfo {year}
  {2018}{\natexlab{a}})},\ \Eprint {http://arxiv.org/abs/1708.01530}
  {arXiv:1708.01530} \BibitemShut {NoStop}%
\bibitem [{\citenamefont {{Elvin-Poole}}\ \emph {et~al.}(2018)\citenamefont
  {{Elvin-Poole}} \emph {et~al.}}]{elvinpoole17}%
  \BibitemOpen
  \bibfield  {author} {\bibinfo {author} {\bibfnamefont {J.}~\bibnamefont
  {{Elvin-Poole}}} \emph {et~al.},\ }\href {\doibase
  10.1103/PhysRevD.98.042006} {\bibfield  {journal} {\bibinfo  {journal}
  {\prd}\ }\textbf {\bibinfo {volume} {98}},\ \bibinfo {pages} {042006}
  (\bibinfo {year} {2018})},\ \Eprint {http://arxiv.org/abs/1708.01536}
  {arXiv:1708.01536} \BibitemShut {NoStop}%
\bibitem [{\citenamefont {{Troxel}}\ \emph {et~al.}(2018)\citenamefont
  {{Troxel}} \emph {et~al.}}]{troxel17}%
  \BibitemOpen
  \bibfield  {author} {\bibinfo {author} {\bibfnamefont {M.~A.}\ \bibnamefont
  {{Troxel}}} \emph {et~al.},\ }\href {\doibase 10.1103/PhysRevD.98.043528}
  {\bibfield  {journal} {\bibinfo  {journal} {\prd}\ }\textbf {\bibinfo
  {volume} {98}},\ \bibinfo {pages} {043528} (\bibinfo {year} {2018})},\
  \Eprint {http://arxiv.org/abs/1708.01538} {arXiv:1708.01538} \BibitemShut
  {NoStop}%
\bibitem [{\citenamefont {{Prat}}\ \emph {et~al.}(2018)\citenamefont {{Prat}}
  \emph {et~al.}}]{prat17}%
  \BibitemOpen
  \bibfield  {author} {\bibinfo {author} {\bibfnamefont {J.}~\bibnamefont
  {{Prat}}} \emph {et~al.},\ }\href {\doibase 10.1103/PhysRevD.98.042005}
  {\bibfield  {journal} {\bibinfo  {journal} {\prd}\ }\textbf {\bibinfo
  {volume} {98}},\ \bibinfo {pages} {042005} (\bibinfo {year} {2018})},\
  \Eprint {http://arxiv.org/abs/1708.01537} {arXiv:1708.01537} \BibitemShut
  {NoStop}%
\bibitem [{\citenamefont {{Abbott}}\ \emph
  {et~al.}(2018{\natexlab{b}})\citenamefont {{Abbott}} \emph {et~al.}}]{dr1}%
  \BibitemOpen
  \bibfield  {author} {\bibinfo {author} {\bibfnamefont {T.~M.~C.}\
  \bibnamefont {{Abbott}}} \emph {et~al.} (\bibinfo {collaboration} {Dark
  Energy Survey Collaboration}),\ }\href@noop {} {\bibfield  {journal}
  {\bibinfo  {journal} {ArXiv e-prints}\ } (\bibinfo {year}
  {2018}{\natexlab{b}})},\ \Eprint {http://arxiv.org/abs/1801.03181}
  {arXiv:1801.03181 [astro-ph.IM]} \BibitemShut {NoStop}%
\bibitem [{\citenamefont {{LSST Dark Energy Science
  Collaboration}}(2012)}]{lsstdesc}%
  \BibitemOpen
  \bibfield  {author} {\bibinfo {author} {\bibnamefont {{LSST Dark Energy
  Science Collaboration}}},\ }\href@noop {} {\bibfield  {journal} {\bibinfo
  {journal} {ArXiv e-prints}\ } (\bibinfo {year} {2012})},\ \Eprint
  {http://arxiv.org/abs/1211.0310} {arXiv:1211.0310 [astro-ph.CO]} \BibitemShut
  {NoStop}%
\bibitem [{\citenamefont {{Laureijs}}\ \emph {et~al.}(2011)\citenamefont
  {{Laureijs}} \emph {et~al.}}]{euclid}%
  \BibitemOpen
  \bibfield  {author} {\bibinfo {author} {\bibfnamefont {R.}~\bibnamefont
  {{Laureijs}}} \emph {et~al.},\ }\href@noop {} {\bibfield  {journal} {\bibinfo
   {journal} {arXiv e-prints}\ } (\bibinfo {year} {2011})},\ \Eprint
  {http://arxiv.org/abs/1110.3193} {arXiv:1110.3193 [astro-ph.CO]} \BibitemShut
  {NoStop}%
\bibitem [{\citenamefont {{Spergel}}\ \emph {et~al.}(2015)\citenamefont
  {{Spergel}} \emph {et~al.}}]{wfirst}%
  \BibitemOpen
  \bibfield  {author} {\bibinfo {author} {\bibfnamefont {D.}~\bibnamefont
  {{Spergel}}} \emph {et~al.},\ }\href@noop {} {\bibfield  {journal} {\bibinfo
  {journal} {arXiv e-prints}\ } (\bibinfo {year} {2015})},\ \Eprint
  {http://arxiv.org/abs/1503.03757} {arXiv:1503.03757 [astro-ph.IM]}
  \BibitemShut {NoStop}%
\bibitem [{\citenamefont {{Ivezi{\'c}}}\ \emph {et~al.}(2008)\citenamefont
  {{Ivezi{\'c}}} \emph {et~al.}}]{izeviclsst}%
  \BibitemOpen
  \bibfield  {author} {\bibinfo {author} {\bibfnamefont {{\v Z}.}~\bibnamefont
  {{Ivezi{\'c}}}} \emph {et~al.},\ }\href@noop {} {\bibfield  {journal}
  {\bibinfo  {journal} {ArXiv e-prints}\ } (\bibinfo {year} {2008})},\ \Eprint
  {http://arxiv.org/abs/0805.2366} {arXiv:0805.2366} \BibitemShut {NoStop}%
\bibitem [{\citenamefont {{The Planck
  Collaboration}}(2006)}]{2006astro.ph..4069T}%
  \BibitemOpen
  \bibfield  {author} {\bibinfo {author} {\bibnamefont {{The Planck
  Collaboration}}},\ }\href@noop {} {\bibfield  {journal} {\bibinfo  {journal}
  {ArXiv Astrophysics e-prints}\ } (\bibinfo {year} {2006})},\ \Eprint
  {http://arxiv.org/abs/astro-ph/0604069} {astro-ph/0604069} \BibitemShut
  {NoStop}%
\bibitem [{\citenamefont {{Smith}}\ \emph {et~al.}(2007)\citenamefont
  {{Smith}}, \citenamefont {{Zahn}},\ and\ \citenamefont
  {{Dor{\'e}}}}]{2007PhRvD..76d3510S}%
  \BibitemOpen
  \bibfield  {author} {\bibinfo {author} {\bibfnamefont {K.~M.}\ \bibnamefont
  {{Smith}}}, \bibinfo {author} {\bibfnamefont {O.}~\bibnamefont {{Zahn}}}, \
  and\ \bibinfo {author} {\bibfnamefont {O.}~\bibnamefont {{Dor{\'e}}}},\
  }\href {\doibase 10.1103/PhysRevD.76.043510} {\bibfield  {journal} {\bibinfo
  {journal} {\prd}\ }\textbf {\bibinfo {volume} {76}},\ \bibinfo {eid} {043510}
  (\bibinfo {year} {2007})},\ \Eprint {http://arxiv.org/abs/0705.3980}
  {arXiv:0705.3980} \BibitemShut {NoStop}%
\bibitem [{\citenamefont {{Giannantonio}}\ \emph {et~al.}(2016)\citenamefont
  {{Giannantonio}} \emph {et~al.}}]{giannantonio16}%
  \BibitemOpen
  \bibfield  {author} {\bibinfo {author} {\bibfnamefont {T.}~\bibnamefont
  {{Giannantonio}}} \emph {et~al.},\ }\href {\doibase 10.1093/mnras/stv2678}
  {\bibfield  {journal} {\bibinfo  {journal} {\mnras}\ }\textbf {\bibinfo
  {volume} {456}},\ \bibinfo {pages} {3213} (\bibinfo {year} {2016})},\ \Eprint
  {http://arxiv.org/abs/1507.05551} {arXiv:1507.05551} \BibitemShut {NoStop}%
\bibitem [{\citenamefont {{Omori}}\ \emph
  {et~al.}(2019{\natexlab{a}})\citenamefont {{Omori}} \emph
  {et~al.}}]{omori18}%
  \BibitemOpen
  \bibfield  {author} {\bibinfo {author} {\bibfnamefont {Y.}~\bibnamefont
  {{Omori}}} \emph {et~al.},\ }\href {\doibase 10.1103/PhysRevD.100.043501}
  {\bibfield  {journal} {\bibinfo  {journal} {\prd}\ }\textbf {\bibinfo
  {volume} {100}},\ \bibinfo {eid} {043501} (\bibinfo {year}
  {2019}{\natexlab{a}})},\ \Eprint {http://arxiv.org/abs/1810.02342}
  {arXiv:1810.02342 [astro-ph.CO]} \BibitemShut {NoStop}%
\bibitem [{\citenamefont {{Peacock}}\ and\ \citenamefont
  {{Bilicki}}(2018)}]{peacock18}%
  \BibitemOpen
  \bibfield  {author} {\bibinfo {author} {\bibfnamefont {J.~A.}\ \bibnamefont
  {{Peacock}}}\ and\ \bibinfo {author} {\bibfnamefont {M.}~\bibnamefont
  {{Bilicki}}},\ }\href {\doibase 10.1093/mnras/sty2314} {\bibfield  {journal}
  {\bibinfo  {journal} {\mnras}\ }\textbf {\bibinfo {volume} {481}},\ \bibinfo
  {pages} {1133} (\bibinfo {year} {2018})},\ \Eprint
  {http://arxiv.org/abs/1805.11525} {arXiv:1805.11525} \BibitemShut {NoStop}%
\bibitem [{\citenamefont {{Carlstrom}}\ \emph {et~al.}(2011)\citenamefont
  {{Carlstrom}} \emph {et~al.}}]{spt}%
  \BibitemOpen
  \bibfield  {author} {\bibinfo {author} {\bibfnamefont {J.~E.}\ \bibnamefont
  {{Carlstrom}}} \emph {et~al.},\ }\href {\doibase 10.1086/659879} {\bibfield
  {journal} {\bibinfo  {journal} {\pasp}\ }\textbf {\bibinfo {volume} {123}},\
  \bibinfo {pages} {568} (\bibinfo {year} {2011})},\ \Eprint
  {http://arxiv.org/abs/0907.4445} {arXiv:0907.4445 [astro-ph.IM]} \BibitemShut
  {NoStop}%
\bibitem [{\citenamefont {{Abazajian}}\ \emph {et~al.}(2015)\citenamefont
  {{Abazajian}} \emph {et~al.}}]{2015APh....63...66A}%
  \BibitemOpen
  \bibfield  {author} {\bibinfo {author} {\bibfnamefont {K.~N.}\ \bibnamefont
  {{Abazajian}}} \emph {et~al.},\ }\href {\doibase
  10.1016/j.astropartphys.2014.05.014} {\bibfield  {journal} {\bibinfo
  {journal} {Astroparticle Physics}\ }\textbf {\bibinfo {volume} {63}},\
  \bibinfo {pages} {66} (\bibinfo {year} {2015})},\ \Eprint
  {http://arxiv.org/abs/1309.5383} {arXiv:1309.5383} \BibitemShut {NoStop}%
\bibitem [{\citenamefont {{Schmittfull}}\ and\ \citenamefont
  {{Seljak}}(2018)}]{SS17}%
  \BibitemOpen
  \bibfield  {author} {\bibinfo {author} {\bibfnamefont {M.}~\bibnamefont
  {{Schmittfull}}}\ and\ \bibinfo {author} {\bibfnamefont {U.}~\bibnamefont
  {{Seljak}}},\ }\href {\doibase 10.1103/PhysRevD.97.123540} {\bibfield
  {journal} {\bibinfo  {journal} {\prd}\ }\textbf {\bibinfo {volume} {97}},\
  \bibinfo {eid} {123540} (\bibinfo {year} {2018})},\ \Eprint
  {http://arxiv.org/abs/1710.09465} {arXiv:1710.09465} \BibitemShut {NoStop}%
\bibitem [{\citenamefont {{Modi}}\ \emph {et~al.}(2017)\citenamefont {{Modi}},
  \citenamefont {{White}},\ and\ \citenamefont {{Vlah}}}]{modi17}%
  \BibitemOpen
  \bibfield  {author} {\bibinfo {author} {\bibfnamefont {C.}~\bibnamefont
  {{Modi}}}, \bibinfo {author} {\bibfnamefont {M.}~\bibnamefont {{White}}}, \
  and\ \bibinfo {author} {\bibfnamefont {Z.}~\bibnamefont {{Vlah}}},\ }\href
  {\doibase 10.1088/1475-7516/2017/08/009} {\bibfield  {journal} {\bibinfo
  {journal} {\jcap}\ }\textbf {\bibinfo {volume} {8}},\ \bibinfo {eid} {009}
  (\bibinfo {year} {2017})},\ \Eprint {http://arxiv.org/abs/1706.03173}
  {arXiv:1706.03173} \BibitemShut {NoStop}%
\bibitem [{\citenamefont {{Dawson}}\ \emph {et~al.}(2013)\citenamefont
  {{Dawson}} \emph {et~al.}}]{boss}%
  \BibitemOpen
  \bibfield  {author} {\bibinfo {author} {\bibfnamefont {K.~S.}\ \bibnamefont
  {{Dawson}}} \emph {et~al.},\ }\href {\doibase 10.1088/0004-6256/145/1/10}
  {\bibfield  {journal} {\bibinfo  {journal} {\aj}\ }\textbf {\bibinfo {volume}
  {145}},\ \bibinfo {eid} {10} (\bibinfo {year} {2013})},\ \Eprint
  {http://arxiv.org/abs/1208.0022} {arXiv:1208.0022 [astro-ph.CO]} \BibitemShut
  {NoStop}%
\bibitem [{\citenamefont {{Levi}}\ \emph {et~al.}(2013)\citenamefont {{Levi}}
  \emph {et~al.}}]{desi13}%
  \BibitemOpen
  \bibfield  {author} {\bibinfo {author} {\bibfnamefont {M.}~\bibnamefont
  {{Levi}}} \emph {et~al.},\ }\href@noop {} {\bibfield  {journal} {\bibinfo
  {journal} {ArXiv e-prints}\ } (\bibinfo {year} {2013})},\ \Eprint
  {http://arxiv.org/abs/1308.0847} {arXiv:1308.0847 [astro-ph.CO]} \BibitemShut
  {NoStop}%
\bibitem [{\citenamefont {{Hoyle}}\ \emph {et~al.}(2018)\citenamefont {{Hoyle}}
  \emph {et~al.}}]{hoyle18des}%
  \BibitemOpen
  \bibfield  {author} {\bibinfo {author} {\bibfnamefont {B.}~\bibnamefont
  {{Hoyle}}} \emph {et~al.},\ }\href {\doibase 10.1093/mnras/sty957} {\bibfield
   {journal} {\bibinfo  {journal} {\mnras}\ }\textbf {\bibinfo {volume}
  {478}},\ \bibinfo {pages} {592} (\bibinfo {year} {2018})},\ \Eprint
  {http://arxiv.org/abs/1708.01532} {arXiv:1708.01532} \BibitemShut {NoStop}%
\bibitem [{\citenamefont {{Bonnett}}\ \emph {et~al.}(2016)\citenamefont
  {{Bonnett}} \emph {et~al.}}]{bonnett2016}%
  \BibitemOpen
  \bibfield  {author} {\bibinfo {author} {\bibfnamefont {C.}~\bibnamefont
  {{Bonnett}}} \emph {et~al.},\ }\href {\doibase 10.1103/PhysRevD.94.042005}
  {\bibfield  {journal} {\bibinfo  {journal} {\prd}\ }\textbf {\bibinfo
  {volume} {94}},\ \bibinfo {eid} {042005} (\bibinfo {year} {2016})},\ \Eprint
  {http://arxiv.org/abs/1507.05909} {arXiv:1507.05909} \BibitemShut {NoStop}%
\bibitem [{\citenamefont {{Newman}}(2008)}]{newman08}%
  \BibitemOpen
  \bibfield  {author} {\bibinfo {author} {\bibfnamefont {J.~A.}\ \bibnamefont
  {{Newman}}},\ }\href {\doibase 10.1086/589982} {\bibfield  {journal}
  {\bibinfo  {journal} {\apj}\ }\textbf {\bibinfo {volume} {684}},\ \bibinfo
  {eid} {88-101} (\bibinfo {year} {2008})},\ \Eprint
  {http://arxiv.org/abs/0805.1409} {arXiv:0805.1409} \BibitemShut {NoStop}%
\bibitem [{\citenamefont {{Cawthon}}\ \emph {et~al.}(2018)\citenamefont
  {{Cawthon}} \emph {et~al.}}]{cawthon17}%
  \BibitemOpen
  \bibfield  {author} {\bibinfo {author} {\bibfnamefont {R.}~\bibnamefont
  {{Cawthon}}} \emph {et~al.},\ }\href {\doibase 10.1093/mnras/sty2424}
  {\bibfield  {journal} {\bibinfo  {journal} {\mnras}\ }\textbf {\bibinfo
  {volume} {481}},\ \bibinfo {pages} {2427} (\bibinfo {year} {2018})},\ \Eprint
  {http://arxiv.org/abs/1712.07298} {arXiv:1712.07298 [astro-ph.CO]}
  \BibitemShut {NoStop}%
\bibitem [{\citenamefont {{Davis}}\ \emph {et~al.}(2017)\citenamefont {{Davis}}
  \emph {et~al.}}]{davis17}%
  \BibitemOpen
  \bibfield  {author} {\bibinfo {author} {\bibfnamefont {C.}~\bibnamefont
  {{Davis}}} \emph {et~al.},\ }\href@noop {} {\bibfield  {journal} {\bibinfo
  {journal} {ArXiv e-prints}\ } (\bibinfo {year} {2017})},\ \Eprint
  {http://arxiv.org/abs/1710.02517} {arXiv:1710.02517} \BibitemShut {NoStop}%
\bibitem [{\citenamefont {{Gatti}}\ \emph {et~al.}(2018)\citenamefont {{Gatti}}
  \emph {et~al.}}]{gatti18}%
  \BibitemOpen
  \bibfield  {author} {\bibinfo {author} {\bibfnamefont {M.}~\bibnamefont
  {{Gatti}}} \emph {et~al.},\ }\href {\doibase 10.1093/mnras/sty466} {\bibfield
   {journal} {\bibinfo  {journal} {\mnras}\ }\textbf {\bibinfo {volume}
  {477}},\ \bibinfo {pages} {1664} (\bibinfo {year} {2018})},\ \Eprint
  {http://arxiv.org/abs/1709.00992} {arXiv:1709.00992} \BibitemShut {NoStop}%
\bibitem [{\citenamefont {{Hoyle}}\ and\ \citenamefont
  {{Rau}}(2019)}]{hoyle18}%
  \BibitemOpen
  \bibfield  {author} {\bibinfo {author} {\bibfnamefont {B.}~\bibnamefont
  {{Hoyle}}}\ and\ \bibinfo {author} {\bibfnamefont {M.~M.}\ \bibnamefont
  {{Rau}}},\ }\href {\doibase 10.1093/mnras/stz502} {\bibfield  {journal}
  {\bibinfo  {journal} {\mnras}\ }\textbf {\bibinfo {volume} {485}},\ \bibinfo
  {pages} {3642} (\bibinfo {year} {2019})},\ \Eprint
  {http://arxiv.org/abs/1802.02581} {arXiv:1802.02581 [astro-ph.CO]}
  \BibitemShut {NoStop}%
\bibitem [{\citenamefont {{Baxter}}\ \emph {et~al.}(2016)\citenamefont
  {{Baxter}} \emph {et~al.}}]{baxter16}%
  \BibitemOpen
  \bibfield  {author} {\bibinfo {author} {\bibfnamefont {E.}~\bibnamefont
  {{Baxter}}} \emph {et~al.},\ }\href {\doibase 10.1093/mnras/stw1584}
  {\bibfield  {journal} {\bibinfo  {journal} {\mnras}\ }\textbf {\bibinfo
  {volume} {461}},\ \bibinfo {pages} {4099} (\bibinfo {year} {2016})},\ \Eprint
  {http://arxiv.org/abs/1602.07384} {arXiv:1602.07384} \BibitemShut {NoStop}%
\bibitem [{\citenamefont {{Kirk}}\ \emph {et~al.}(2016)\citenamefont {{Kirk}}
  \emph {et~al.}}]{kirk16}%
  \BibitemOpen
  \bibfield  {author} {\bibinfo {author} {\bibfnamefont {D.}~\bibnamefont
  {{Kirk}}} \emph {et~al.},\ }\href {\doibase 10.1093/mnras/stw570} {\bibfield
  {journal} {\bibinfo  {journal} {\mnras}\ }\textbf {\bibinfo {volume} {459}},\
  \bibinfo {pages} {21} (\bibinfo {year} {2016})},\ \Eprint
  {http://arxiv.org/abs/1512.04535} {arXiv:1512.04535} \BibitemShut {NoStop}%
\bibitem [{\citenamefont {{Omori}}\ \emph
  {et~al.}(2019{\natexlab{b}})\citenamefont {{Omori}} \emph
  {et~al.}}]{omori182}%
  \BibitemOpen
  \bibfield  {author} {\bibinfo {author} {\bibfnamefont {Y.}~\bibnamefont
  {{Omori}}} \emph {et~al.},\ }\href {\doibase 10.1103/PhysRevD.100.043517}
  {\bibfield  {journal} {\bibinfo  {journal} {\prd}\ }\textbf {\bibinfo
  {volume} {100}},\ \bibinfo {eid} {043517} (\bibinfo {year}
  {2019}{\natexlab{b}})},\ \Eprint {http://arxiv.org/abs/1810.02441}
  {arXiv:1810.02441 [astro-ph.CO]} \BibitemShut {NoStop}%
\bibitem [{\citenamefont {{Schaan}}\ \emph {et~al.}(2017)\citenamefont
  {{Schaan}}, \citenamefont {{Krause}}, \citenamefont {{Eifler}}, \citenamefont
  {{Dor{\'e}}}, \citenamefont {{Miyatake}}, \citenamefont {{Rhodes}},\ and\
  \citenamefont {{Spergel}}}]{schaan17}%
  \BibitemOpen
  \bibfield  {author} {\bibinfo {author} {\bibfnamefont {E.}~\bibnamefont
  {{Schaan}}}, \bibinfo {author} {\bibfnamefont {E.}~\bibnamefont {{Krause}}},
  \bibinfo {author} {\bibfnamefont {T.}~\bibnamefont {{Eifler}}}, \bibinfo
  {author} {\bibfnamefont {O.}~\bibnamefont {{Dor{\'e}}}}, \bibinfo {author}
  {\bibfnamefont {H.}~\bibnamefont {{Miyatake}}}, \bibinfo {author}
  {\bibfnamefont {J.}~\bibnamefont {{Rhodes}}}, \ and\ \bibinfo {author}
  {\bibfnamefont {D.~N.}\ \bibnamefont {{Spergel}}},\ }\href {\doibase
  10.1103/PhysRevD.95.123512} {\bibfield  {journal} {\bibinfo  {journal}
  {\prd}\ }\textbf {\bibinfo {volume} {95}},\ \bibinfo {eid} {123512} (\bibinfo
  {year} {2017})},\ \Eprint {http://arxiv.org/abs/1607.01761}
  {arXiv:1607.01761} \BibitemShut {NoStop}%
\bibitem [{\citenamefont {{Font-Ribera}}\ \emph {et~al.}(2014)\citenamefont
  {{Font-Ribera}}, \citenamefont {{McDonald}}, \citenamefont {{Mostek}},
  \citenamefont {{Reid}}, \citenamefont {{Seo}},\ and\ \citenamefont
  {{Slosar}}}]{nzsource}%
  \BibitemOpen
  \bibfield  {author} {\bibinfo {author} {\bibfnamefont {A.}~\bibnamefont
  {{Font-Ribera}}}, \bibinfo {author} {\bibfnamefont {P.}~\bibnamefont
  {{McDonald}}}, \bibinfo {author} {\bibfnamefont {N.}~\bibnamefont
  {{Mostek}}}, \bibinfo {author} {\bibfnamefont {B.~A.}\ \bibnamefont
  {{Reid}}}, \bibinfo {author} {\bibfnamefont {H.-J.}\ \bibnamefont {{Seo}}}, \
  and\ \bibinfo {author} {\bibfnamefont {A.}~\bibnamefont {{Slosar}}},\ }\href
  {\doibase 10.1088/1475-7516/2014/05/023} {\bibfield  {journal} {\bibinfo
  {journal} {\jcap}\ }\textbf {\bibinfo {volume} {5}},\ \bibinfo {eid} {023}
  (\bibinfo {year} {2014})},\ \Eprint {http://arxiv.org/abs/1308.4164}
  {arXiv:1308.4164} \BibitemShut {NoStop}%
\bibitem [{\citenamefont {{Gorecki}}\ \emph {et~al.}(2014)\citenamefont
  {{Gorecki}}, \citenamefont {{Abate}}, \citenamefont {{Ansari}}, \citenamefont
  {{Barrau}}, \citenamefont {{Baumont}}, \citenamefont {{Moniez}},\ and\
  \citenamefont {{Ricol}}}]{gorecki14}%
  \BibitemOpen
  \bibfield  {author} {\bibinfo {author} {\bibfnamefont {A.}~\bibnamefont
  {{Gorecki}}}, \bibinfo {author} {\bibfnamefont {A.}~\bibnamefont {{Abate}}},
  \bibinfo {author} {\bibfnamefont {R.}~\bibnamefont {{Ansari}}}, \bibinfo
  {author} {\bibfnamefont {A.}~\bibnamefont {{Barrau}}}, \bibinfo {author}
  {\bibfnamefont {S.}~\bibnamefont {{Baumont}}}, \bibinfo {author}
  {\bibfnamefont {M.}~\bibnamefont {{Moniez}}}, \ and\ \bibinfo {author}
  {\bibfnamefont {J.-S.}\ \bibnamefont {{Ricol}}},\ }\href {\doibase
  10.1051/0004-6361/201321102} {\bibfield  {journal} {\bibinfo  {journal}
  {\aap}\ }\textbf {\bibinfo {volume} {561}},\ \bibinfo {eid} {A128} (\bibinfo
  {year} {2014})},\ \Eprint {http://arxiv.org/abs/1301.3010} {arXiv:1301.3010}
  \BibitemShut {NoStop}%
\bibitem [{\citenamefont {{Ono}}\ \emph {et~al.}(2018)\citenamefont {{Ono}}
  \emph {et~al.}}]{goldrush}%
  \BibitemOpen
  \bibfield  {author} {\bibinfo {author} {\bibfnamefont {Y.}~\bibnamefont
  {{Ono}}} \emph {et~al.},\ }\href {\doibase 10.1093/pasj/psx103} {\bibfield
  {journal} {\bibinfo  {journal} {\pasj}\ }\textbf {\bibinfo {volume} {70}},\
  \bibinfo {eid} {S10} (\bibinfo {year} {2018})},\ \Eprint
  {http://arxiv.org/abs/1704.06004} {arXiv:1704.06004} \BibitemShut {NoStop}%
\bibitem [{\citenamefont {{Steidel}}\ and\ \citenamefont
  {{Hamilton}}(1992)}]{steidel92}%
  \BibitemOpen
  \bibfield  {author} {\bibinfo {author} {\bibfnamefont {C.~C.}\ \bibnamefont
  {{Steidel}}}\ and\ \bibinfo {author} {\bibfnamefont {D.}~\bibnamefont
  {{Hamilton}}},\ }\href {\doibase 10.1086/116287} {\bibfield  {journal}
  {\bibinfo  {journal} {\aj}\ }\textbf {\bibinfo {volume} {104}},\ \bibinfo
  {pages} {941} (\bibinfo {year} {1992})}\BibitemShut {NoStop}%
\bibitem [{\citenamefont {{Story}}\ \emph {et~al.}(2013)\citenamefont {{Story}}
  \emph {et~al.}}]{story13}%
  \BibitemOpen
  \bibfield  {author} {\bibinfo {author} {\bibfnamefont {K.~T.}\ \bibnamefont
  {{Story}}} \emph {et~al.},\ }\href {\doibase 10.1088/0004-637X/779/1/86}
  {\bibfield  {journal} {\bibinfo  {journal} {\apj}\ }\textbf {\bibinfo
  {volume} {779}},\ \bibinfo {eid} {86} (\bibinfo {year} {2013})},\ \Eprint
  {http://arxiv.org/abs/1210.7231} {arXiv:1210.7231} \BibitemShut {NoStop}%
\bibitem [{\citenamefont {{van Engelen}}\ \emph {et~al.}(2012)\citenamefont
  {{van Engelen}} \emph {et~al.}}]{vanengelen12}%
  \BibitemOpen
  \bibfield  {author} {\bibinfo {author} {\bibfnamefont {A.}~\bibnamefont {{van
  Engelen}}} \emph {et~al.},\ }\href {\doibase 10.1088/0004-637X/756/2/142}
  {\bibfield  {journal} {\bibinfo  {journal} {\apj}\ }\textbf {\bibinfo
  {volume} {756}},\ \bibinfo {eid} {142} (\bibinfo {year} {2012})},\ \Eprint
  {http://arxiv.org/abs/1202.0546} {arXiv:1202.0546} \BibitemShut {NoStop}%
\bibitem [{\citenamefont {{Omori}}\ \emph {et~al.}(2017)\citenamefont {{Omori}}
  \emph {et~al.}}]{omori17}%
  \BibitemOpen
  \bibfield  {author} {\bibinfo {author} {\bibfnamefont {Y.}~\bibnamefont
  {{Omori}}} \emph {et~al.},\ }\href {\doibase 10.3847/1538-4357/aa8d1d}
  {\bibfield  {journal} {\bibinfo  {journal} {\apj}\ }\textbf {\bibinfo
  {volume} {849}},\ \bibinfo {eid} {124} (\bibinfo {year} {2017})},\ \Eprint
  {http://arxiv.org/abs/1705.00743} {arXiv:1705.00743} \BibitemShut {NoStop}%
\bibitem [{\citenamefont {{Adam}}\ \emph {et~al.}(2016)\citenamefont {{Adam}}
  \emph {et~al.}}]{planckhfi}%
  \BibitemOpen
  \bibfield  {author} {\bibinfo {author} {\bibfnamefont {R.}~\bibnamefont
  {{Adam}}} \emph {et~al.} (\bibinfo {collaboration} {Planck Collaboration}),\
  }\href {\doibase 10.1051/0004-6361/201525820} {\bibfield  {journal} {\bibinfo
   {journal} {\aap}\ }\textbf {\bibinfo {volume} {594}},\ \bibinfo {eid} {A8}
  (\bibinfo {year} {2016})},\ \Eprint {http://arxiv.org/abs/1502.01587}
  {arXiv:1502.01587} \BibitemShut {NoStop}%
\bibitem [{\citenamefont {{Okamoto}}\ and\ \citenamefont
  {{Hu}}(2003)}]{okamotohu}%
  \BibitemOpen
  \bibfield  {author} {\bibinfo {author} {\bibfnamefont {T.}~\bibnamefont
  {{Okamoto}}}\ and\ \bibinfo {author} {\bibfnamefont {W.}~\bibnamefont
  {{Hu}}},\ }\href {\doibase 10.1103/PhysRevD.67.083002} {\bibfield  {journal}
  {\bibinfo  {journal} {\prd}\ }\textbf {\bibinfo {volume} {67}},\ \bibinfo
  {eid} {083002} (\bibinfo {year} {2003})},\ \Eprint
  {http://arxiv.org/abs/astro-ph/0301031} {astro-ph/0301031} \BibitemShut
  {NoStop}%
\bibitem [{\citenamefont {{Benson}}\ \emph {et~al.}(2014)\citenamefont
  {{Benson}} \emph {et~al.}}]{2014SPIE.9153E..1PB}%
  \BibitemOpen
  \bibfield  {author} {\bibinfo {author} {\bibfnamefont {B.~A.}\ \bibnamefont
  {{Benson}}} \emph {et~al.},\ }in\ \href {\doibase 10.1117/12.2057305} {\emph
  {\bibinfo {booktitle} {Millimeter, Submillimeter, and Far-Infrared Detectors
  and Instrumentation for Astronomy VII}}},\ \bibinfo {series} {\procspie},
  Vol.\ \bibinfo {volume} {9153}\ (\bibinfo {year} {2014})\ p.\ \bibinfo
  {pages} {91531P},\ \Eprint {http://arxiv.org/abs/1407.2973} {arXiv:1407.2973
  [astro-ph.IM]} \BibitemShut {NoStop}%
\bibitem [{\citenamefont {{Austermann}}\ \emph {et~al.}(2012)\citenamefont
  {{Austermann}} \emph {et~al.}}]{sptpol}%
  \BibitemOpen
  \bibfield  {author} {\bibinfo {author} {\bibfnamefont {J.~E.}\ \bibnamefont
  {{Austermann}}} \emph {et~al.},\ }in\ \href {\doibase 10.1117/12.927286}
  {\emph {\bibinfo {booktitle} {Millimeter, Submillimeter, and Far-Infrared
  Detectors and Instrumentation for Astronomy VI}}},\ \bibinfo {series}
  {\procspie}, Vol.\ \bibinfo {volume} {8452}\ (\bibinfo {year} {2012})\ p.\
  \bibinfo {pages} {84521E},\ \Eprint {http://arxiv.org/abs/1210.4970}
  {arXiv:1210.4970 [astro-ph.IM]} \BibitemShut {NoStop}%
\bibitem [{\citenamefont {{Abazajian}}\ \emph {et~al.}(2016)\citenamefont
  {{Abazajian}} \emph {et~al.}}]{cmbs4}%
  \BibitemOpen
  \bibfield  {author} {\bibinfo {author} {\bibfnamefont {K.~N.}\ \bibnamefont
  {{Abazajian}}} \emph {et~al.} (\bibinfo {collaboration} {CMB-S4
  Collaboration}),\ }\href@noop {} {\bibfield  {journal} {\bibinfo  {journal}
  {ArXiv e-prints}\ } (\bibinfo {year} {2016})},\ \Eprint
  {http://arxiv.org/abs/1610.02743} {arXiv:1610.02743} \BibitemShut {NoStop}%
\bibitem [{\citenamefont {{Abell}}\ \emph {et~al.}(2009)\citenamefont {{Abell}}
  \emph {et~al.}}]{lsstsciencebook}%
  \BibitemOpen
  \bibfield  {author} {\bibinfo {author} {\bibfnamefont {P.~A.}\ \bibnamefont
  {{Abell}}} \emph {et~al.} (\bibinfo {collaboration} {LSST Science
  Collaboration}),\ }\href@noop {} {\bibfield  {journal} {\bibinfo  {journal}
  {ArXiv e-prints}\ } (\bibinfo {year} {2009})},\ \Eprint
  {http://arxiv.org/abs/0912.0201} {arXiv:0912.0201 [astro-ph.IM]} \BibitemShut
  {NoStop}%
\bibitem [{\citenamefont {{Bleem}}\ \emph {et~al.}(2012)\citenamefont {{Bleem}}
  \emph {et~al.}}]{2012ApJ...753L...9B}%
  \BibitemOpen
  \bibfield  {author} {\bibinfo {author} {\bibfnamefont {L.~E.}\ \bibnamefont
  {{Bleem}}} \emph {et~al.},\ }\href {\doibase 10.1088/2041-8205/753/1/L9}
  {\bibfield  {journal} {\bibinfo  {journal} {\apjl}\ }\textbf {\bibinfo
  {volume} {753}},\ \bibinfo {eid} {L9} (\bibinfo {year} {2012})},\ \Eprint
  {http://arxiv.org/abs/1203.4808} {arXiv:1203.4808 [astro-ph.CO]} \BibitemShut
  {NoStop}%
\bibitem [{\citenamefont {{Limber}}(1953)}]{1953ApJ...117..134L}%
  \BibitemOpen
  \bibfield  {author} {\bibinfo {author} {\bibfnamefont {D.~N.}\ \bibnamefont
  {{Limber}}},\ }\href {\doibase 10.1086/145672} {\bibfield  {journal}
  {\bibinfo  {journal} {\apj}\ }\textbf {\bibinfo {volume} {117}},\ \bibinfo
  {pages} {134} (\bibinfo {year} {1953})}\BibitemShut {NoStop}%
\bibitem [{\citenamefont {{Kaiser}}(1992)}]{1992ApJ...388..272K}%
  \BibitemOpen
  \bibfield  {author} {\bibinfo {author} {\bibfnamefont {N.}~\bibnamefont
  {{Kaiser}}},\ }\href {\doibase 10.1086/171151} {\bibfield  {journal}
  {\bibinfo  {journal} {\apj}\ }\textbf {\bibinfo {volume} {388}},\ \bibinfo
  {pages} {272} (\bibinfo {year} {1992})}\BibitemShut {NoStop}%
\bibitem [{\citenamefont {{Ade}}\ \emph {et~al.}(2016)\citenamefont {{Ade}}
  \emph {et~al.}}]{planck15}%
  \BibitemOpen
  \bibfield  {author} {\bibinfo {author} {\bibfnamefont {P.~A.~R.}\
  \bibnamefont {{Ade}}} \emph {et~al.} (\bibinfo {collaboration} {Planck
  Collaboration}),\ }\href {\doibase 10.1051/0004-6361/201525830} {\bibfield
  {journal} {\bibinfo  {journal} {\aap}\ }\textbf {\bibinfo {volume} {594}},\
  \bibinfo {eid} {A13} (\bibinfo {year} {2016})},\ \Eprint
  {http://arxiv.org/abs/1502.01587} {arXiv:1502.01587} \BibitemShut {NoStop}%
\bibitem [{\citenamefont {{Howlett}}\ \emph {et~al.}(2012)\citenamefont
  {{Howlett}}, \citenamefont {{Lewis}}, \citenamefont {{Hall}},\ and\
  \citenamefont {{Challinor}}}]{2012JCAP...04..027H}%
  \BibitemOpen
  \bibfield  {author} {\bibinfo {author} {\bibfnamefont {C.}~\bibnamefont
  {{Howlett}}}, \bibinfo {author} {\bibfnamefont {A.}~\bibnamefont {{Lewis}}},
  \bibinfo {author} {\bibfnamefont {A.}~\bibnamefont {{Hall}}}, \ and\ \bibinfo
  {author} {\bibfnamefont {A.}~\bibnamefont {{Challinor}}},\ }\href {\doibase
  10.1088/1475-7516/2012/04/027} {\bibfield  {journal} {\bibinfo  {journal}
  {\jcap}\ }\textbf {\bibinfo {volume} {4}},\ \bibinfo {eid} {027} (\bibinfo
  {year} {2012})},\ \Eprint {http://arxiv.org/abs/1201.3654} {arXiv:1201.3654
  [astro-ph.CO]} \BibitemShut {NoStop}%
\bibitem [{\citenamefont {{Lewis}}\ \emph {et~al.}(2000)\citenamefont
  {{Lewis}}, \citenamefont {{Challinor}},\ and\ \citenamefont
  {{Lasenby}}}]{2000ApJ...538..473L}%
  \BibitemOpen
  \bibfield  {author} {\bibinfo {author} {\bibfnamefont {A.}~\bibnamefont
  {{Lewis}}}, \bibinfo {author} {\bibfnamefont {A.}~\bibnamefont
  {{Challinor}}}, \ and\ \bibinfo {author} {\bibfnamefont {A.}~\bibnamefont
  {{Lasenby}}},\ }\href {\doibase 10.1086/309179} {\bibfield  {journal}
  {\bibinfo  {journal} {\apj}\ }\textbf {\bibinfo {volume} {538}},\ \bibinfo
  {pages} {473} (\bibinfo {year} {2000})},\ \Eprint
  {http://arxiv.org/abs/astro-ph/9911177} {astro-ph/9911177} \BibitemShut
  {NoStop}%
\bibitem [{\citenamefont {{Smith}}\ \emph {et~al.}(2003)\citenamefont
  {{Smith}}, \citenamefont {{Peacock}}, \citenamefont {{Jenkins}},
  \citenamefont {{White}}, \citenamefont {{Frenk}}, \citenamefont {{Pearce}},
  \citenamefont {{Thomas}}, \citenamefont {{Efstathiou}},\ and\ \citenamefont
  {{Couchman}}}]{2003MNRAS.341.1311S}%
  \BibitemOpen
  \bibfield  {author} {\bibinfo {author} {\bibfnamefont {R.~E.}\ \bibnamefont
  {{Smith}}}, \bibinfo {author} {\bibfnamefont {J.~A.}\ \bibnamefont
  {{Peacock}}}, \bibinfo {author} {\bibfnamefont {A.}~\bibnamefont
  {{Jenkins}}}, \bibinfo {author} {\bibfnamefont {S.~D.~M.}\ \bibnamefont
  {{White}}}, \bibinfo {author} {\bibfnamefont {C.~S.}\ \bibnamefont
  {{Frenk}}}, \bibinfo {author} {\bibfnamefont {F.~R.}\ \bibnamefont
  {{Pearce}}}, \bibinfo {author} {\bibfnamefont {P.~A.}\ \bibnamefont
  {{Thomas}}}, \bibinfo {author} {\bibfnamefont {G.}~\bibnamefont
  {{Efstathiou}}}, \ and\ \bibinfo {author} {\bibfnamefont {H.~M.~P.}\
  \bibnamefont {{Couchman}}},\ }\href {\doibase
  10.1046/j.1365-8711.2003.06503.x} {\bibfield  {journal} {\bibinfo  {journal}
  {\mnras}\ }\textbf {\bibinfo {volume} {341}},\ \bibinfo {pages} {1311}
  (\bibinfo {year} {2003})},\ \Eprint {http://arxiv.org/abs/astro-ph/0207664}
  {astro-ph/0207664} \BibitemShut {NoStop}%
\bibitem [{\citenamefont {{Krause}}\ and\ \citenamefont
  {{Eifler}}(2017)}]{krauseeifler}%
  \BibitemOpen
  \bibfield  {author} {\bibinfo {author} {\bibfnamefont {E.}~\bibnamefont
  {{Krause}}}\ and\ \bibinfo {author} {\bibfnamefont {T.}~\bibnamefont
  {{Eifler}}},\ }\href {\doibase 10.1093/mnras/stx1261} {\bibfield  {journal}
  {\bibinfo  {journal} {\mnras}\ }\textbf {\bibinfo {volume} {470}},\ \bibinfo
  {pages} {2100} (\bibinfo {year} {2017})},\ \Eprint
  {http://arxiv.org/abs/1601.05779} {arXiv:1601.05779 [astro-ph.CO]}
  \BibitemShut {NoStop}%
\bibitem [{\citenamefont {{Motloch}}\ \emph {et~al.}(2017)\citenamefont
  {{Motloch}}, \citenamefont {{Hu}},\ and\ \citenamefont
  {{Benoit-L{\'e}vy}}}]{motlochhulevy}%
  \BibitemOpen
  \bibfield  {author} {\bibinfo {author} {\bibfnamefont {P.}~\bibnamefont
  {{Motloch}}}, \bibinfo {author} {\bibfnamefont {W.}~\bibnamefont {{Hu}}}, \
  and\ \bibinfo {author} {\bibfnamefont {A.}~\bibnamefont
  {{Benoit-L{\'e}vy}}},\ }\href {\doibase 10.1103/PhysRevD.95.043518}
  {\bibfield  {journal} {\bibinfo  {journal} {\prd}\ }\textbf {\bibinfo
  {volume} {95}},\ \bibinfo {eid} {043518} (\bibinfo {year} {2017})},\ \Eprint
  {http://arxiv.org/abs/1612.05637} {arXiv:1612.05637 [astro-ph.CO]}
  \BibitemShut {NoStop}%
\bibitem [{\citenamefont {{Crocce}}\ \emph {et~al.}(2016)\citenamefont
  {{Crocce}} \emph {et~al.}}]{2016MNRAS.455.4301C}%
  \BibitemOpen
  \bibfield  {author} {\bibinfo {author} {\bibfnamefont {M.}~\bibnamefont
  {{Crocce}}} \emph {et~al.},\ }\href {\doibase 10.1093/mnras/stv2590}
  {\bibfield  {journal} {\bibinfo  {journal} {\mnras}\ }\textbf {\bibinfo
  {volume} {455}},\ \bibinfo {pages} {4301} (\bibinfo {year} {2016})},\ \Eprint
  {http://arxiv.org/abs/1507.05360} {arXiv:1507.05360} \BibitemShut {NoStop}%
\bibitem [{\citenamefont {{Baxter}}\ \emph {et~al.}(2019)\citenamefont
  {{Baxter}} \emph {et~al.}}]{baxter18}%
  \BibitemOpen
  \bibfield  {author} {\bibinfo {author} {\bibfnamefont {E.~J.}\ \bibnamefont
  {{Baxter}}} \emph {et~al.},\ }\href {\doibase 10.1103/PhysRevD.99.023508}
  {\bibfield  {journal} {\bibinfo  {journal} {\prd}\ }\textbf {\bibinfo
  {volume} {99}},\ \bibinfo {eid} {023508} (\bibinfo {year} {2019})},\ \Eprint
  {http://arxiv.org/abs/1802.05257} {arXiv:1802.05257 [astro-ph.CO]}
  \BibitemShut {NoStop}%
\bibitem [{\citenamefont {{Mandelbaum}}\ \emph {et~al.}(2018)\citenamefont
  {{Mandelbaum}} \emph {et~al.}}]{newlsst}%
  \BibitemOpen
  \bibfield  {author} {\bibinfo {author} {\bibfnamefont {R.}~\bibnamefont
  {{Mandelbaum}}} \emph {et~al.} (\bibinfo {collaboration} {LSST Dark Energy
  Science Collaboration}),\ }\href@noop {} {\bibfield  {journal} {\bibinfo
  {journal} {ArXiv e-prints}\ } (\bibinfo {year} {2018})},\ \Eprint
  {http://arxiv.org/abs/1809.01669} {arXiv:1809.01669} \BibitemShut {NoStop}%
\bibitem [{\citenamefont {{Newman}}\ \emph {et~al.}(2015)\citenamefont
  {{Newman}} \emph {et~al.}}]{newman15}%
  \BibitemOpen
  \bibfield  {author} {\bibinfo {author} {\bibfnamefont {J.~A.}\ \bibnamefont
  {{Newman}}} \emph {et~al.},\ }\href {\doibase
  10.1016/j.astropartphys.2014.06.007} {\bibfield  {journal} {\bibinfo
  {journal} {Astroparticle Physics}\ }\textbf {\bibinfo {volume} {63}},\
  \bibinfo {pages} {81} (\bibinfo {year} {2015})},\ \Eprint
  {http://arxiv.org/abs/1309.5384} {arXiv:1309.5384} \BibitemShut {NoStop}%
\bibitem [{\citenamefont {{Rozo}}\ \emph {et~al.}(2016)\citenamefont {{Rozo}}
  \emph {et~al.}}]{rozo16}%
  \BibitemOpen
  \bibfield  {author} {\bibinfo {author} {\bibfnamefont {E.}~\bibnamefont
  {{Rozo}}} \emph {et~al.},\ }\href {\doibase 10.1093/mnras/stw1281} {\bibfield
   {journal} {\bibinfo  {journal} {\mnras}\ }\textbf {\bibinfo {volume}
  {461}},\ \bibinfo {pages} {1431} (\bibinfo {year} {2016})},\ \Eprint
  {http://arxiv.org/abs/1507.05460} {arXiv:1507.05460 [astro-ph.IM]}
  \BibitemShut {NoStop}%
\bibitem [{\citenamefont {{Tanoglidis}}\ \emph {et~al.}(2020)\citenamefont
  {{Tanoglidis}}, \citenamefont {{Chang}},\ and\ \citenamefont
  {{Frieman}}}]{tanoglidis}%
  \BibitemOpen
  \bibfield  {author} {\bibinfo {author} {\bibfnamefont {D.}~\bibnamefont
  {{Tanoglidis}}}, \bibinfo {author} {\bibfnamefont {C.}~\bibnamefont
  {{Chang}}}, \ and\ \bibinfo {author} {\bibfnamefont {J.}~\bibnamefont
  {{Frieman}}},\ }\href {\doibase 10.1093/mnras/stz3281} {\bibfield  {journal}
  {\bibinfo  {journal} {\mnras}\ }\textbf {\bibinfo {volume} {491}},\ \bibinfo
  {pages} {3535} (\bibinfo {year} {2020})},\ \Eprint
  {http://arxiv.org/abs/1908.07150} {arXiv:1908.07150 [astro-ph.CO]}
  \BibitemShut {NoStop}%
\bibitem [{\citenamefont {{Dvali}}\ \emph {et~al.}(2000)\citenamefont
  {{Dvali}}, \citenamefont {{Gabadadze}},\ and\ \citenamefont
  {{Porrati}}}]{dgp}%
  \BibitemOpen
  \bibfield  {author} {\bibinfo {author} {\bibfnamefont {G.}~\bibnamefont
  {{Dvali}}}, \bibinfo {author} {\bibfnamefont {G.}~\bibnamefont
  {{Gabadadze}}}, \ and\ \bibinfo {author} {\bibfnamefont {M.}~\bibnamefont
  {{Porrati}}},\ }\href {\doibase 10.1016/S0370-2693(00)00669-9} {\bibfield
  {journal} {\bibinfo  {journal} {Physics Letters B}\ }\textbf {\bibinfo
  {volume} {485}},\ \bibinfo {pages} {208} (\bibinfo {year} {2000})},\ \Eprint
  {http://arxiv.org/abs/hep-th/0005016} {hep-th/0005016} \BibitemShut {NoStop}%
\bibitem [{\citenamefont {{Hearin}}\ \emph {et~al.}(2010)\citenamefont
  {{Hearin}}, \citenamefont {{Zentner}}, \citenamefont {{Ma}},\ and\
  \citenamefont {{Huterer}}}]{hearin2010}%
  \BibitemOpen
  \bibfield  {author} {\bibinfo {author} {\bibfnamefont {A.~P.}\ \bibnamefont
  {{Hearin}}}, \bibinfo {author} {\bibfnamefont {A.~R.}\ \bibnamefont
  {{Zentner}}}, \bibinfo {author} {\bibfnamefont {Z.}~\bibnamefont {{Ma}}}, \
  and\ \bibinfo {author} {\bibfnamefont {D.}~\bibnamefont {{Huterer}}},\ }\href
  {\doibase 10.1088/0004-637X/720/2/1351} {\bibfield  {journal} {\bibinfo
  {journal} {\apj}\ }\textbf {\bibinfo {volume} {720}},\ \bibinfo {pages}
  {1351} (\bibinfo {year} {2010})},\ \Eprint {http://arxiv.org/abs/1002.3383}
  {arXiv:1002.3383} \BibitemShut {NoStop}%
\bibitem [{\citenamefont {{Huterer}}\ \emph {et~al.}(2013)\citenamefont
  {{Huterer}}, \citenamefont {{Cunha}},\ and\ \citenamefont
  {{Fang}}}]{huterer13}%
  \BibitemOpen
  \bibfield  {author} {\bibinfo {author} {\bibfnamefont {D.}~\bibnamefont
  {{Huterer}}}, \bibinfo {author} {\bibfnamefont {C.~E.}\ \bibnamefont
  {{Cunha}}}, \ and\ \bibinfo {author} {\bibfnamefont {W.}~\bibnamefont
  {{Fang}}},\ }\href {\doibase 10.1093/mnras/stt653} {\bibfield  {journal}
  {\bibinfo  {journal} {\mnras}\ }\textbf {\bibinfo {volume} {432}},\ \bibinfo
  {pages} {2945} (\bibinfo {year} {2013})},\ \Eprint
  {http://arxiv.org/abs/1211.1015} {arXiv:1211.1015} \BibitemShut {NoStop}%
\bibitem [{\citenamefont {{Crocce}}\ \emph {et~al.}(2011)\citenamefont
  {{Crocce}}, \citenamefont {{Cabr{\'e}}},\ and\ \citenamefont
  {{Gazta{\~n}aga}}}]{crocce11}%
  \BibitemOpen
  \bibfield  {author} {\bibinfo {author} {\bibfnamefont {M.}~\bibnamefont
  {{Crocce}}}, \bibinfo {author} {\bibfnamefont {A.}~\bibnamefont
  {{Cabr{\'e}}}}, \ and\ \bibinfo {author} {\bibfnamefont {E.}~\bibnamefont
  {{Gazta{\~n}aga}}},\ }\href {\doibase 10.1111/j.1365-2966.2011.18393.x}
  {\bibfield  {journal} {\bibinfo  {journal} {\mnras}\ }\textbf {\bibinfo
  {volume} {414}},\ \bibinfo {pages} {329} (\bibinfo {year} {2011})},\ \Eprint
  {http://arxiv.org/abs/1004.4640} {arXiv:1004.4640} \BibitemShut {NoStop}%
\end{thebibliography}%

\appendix
\section{Redshift Parameters For Each Bin}
\label{sec:appendix0}

Throughout this work (beginning in Section \ref{sec:fiducial}) we use a Gaussian redshift distribution in each redshift bin, with mean, $z_{0,i}$, and width, $\sigma_{\text{z},i}$. As described in Section \ref{sec:makingdndz}, to estimate realistic parameters for each Gaussian distribution, we start with the LSST and DES redshift distributions described in Section \ref{sec:datasets}, then apply a photometric redshift error of $\sigma_{\text{ph}}=0.05(1+z)$ and assign each galaxy a photometric redshift. We then bin the galaxies by this photo z in the range listed in Table \ref{table1}. Then, we estimate the mean and standard deviation of the resulting true redshift distribution of each bin. We use this mean and standard deviation as the Gaussian parameters $z_{0,i}$ and $\sigma_{\text{z},i}$ for each redshift bin. We list these parameters as well as the resulting galaxy density of each bin in Table \ref{table1}.


\begin{table}[h!]
\centering
 \caption{Redshift parameters for the assumed Gaussian redshift distributions of LSST and DES used throughout this work.}
 \label{table1}
 \begin{tabularx}{.5\textwidth}{|X|X|X|X|X|}
 \hline
 \multicolumn{5}{| c |}{LSST Redshift Parameters} \\
 \hline
 Bin No. & $z$ range & $z_0$ & $\sigma_z$ & $n(z)$ ($\text{arcmin}^{-2}$)  \\ 
 \hline
 1 & 0-0.25 & 0.207 & 0.0746 & 2.80 \\ 
 \hline
 2 & 0.25-0.5 & 0.401 & 0.0967 & 9.55  \\ 
 \hline
 3 & 0.5-0.75 & 0.631 & 0.107 & 11.6 \\ 
 \hline
 4 & 0.75-1.0 & 0.871 & 0.117 & 9.97 \\ 
 \hline
 5 & 1.0-1.5 & 1.221 & 0.179 & 12.8 \\ 
 \hline
 6 & 1.5-2.0 & 1.689 & 0.184 & 6.64 \\ 
 \hline
 7 & 2.0-2.5 & 2.178 & 0.216 & 2.14 \\ 
 \hline
 8 & 2.5-3.0 & 2.721 & 0.240 & 1.11 \\ 
 \hline
 9 & 3.0-3.5 & 3.213 & 0.251 & 0.781 \\ 
 \hline
 10 & 3.5-4.0 & 3.706 & 0.271 & 0.478 \\ 
 \hline
 11 & 4.0-5.5 & 4.434 & 0.446 & 0.512 \\ 
 \hline
 12 & 5.5-7.0 & 5.736 & 0.391 & 0.0523 \\ 
 \hline
 \multicolumn{5}{c}{} \\
 \hline
 \multicolumn{5}{| c |}{DES Redshift Parameters} \\
 \hline
 Bin No. & $z$ range & $z_0$ & $\sigma_z$ & $n(z)$ ($\text{arcmin}^{-2}$)  \\ 
 \hline
 1 & 0-0.25 & 0.199 & 0.0770 & 0.820 \\ 
 \hline
 2 & 0.25-0.5 & 0.403 & 0.0975 & 2.79  \\ 
 \hline
 3 & 0.5-0.75 & 0.632 & 0.105 & 3.40 \\ 
 \hline
 4 & 0.75-1.0 & 0.859 & 0.112 & 2.92 \\ 
 \hline
 5 & 1.0-1.5 & 1.145 & 0.155 & 2.07 \\ 
 \hline
\end{tabularx}
\end{table}

\section{Power Spectra Dependence on Parameters}
\label{sec:appendix}
\subsection{Fiducial Parameters}

\begin{figure*}
\begin{center}
\includegraphics[width=1.0 \textwidth]{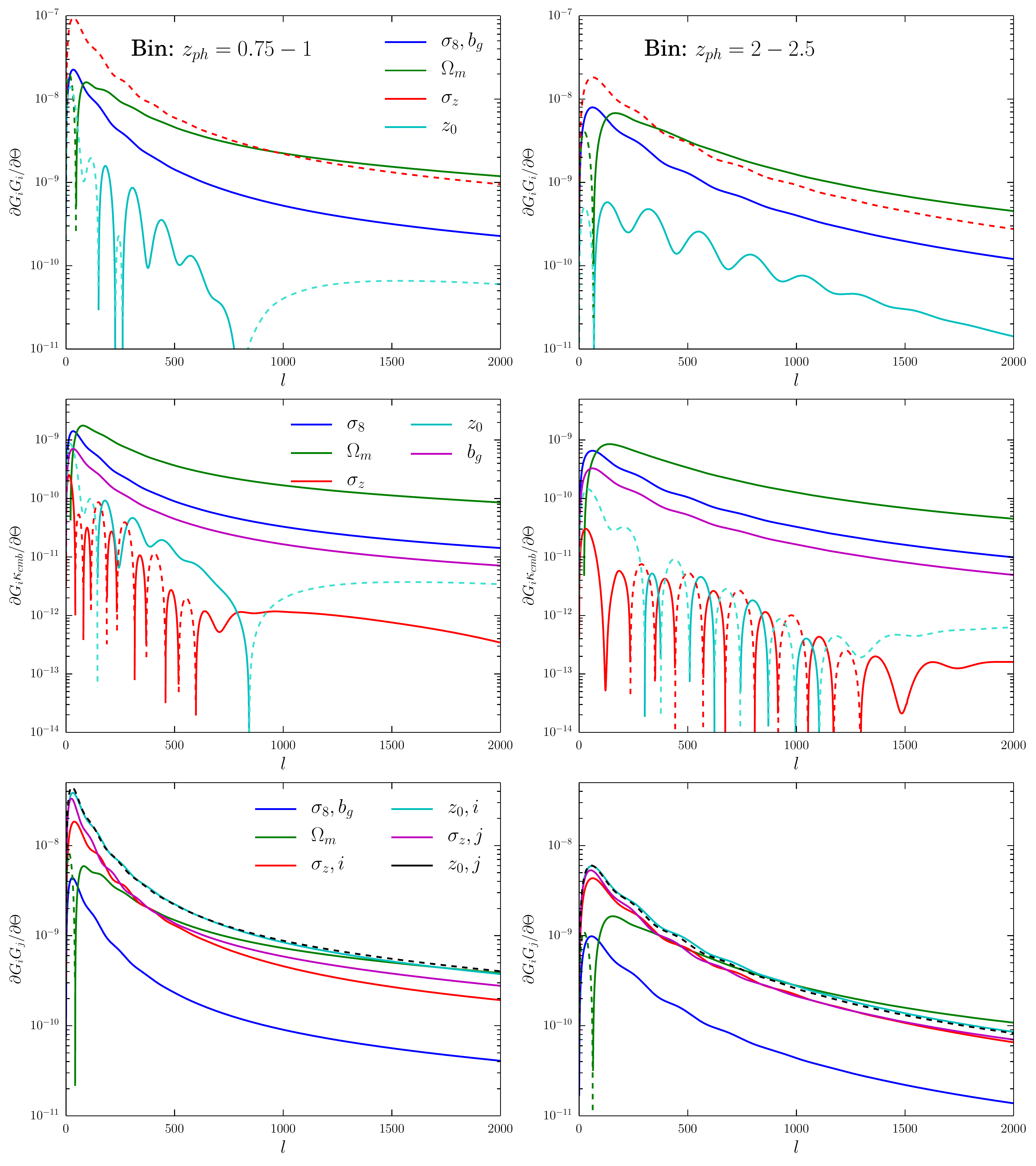}
\end{center}
\caption{Parameter dependence for the galaxy autopower spectra (top), galaxy-CMB lensing cross spectra (middle row), and the cross-spectra of two adjacent galaxy bins (bottom row). Shown are the spectra for the redshift bin $0.75<z_{\text{ph}}<1.0$ (left) and $2.0<z_{\text{ph}}<2.5$ (right). Dotted lines signify a negative correlation with the parameter and the spectra. For the galaxy cross-spectra, the listed bins are the $j$th bin and are cross-correlated with their neighboring lower redshift bins ($0.5<z<0.75$ and $1.5<z<2.0$, respectively) which are the $i$th bins in the bottom row.}
\label{fig:dggauto1}
\end{figure*}

To get a better intuition of which power spectra constrain which parameters, Figure \ref{fig:dggauto1} shows $d C_l/d \theta$ for the various combinations of spectra and parameters for the photometric redshift bins $0.75<z_{\text{ph}}<1.0$ and $2.0<z_{\text{ph}}<2.5$, with the redshift parameters listed in Table \ref{table1}. We show two redshift bins to broadly see trends of how dependence on different parameters changes with redshift. 

We can see for the galaxy autopower spectra (top row), which are also the highest S/N spectra, the parameters $b_{\text{g}}$ and $\sigma_8$ equivalently scale the spectra. We also see that increasing $b_{\text{g}}$ and $\sigma_{\text{z}}$ both directly scale the galaxy autopower spectra at all scales (in our modeling of no scale-dependent galaxy bias). Other than a normalization factor of the step sizes in the plot, for the galaxy autospectra, $b_{\text{g}}$, $\sigma_8$ and $\sigma_{\text{z}}$ are degenerate. Adding the galaxy-CMB lensing cross-spectra (middle row) can break the degeneracy of $b_{\text{g}}$ and $\sigma_8$, but has little dependence on $\sigma_{\text{z}}$. The cross-spectra of adjacent galaxy redshift bins (bottom row) have a large dependence on $\sigma_{\text{z}}$, in a way that is not degenerate with other parameters. These plots show that both the galaxy-CMB lensing cross-spectra and galaxy-galaxy cross-spectra are necessary to break the degeneracy between $b_{\text{g}}$, $\sigma_8$, and $\sigma_{\text{z}}$ that arises in the galaxy autospectra when incorporating redshift uncertainties. 

We can also see that the parameters $z_0$ and $\Omega_{\text{m}}$ are largely not degenerate with other parameters in the galaxy autospectra (top row). For this reason, constraints on these parameters are less correlated with, e.g., $\sigma_8$ improvements (Figure \ref{fig:s4_z0}).

\subsection{Extra Parameters}
\label{sec:appendixextra}

\begin{figure*}
\begin{center}
\includegraphics[width=1.0 \textwidth]{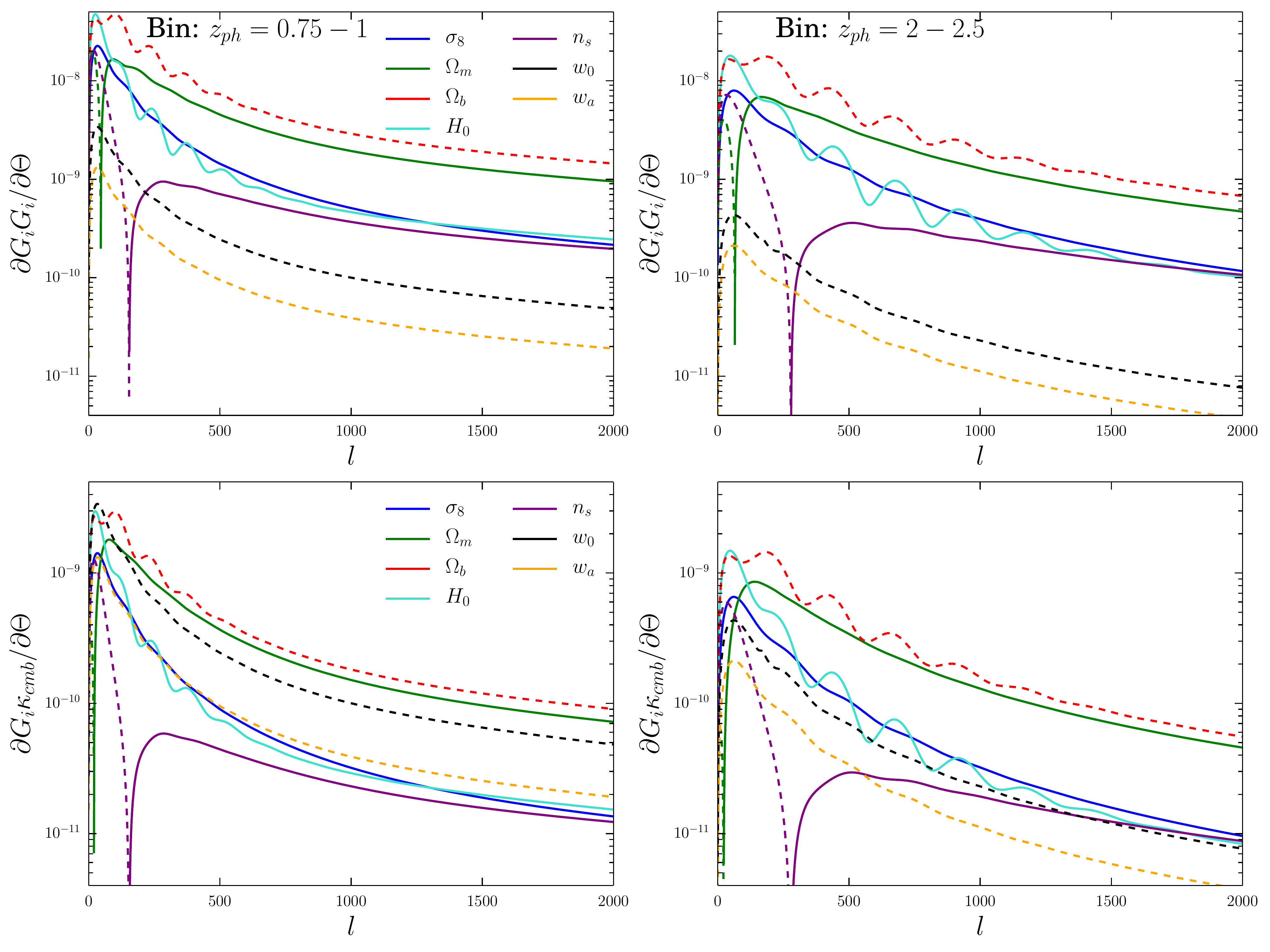}
\end{center}
\caption{Parameter dependence for the galaxy autopower spectra (top) and galaxy-CMB lensing cross spectra (bottom) of the redshift bins $0.75<z_{\text{ph}}<1.0$ (left) and $2.0<z_{\text{ph}}<2.5$ (right). Dotted lines signify a negative correlation with the parameter and the spectra. The cross-spectra of two adjacent galaxy bins (not shown) look very similar to the galaxy autospectra since the dependence on these parameters is from $P(k)$, thus changing each power spectra similarly. Although the $l$ dependence of $\sigma_8$ looks similar to the inverse $l$ dependence of $w_0$ and $w_{\text{a}}$, they are different, unlike, e.g., $\sigma_8$ and $\sigma_{\text{z}}$ in the autospectra of Figure \ref{fig:dggauto1}.}
\label{fig:dclextra}
\end{figure*}

In Figure \ref{fig:dclextra}, we show how the power spectra depend on the extra parameters used in Section \ref{sec:altmodels}, namely,  $w_0, w_{\text{a}}, H_0, \Omega_{\text{b}}$, and $n_{\text{s}}$, along with the fiducial parameters, $\sigma_8$ and $\Omega_{\text{m}}$. We again show $d C_l/d \theta$ for the various parameters in two of the redshift bins, $0.75<z_{\text{ph}}<1.0$ and $2.0<z_{\text{ph}}<2.5$.

We can see that the spectra do not depend on these parameters with the same (constant) scale dependence as they do with $\sigma_8$. Thus, these parameters are not degenerate with $b_{\text{g}}$, $\sigma_8$, and $\sigma_{\text{z}}$. This explains why the extra parameters in Section \ref{sec:altmodels} only minimally add to the projected uncertainty on $\sigma_8$. The similarity in how these parameters affect each of the power spectra in Figure \ref{fig:dclextra} is due to the fact that most of these parameters only affect $P(k)$, and all of the spectra have similar dependence on $P(k)$ (Equation \ref{clequation}).

\section{Impact of Low-$l$ Limit}
\label{sec:appendix2}

In this work, we use the Limber approximation (Equation \ref{clequation}) throughout for computational speed. However, it is known that the approximation breaks down at large scales (low $l$, e.g., \cite{huterer13}, \cite{crocce11}). \cite{SS17} uses the Limber approximation for only $l>50$. We repeated our fiducial analysis (12 bins, 49 parameters, no prior information, LSST/CMB-S4) using $l_{\text{min}}=50$ instead of 20. We found that our constraints on all parameters for both the $l_{\text{max}}=1000$ and 2000 cases degraded by $5 \%$ or less, with the exception of parameters in the three redshift bins separated by $z=[0.75,1,1.5,2]$ where much of the peak of the power spectra is in the cut-out range of $l=20-50$. In these bins, the constraints on $\sigma_8$ degraded by $[19 \%, 26 \%, 9 \%]$ and $[7 \%, 14 \%, 3 \%]$ for the $l_{\text{max}}=1000$ and 2000 cases, respectively. These numbers are thus an upper limit to how much our results may degrade due to the likely inaccurate use of the Limber approximation at $l=20-50$.

\end{document}